\documentclass[usenatbib]{mn2e}
\usepackage{graphicx}
\usepackage{times}
\usepackage{amsmath}
\usepackage{appendix}
\usepackage{deluxetable}
\usepackage{pdflscape}
\usepackage{xfrac}
\usepackage[letterpaper,margin=1.05in]{geometry}
\usepackage{amsfonts}
\usepackage{amssymb}
\usepackage{bm}
\usepackage{epstopdf}
\usepackage{ulem} 

\newcommand{\apj}{ApJ}
\newcommand{\apjl}{ApJ}

\newcommand{\aj}{AJ}

\newcommand{\aap}{A\&A}
\newcommand{\mnras}{MNRAS}

\newcommand{\abs}[1]{\ensuremath{\mid #1 \mid}}

\newcommand{\degree}{\ensuremath{^\circ}}

\newcommand{\muasperyr}  {\ensuremath{\mu\mbox{as\,yr}^{-1}}}

\newcommand{\kms}     {\ensuremath{\mbox{km\,s}^{-1}}}

\newcommand{\Lsun}   {\ensuremath{L_{\odot}}}
\newcommand{\Msun}   {\ensuremath{M_{\odot}}}

\newcommand{\half}{\sfrac{1}{2}}
\newcommand{\threehalves}{\sfrac{3}{2}}

\begin{document}
\title[Local Group Planes]{The Formation of Local Group Planes of Galaxies\\}

\author[E.J. Shaya and R.B. Tully]
{ Ed J. Shaya$^{1}$\thanks{E-mail: eshaya@umd.edu} 
and R. Brent Tully$^{2}$ \\
$^{1}$Department of Astronomy, University of Maryland, College Park, MD 20742\\
$^{2}$Institute for Astronomy, University of Hawaii, Honolulu, Hawaii 96822\\}

\maketitle

\begin{abstract}
The confinement of most satellite galaxies in the Local Group to thin planes presents a challenge to the theory of hierarchical galaxy clustering.  The PAndAS collaboration has identified a particularly thin configuration with kinematic coherence among companions of M\,31 and there have been long standing claims that the dwarf companions to the Milky Way lie in a plane roughly orthogonal to the disk of our galaxy.  This discussion investigates the possible origins of four Local Group planes: the plane similar, but not identical to that identified by the PAndAS collaboration, an adjacent slightly tilted plane, and two planes in the vicinity of the Milky Way: one with very nearby galaxies and the other with more distant ones.  Plausible orbits are found by using a combination of Numerical Action methods and a backward in time integration procedure.  This investigation assumes that the companion galaxies formed at an early time in accordance with the standard cosmological model.  For M\,31, M\,33, IC\,10, and Leo\,I,  solutions are found that are consistent with measurements of their proper motions.  For galaxies in planes, there must be commonalities in their proper motions, and this constraint greatly limits the number of physically plausible solutions.  Key to the formation of the planar structures has been the evacuation of the Local Void and consequent build-up of the Local Sheet, a wall of this void.  Most of the M\,31 companion galaxies were born in early-forming filamentary or sheet-like substrata that chased M\,31 out of the void.  M\,31 is a moving target because of its attraction toward the Milky Way, and the result has been alignments stretched toward our galaxy.  In the case of the configuration around the Milky Way, it appears that our galaxy was in a three-way competition for companions with M\,31 and Centaurus A.  Only those within a modest band fell our way. The Milky Ways' attraction toward the Virgo Cluster resulted in alignment along the Milky Way$-$Virgo Cluster line.
 
\end{abstract}


\section{Introduction}
\label{sec:Introduction}
The surprising discovery that half of the satellites of the Andromeda Galaxy lie in a thin plane with ordered motion
 \citep{Ibata_etal2013, Conn_etal2013} introduces a new challenge to theories of galaxy formation.
In a Canada-France-Hawaii Telescope survey of 400 sq. deg around M\,31, called PAndAS, the number of known M\,31 satellite
 galaxies was increased to 34, of which 27 are dSph.
Fifteen of these satellites were found to lie in a thin plane $\sim 13$ kpc thick that stretches up to 600 kpc along the plane
 \citep{Ibata_etal2013}.
Sharing the plane are M\,31 and, beyond the PAndAS footprint,  LGS\,3, IC1613 \citep{Tully2013}, and our own Milky Way (MW).
Martin et al. (2013) have found yet another satellite in the plane from imaging with Pan-STARRS.  
The plane is seen nearly edge on to us, so there is little doubt about its validity.
Galaxies to the Galactic north of M\,31 are redshifted with respect to M\,31 and on the south side they are blueshifted.
The distribution of the dwarfs has a front to back asymmetry with the majority  in front of M\,31.

\citet{Tully2013} pointed out that all the M\,31 satellites at higher Galactic longitude than the PAndAS plane, including 
the spiral galaxy M\,33, lie in a second plane slightly tilted from the first.  
As it turns out, this plane has a roughly  similar  width and extent as the first plane.
He also noted that almost all galaxies in the adjacent Centaurus Group lie in two parallel planes.

It has long been known that most dwarf galaxies near the MW are aligned on a great circle in the
sky  that passes close to the Galactic Pole \citep{Kunkel_Demers_1976, LyndenBell1976, Metz_etal2007, 
Pawlowski_etal2012}.   
Altogether then,  we now find that 43 of the 50 Local Group (LG) satellites within 1.1 Mpc, lie within the 4 planar structures: 
the plane reported by Ibata et al., the adjacent plane noted by Tully, and two Milky Way Satellite planes that we delineate here.

These discoveries stress the standard theory of hierarchical clustering by Cold Dark Matter (CDM).  
It was already appreciated that CDM theory predicts that there should be more satellite galaxies than observed \citep{Klypin_etal1999, Moore_etal1999}.  
Attempts to reconcile this include assuming that gas was lost from low mass halos due to supernovae driven winds, or that
collapse of gas into low mass halos was suppressed by reionization heating, 
rendering most of them invisible.  
Now it appears, that the few visible dwarf galaxies lie in distinctive structures so thin that their formation may have required dissipational processes.

It has been suggested that the thinness of the plane of Andromeda satellites could be explained by a tidal tail from a merger with M\,31 (Hammer et al 2013). 
In this scenario, a medium sized galaxy merged with M\,31 about 9 Gyr ago.   
The gas and dust from one of the tidal tails dissipated into a thin disk, and then fragmented into dSph galaxies.
However if this is a common means of forming dSphs it is difficult to understand the typically elevated velocity dispersions in dSphs
 which have been taken to indicate high mass-to-light ratios and thus significant amounts of dark matter. 
The velocity dispersion of the dark matter in a galaxy is too high to remain bound to a small lopped off piece of it.  
Yet another difficulty with the idea of dSph originating in tidal tails is explaining why the luminosity-metallicity relation
 continues into the dSphs luminosity range (Cheng et al. 2012).  If they are remnants of larger galaxies, then their metallicities should be more akin to those of more luminous galaxies.

However, cold dark matter simulations have not been totally stumped by highly anisotropic distributions of dwarf galaxies in groups.  To understand the odd planar distribution near the MW \citet{Libeskind_etal2011}  conducted constrained N-body simulations that formed, at high resolution, a group of galaxies with properties matching the Local Group.  
At lower resolution, the simulation extended 64 $h^{-1}$ Mpc  and included external structures like the Virgo Cluster  and Coma Cluster.  
For each of the MW and M\,31 analogues, there was a stream of matter out to 5 virial radii falling toward the galaxy and the direction of emanation was close to the Virgo Cluster.  
This matches the observed MW anisotropy of dwarf galaxies because the direction to the Virgo Cluster happens to be only $12\degree$ from the Galactic Pole.

More generally, simulations including both dark matter particles and gaseous hydrodynamics \citep{Danovich_etal2012}   have found that it is typical for 70\% of the material influx into  $10^{12} \Msun$ haloes at z = 0 to come in along 3 dominant streams.
These streams are embedded in sheets which bring another 20\% of the influx. 
Unlike the larger scale filaments, these filaments are not necessarily at the intersections of sheets;
they are just within the sheets.
\citet{Goerdt_Burkert2013} find that when there are 3 or more sheets, two are likely to be at small angles from each other forming a plane extending on both sides of the halo.
Thus, simulations predict that most of the material falling into giant galaxies follow along thin streams and sheets, and anisotropic distributions of satellites are now expected within the standard dark matter scenario.
But questions remains as to how galaxies that have fallen through the potential of the giant galaxy, some repeatedly,
remain in these coherent planes. 
If it is even slightly out of the plane, a dwarf passing a giant galaxy would typically be thrown well out of the plane. 
Unless all of the dwarf galaxies in planes are falling in for the first time, there remains an issue of maintaining them in thin planes.
Even more puzzling, the second plane near M\,31 is 100 kpc offset from the M\,31 and the timescale for buckling in toward M\,31 would only by a few $\times$ 100 Myr.  
It is hard to see how this plane can be interpreted as one of  the persistent filaments of dark matter formed in N-body simulations.

These discoveries come upon us as we are immersed in a program to study the orbits of Local Group galaxies using Numerical Action
 Methods (NAM).
The NAM technique solves for the orbits of many bodies under mutual gravitational attraction in a cosmological context and makes use of  mixed boundary value constraints.  
It combines present day positions with early time constraints on the peculiar velocities to attain the necessary number of phase space constraints.
Mutual trajectories are sought that have peculiar velocities consistent with linear perturbation theory at a chosen early epoch  (see the Appendix for more details).    
The NAM technique, also know as Least Action, has been used to calculate the orbits of:  the Local Group  \citep{Peebles1989, Peebles1990, Peebles1994, Peebles1995, Peebles_etal2001, PeeblesTully2013}, galaxies out to 2,000 \kms\ \citep{Sharpe_etal2001}, and
galaxies out to 3,000 \kms\ \citep{Shaya_etal1995}.  It has been shown to reconstruct the orbits of halos in N-body simulations in reliability tests  \citep{ BranchiniCarlberg1994, NusserBranchini2000, Branchini_etal2002, Phelps_etal2006}, although it is noted that on scales of a group,  a confusion from multiple solutions can arise that may require additional constraints like proper motions to resolve.

We focus here on the part of the model that includes the Milky Way and other galaxies within about 1.5 Mpc.  
The arrangement of galaxies into planar structures however means that there must be similarities in proper motions among members of a
 plane and that could provide the additional constraints needed to eliminate many of the possible solutions.
The existence of planes implies commonalities in velocity directions, giving essential clues for reconstructing the formation history of
 the Local Group.
If we discover how to make use of these convoluted constraints, we could reconstruct the development of the planes and the present
 configuration of the LG.
If the effort fails, it could point to the need for additional physics that has been left out of the standard cosmological picture. 

Despite the daunting complexity of the task,  we pushed forth in search of simple dynamical orbits that fit the excellent new data and
 were consistent with a natural explanation for these thin planes.
Since there were many parameters that could potentially be varied to get a fit (distances could vary by 5\%, masses could vary by
 large factors, and redshifts could vary by several \kms\ to allow for sloshing between the dark matter and the baryons), we were
 worried that a good fit might result because of clever adjustments. Therefore we chose to vary essentially nothing in this first attempt.
   Some freedom comes from the selection of one orbit from the handful of acceptable solutions.
Our motivation, after all, is not necessarily to get excellent fits, but to get ideas on what physics could have played a role in
developing these thin planes and to provide guidelines for initial density distributions for future constrained N-body simulations of the LG. 

On the observational side, there are now good constraints from accurate distances, velocities, and now some relevant proper motions (M\,31, M\,33, IC\,10, and Leo\,I) for the Local Group.   
The dwarfs Carina , Fornax , Sculptor, Leo\,II and Ursa Minor also have proper motion measurements, but  their uncertainties at this time are too large to usefully constrain the multiple orbits.
The LMC and SMC also have proper motion measurements, and we briefly discuss our skepticism about finding definitive trajectories
for them.

With our orbit reconstruction effort, it isl important that we incorporate the tidal influences of structures at larger scales. 
In addition to the familiar components like nearby large galaxies, the Virgo Cluster \citep{Aaronson_etal1982}, and the distant but massive Great Attractor region \citep{Shaya1984, Dressler_etal1987}, there is the {\it absence} of pull from the Local Void \citep{TullyFisher1987}. 
These large scale components, although not directly the topic of this article,  turn out to be necessary support players in the development of the planes, therefore the models at larger scales are briefly presented.

In \S \ref{sec:Data} we briefly discuss the data on LG galaxies that we took both from the literature and from our own large projects to obtain high precision distances.  
In \S \ref{sec:Methodology} we present our new hybrid numerical methodology which combines NAM for giant galaxies with a new "backtracking" technique for dwarf galaxies that iterates backwards in time from the present positions. 
 In \S \ref{sec:Results} we show solutions for the MW/M\,31/M\,33 orbits extracted from our large scale solution that includes about 180 galaxies/groups/clusters out to $\sim$ 50 Mpc. 
Then, we use backtracking to add the orbits of the LG dwarfs by finding which directions of the their present day velocities are consistent with linear perturbation theory and with residing in its plane, if it is in a plane.
After selecting the one `best' velocity direction for each galaxy, we look for properties of the z=4 distribution that contributed to the formation of the observed planes.
\S \ref{sec:Discussion} provides an overview of the different planes and describes what possible physical processes were involved in the plane formations.
\S \ref{sec:Conclusions} reviews the results and discusses future tests.
The Appendix presents mathematical details of the current implementation of NAM and the backtracking routine.

\section{Data}
 \label{sec:Data}

 To begin dynamical analyses of the LG, we need good distance and redshift measurements to constrain the present position-velocity vectors.  
We also use total luminosities to infer masses, but accuracy in luminosity is needed only for the three most massive galaxies because they dominate the gravitational potential so completely. 
Since the Local Group contains the closest galaxies to us, there is a wealth of excellent data on most of its members that includes multiple, highly reliable distance determinations and redshifts based on well resolved stars.  
The newly discovered dwarf galaxies near M\,31 are the worst cases in that they each only have one Tip of the Red Giant Branch (TRGB) measurement with distance errors in the range of 5\% \citep{Conn_etal2012}. 
In Table\,\ref{table:input}, we present the input data for each galaxy in the Local Group.  This includes distance, K-band luminosity, position, and recessional velocity.  
It also provides which of 4 LG planes it resides in,  numbered 1 - 4, and 0 means it is not in one of the planes.
K-band luminosities are from the 2MASS All-Sky Extended Source Catalog  \citep{XSC}
or 2MASS Large Galaxy Atlas \citep{LGA}, or in the case of dwarfs, from B to Ks color
transfers as given by \citet{Karachentsev_etal2013}.  Distances are collected from a multitude of
authors.  The methods include the Cepheid Period-Luminosity relation \citep{Freedman_etal2001},
the related RR Lyrae relation \citep{Pietrzynski_etal2009}, the Horizontal Branch luminosity
\citep{Richardson_etal2011}, Eclipsing Binaries \citep{Bonanos_etal2006}, and the standard candle of the TRGB \citep{Dolphin_etal2002, Dolphin_etal2003, Dalcanton_etal2009}.  The large number
of new dwarfs near M\,31 have homogeneously derived TRGB distances based on a Bayesian statistical
analysis of the color-magnitude diagram \citep{Conn_etal2012}.  Other distances are computed from a TRGB maximum likelihood
technique \citep{Makarov_etal2006, Jacobs_etal2009}.

Measurement of proper motions of nearby galaxies has recently seen its dawn and is a great boon to dynamical analyses. 
In \citet{Brunthaler_etal2005}, the proper motion of two H$_2$0 masers of M\,33 provided the distance and proper motion of the galaxy.  
The error in proper motion with a baseline of just 3 years was 7.3 $\muasperyr$.  
M\,33 appears to be moving upward in supergalactic latitude, SGB, toward M\,31.  
Very Long Baseline Interferometer (VLBI) measurements of an H$_2$0 maser have also provided a proper motion of IC\,10 \citep{Brunthaler_etal2007} with errors of $ \sim9 \muasperyr$ in each component or a total error of 42 \kms.
The resulting total velocity is upward in SGZ and away from  M\,31. 
A Hubble Space Telescope (HST) measurement of the proper motion of M\,31 \citep{vdM_etal2012} is based on observations of 3 regions within the galaxy  
with $\sim$3 year baselines. 
Resultant errors are 12 $\muasperyr$ in each orthogonal component. 
The transverse velocity of M\,31 is found to be 33 $\pm$ 34 \kms, assuming the orbital velocity of the local standard of rest  is 240 \kms,  and indicates motion almost directly toward the MW.

\begin{onecolumn}
\begin{deluxetable}{lrrrrrrr}
\tablewidth{0pt}
\tabletypesize{\scriptsize}

\tablecaption{Input Data for Local Group Galaxies  \label{table:input}}
\tablehead{
\colhead{Galaxy} & 
\colhead{SGL\tablenotemark{a}\tablenotetext{a}{Supergalactic Longitude}} & 
\colhead{SGB\tablenotemark{b}\tablenotetext{b}{Supergalactic Latitude}}& 
\colhead{Distance} & 
\colhead{$cz_{helio}$\tablenotemark{c}\tablenotetext{c}{Observed redshift in heliocentric frame}} & \colhead{$CZ_{MW}$\tablenotemark{d}\tablenotetext{d}{Observed redshift in MW frame}} & \colhead{Luminosity\tablenotemark{e}\tablenotetext{e}{K-band luminosity from 2MASS}}& 
\colhead{Plane}\\

\colhead{$\cdots$} & \colhead{deg} & \colhead{deg} &\colhead{Mpc} &  \colhead{$\kms$} & \colhead{$\kms$} & \colhead{$\Lsun$(K-band)} &  \colhead{$\cdots$} 
}
\startdata
MW &185.7861 & 42.3103 & 0.01 &  -11 &    0 & 4.50E+10 & 3\\
UMi & 47.7130 & 27.1060 & 0.07 & -235 &  -54 & 2.45E+03 & 4\\
Draco & 43.7891 & 44.2344 & 0.08 & -276 &  -58 & 2.18E+04 & 4\\
Sculptor &263.9810 & -9.6841 & 0.08 &  108 &   71 & 8.97E+04 & 4\\
Sextans &105.3886 &-39.2323 & 0.09 &  227 &   56 & 7.82E+03 & 4\\
Carina &210.0974 &-54.6399 & 0.11 &  237 &   -3 & 3.01E+04 & 4\\
Fornax &265.3703 &-30.2764 & 0.15 &   49 &  -49 & 9.46E+05 & 4\\
Leo\,II & 87.1012 &-16.2696 & 0.22 &   79 &   17 & 3.48E+04 & 4\\
Leo\,I & 88.8989 &-34.5552 & 0.26 &  284 &  162 & 4.59E+05 & 3\\
Leo\,T & 78.4013 &-38.7519 & 0.41 &   35 &  -73 & 5.37E+05 & 3\\
Phoenix &254.2878 &-20.8615 & 0.41 &  -13 & -112 & 1.47E+05 & 4\\
And\,16 &328.0742 &  7.4170 & 0.48 & -367 & -192 & 1.41E+06 & 2\\
NGC\,6822 &229.0764 & 57.0883 & 0.51 &  -57 &   55 & 1.19E+08 & 3\\
And\,9 &338.4811 & 11.0613 & 0.60 & -209 &  -12 & 8.87E+05 & 2\\
And\,15 &334.4339 &  5.9373 & 0.62 & -323 & -146 & 1.84E+06 & 2\\
NGC\,185 &343.2679 & 14.3030 & 0.64 & -227 &  -18 & 2.13E+08 & 1\\
LGS\,3 &318.1259 &  3.8057 & 0.65 & -281 & -133 & 7.60E+03 & 1\\
And\,2 &330.0131 &  4.2393 & 0.66 & -194 &  -27 & 4.34E+06 & 2\\
And\,10 &340.6064 &  8.9981 & 0.67 & -164 &   27 & 1.33E+06 & 2\\
And\,30 &344.5582 & 14.8572 & 0.68 & -141 &   69 & 1.00E+05 & 1\\
And\,3 &331.1259 & 13.0998 & 0.73 & -344 & -149 & 4.47E+06 & 1\\
And\,11 &328.8525 & 10.4235 & 0.73 & -427 & -242 & 2.55E+05 & 1\\
And\,17 &339.1210 & 14.0581 & 0.73 & -251 &  -46 & 5.20E+05 & 1\\
Cetus\,dSph &283.8351 &  3.8215 & 0.73 &  -87 &  -19 & 1.57E+05 & 3\\
NGC\,147 &343.3205 & 15.2734 & 0.73 & -193 &   17 & 1.53E+08 & 1\\
And\,5 &343.5386 &  9.0233 & 0.74 & -397 & -203 & 3.40E+06 & 2\\
And\,20 &328.8762 & 18.5316 & 0.74 & -456 & -250 & 1.49E+05 & 0\\
And\,25 &341.5522 & 15.5946 & 0.74 & -108 &  102 & 3.05E+06 & 1\\
Leo\,A & 69.9139 &-25.7978 & 0.74 &   28 &  -18 & 7.87E+05 & 0\\
And\,1 &333.0550 & 11.4069 & 0.75 & -376 & -184 & 1.89E+07 & 1\\
And\,23 &335.8494 &  3.2797 & 0.75 & -243 &  -73 & 6.03E+06 & 2\\
And\,26 &342.5412 & 16.7777 & 0.75 & -261 &  -46 & 3.22E+06 & 1\\
IC\,1613 &299.1743 & -1.7997 & 0.75 & -238 & -149 & 5.70E+07 & 1\\
M\,31 &336.1930 & 12.5521 & 0.77 & -300 & -101 & 6.15E+10 & 1\\
Cas\,dSph &345.6484 & 26.0511 & 0.79 & -307 &  -71 & 4.25E+07 & 0\\
And\,14 &325.0418 &  8.4334 & 0.79 & -481 & -307 & 1.18E+06 & 1\\
IC\,10 &354.4176 & 17.8657 & 0.79 & -348 & -127 & 8.69E+08 & 1\tablenotemark{f}\\
PegdSph &317.3085 & 20.5577 & 0.80 & -341 & -147 & 1.62E+07 & 0\\
And\,19 &329.1129 & 16.0809 & 0.82 & -111 &   89 & 1.38E+06 & 0\\
And\,21 &336.4705 & 21.5466 & 0.83 & -363 & -142 & 4.25E+06 & 0\\
And\,22 &325.6071 &  0.3957 & 0.85 & -130 &   17 & 2.60E+05 & 2\\
M\,33 &328.4673 & -0.0899 & 0.86 & -180 &  -29 & 4.04E+09 & 2\\
And\,13 &328.3145 &  9.1275 & 0.86 & -185 &   -5 & 4.34E+05 & 1\\
And\,12 &329.4716 & 10.3170 & 0.91 & -556 & -371 & 3.44E+05 & 1\\
And\,24 &342.6520 &  7.3546 & 0.91 & -134 &   54 & 1.26E+06 & 2\\
Tucana &227.6081 & -0.9182 & 0.92 &  194 &   88 & 3.91E+04 & 0\\
PegDIG &305.8332 & 24.3091 & 0.95 & -178 &    3 & 9.14E+05 & 0\\
DDO\,210 &252.0792 & 50.2450 & 0.97 & -132 &   -3 & 9.23E+05 & 3\\
WLM &277.8076 &  8.0847 & 0.98 & -124 &  -59 & 2.67E+07 & 3\\
SagDIG &221.2695 & 55.5181 & 1.05 &  -80 &   17 & 9.27E+05 & 3\\
And\,18 &339.3224 & 20.2960 & 1.21 & -332 & -111 & 3.02E+06 & 0\\
And\,27 &340.2162 & 14.1524 & 1.25 & -535 & -329 & 1.69E+06 & 1\tablenotemark{f} \\
NGC\,3109 &137.9541 &-45.1035 & 1.33 &  403 &  168 & 3.58E+07 & 0\\
Antlia &139.5869 &-44.8032 & 1.35 &  361 &  124 & 8.75E+04 & 0\\
UGC\,4879 & 47.6113 &-15.0140 & 1.36 &  -27 &   20 & 3.55E+06 & 0\\
Sextans\,B & 95.4628 &-39.6199 & 1.43 &  302 &  152 & 2.70E+07 & 0\\
Sextans\,A &109.0063 &-40.6585 & 1.43 &  325 &  144 & 2.61E+07 & 0\\
\enddata
\tablenotetext{f}{IC\,10 and And\,27 were removed from \textbf{Plane\,1}  analysis because the modeling indicated they had to be moving through the plane quickly.  For IC\,10, the proper motion directly indicates that it is moving at high inclination to the plane.  The highly negative redshift of And\,27 has no solution if it  interacted with only  LG galaxies (see Figure~\ref{fig:And27} and accompanying text).}
\end{deluxetable}
\end{onecolumn}
\begin{twocolumn}
\section{Methodology}
\label{sec:Methodology}

\subsection{Numerical Action Method}
\label{sec:NAM}
	 
There are two approaches to an understanding of the dynamics that led the Local Group to its present configuration: 
1) run sufficient N-body simulations, with adaptive mesh refinement for precision on small scales,  until one finds in one of them a group that closely resembles the Local Group, or 2) use a celestial mechanics like technique  that solves for possible orbits constrained by the known position, velocity, and masses going backwards in time.  
The former has the disadvantages that incorrect assumptions in the initial perturbation spectrum at small scales and other errors in the very early stages of small scale structure could lead to problems in developing the LG planes.
In this work, we use the latter method to ensure solutions are fully compatible with the observed distribution.  
However, by going this route,  we miss out on a quantitative measure of the probability of occurrence for the LG planes.

The Numerical Action methodology assumes that galaxies assembled at high redshift, and that the subsequent assembly of groups of galaxies is a 
consequence of the motions induced in galaxies by other galaxies in the vicinity.   
At early times the deviations from Hubble flow were small and follow linear perturbation theory.  
The orbits are parameterized by their positions at a discrete set of times and represent the center of mass motions of all of the matter that is present 
today in a galaxy.  
The NAM derivation \citep{Peebles1995} does not refer to mass excesses but rather depends on total masses associated today with the bound systems and therefore masses are held constant.
An orbit should therefore faithfully reproduce the true center of mass motion as long as the majority of the associated matter was relatively nearby at the first time step which we have chosen to be z=4 (a = 0.2).  
For this work, we used the original version of NAM in which positions are fixed and redshifts are free.  
An alternate version has been devised in which redshift is fixed and distances are free \citep{Phelps2002, Phelps_etal2006}.
The cosmological parameters $H_0 = 73$, $\Omega_m = 0.24$, $\Omega_\Lambda = 0.76$ are assumed.  
Separate mass-to-light constants are taken for ellipticals and spirals.  
Mass distributions are taken to grow linearly with distance from the centers until outermost radii which are set to be proportional to $M^{\half}$.   

Each trial of the NAM procedure begins with random walk paths for every galaxy as initial guesses.  
Each path is `relaxed' to a physically plausible path by making an adjustment to positions at each timestep that decrease the derivatives of the action until the path is at an extremum of the action (see Appendix).
An iterative procedure takes turns relaxing  each galaxy and moving on to the galaxy furthest from relaxed as determined by the sum of the squares of the derivatives.
The derivatives are multiplied by the mass of the object to give priority to the higher mass systems which need to be settled before the lower mass systems can be correctly relaxed.
After many iterations one arrives at extrema of the Action. 
Since the problem is inherently multi-valued, this entire procedure, starting from random walk paths, is repeated many times until one is confident that all solutions have been found. 

Early on, when modeling regions larger than 10 Mpc, the matter density that provided the best fit on small scales caused general outflow in the outer parts of the model and fell well short of providing the WMAP value for $\Omega_m$ 
Therefore, we introduced a smooth density component $\Omega_s$ to supplement the density in observable structures.   
There are physical reasons to justify this addition to the model. 
N-body simulations show that there is a 25 - 35\% component of dark matter particles that is not in galaxies, groups, or clusters. 
Much of this matter remains along filaments or sheets.  
In addition,  much dark matter escapes entirely from its parent structure formations as a result of tidal forces during encounters and mergers.  
This makes sense if one considers that much of the dark matter in galaxies is less bound than any of the baryonic matter, and some of the baryonic material becomes unbound in tidal interactions. 
The overall fitting improves considerably with the addition of this parameter.

\subsection{Backtracking}
 \label{sec:Backtracking}

There were several weaknesses with the NAM procedure which needed to be addressed before tackling the more complex orbits that we would face at spatial scales $\ll$ 1 Mpc.  
First, there was no guaranty that, within a reasonable number of trials, any single trial would have all galaxies 
with the correctly chosen trajectories of all of the multiple solutions 
There is no known relation between the reality of a trajectory and how often it is realized in NAM.  
The procedure is to continue to make runs until no new solutions arise after a ``large number" of additional runs. 
While this sounds reasonable, it presumes that there are no real trajectories that have vanishing low probability of being reached when starting with random initial paths.  
However, an actual trajectory that passed through deep inside the potential of a giant galaxy would be realized rarely by relaxing from random initial guesses because if, in the calculation, it passed through the giant galaxy at a slightly different path, then it could have a very different position at the first time step from the real one. 
At this remote location, the local gradients are highly unlikely to lead toward the correct initial position.

Another problem is that it is difficult to make use of proper motions since this is an output parameter and one may need to make many trials to get a match for just one proper motion.   
If it is a giant galaxy, then inevitably all of the orbits of nearby galaxies are also wrong until the match is found.
It would be preferable if one could constrain trial orbits by the known proper motions to begin with.

Lastly, the NAM procedure has no way to accept solutions that are `good enough' as it always searches for 
solutions in which each galaxy reaches its extrema with essentially machine accuracy.  
At late times this is a reasonable criterion, but at the first time step it seems reasonable to believe that the potential field from NAM 
is never exactly right because its source is only a small number of discrete points.  
Therefore, the initial velocity need not be exactly what perturbation theory requires based on this potential.  
It is nice to have complete self-consistency, but inflexibility in the initial velocity could result in throwing out solutions that are actually fairly close to reality.
   
There is a way to address these issues for the dwarf galaxies. 
Once the general potential is solved based solely on a NAM procedure with the giant galaxies, a dwarf galaxy's path can be calculated by time reversed integration of the equations of motion, starting from its present position and velocity.  
The fact that we only know redshifts for these galaxies, and thus only radial velocities, translates into a problem of finding the
best direction for each galaxy's velocity, a 2 dimensional problem with parameters, $SGL_V, SGB_V$, the supergalactic L and B directions of the velocity in the MW frame of rest.
As long as the dwarf galaxy did not interact with another dwarf galaxy, altering its orbit probably has no significant affects on other galaxies or the time dependent potential of the NAM solution.

For a given direction of current motion one ends up with a velocity at the earliest time step that can be compared with what is expected from linear theory given the
potential at that position. 
In practice we use the sum of the gradient in the action at the first time step as our criteria, but the two constraints are closely related (see A3 of the appendix). 
One can then map out the entire range of possible directions in the plane of the sky of the present velocity vector by stepping by 1\degree\ or 2\degree\ intervals.
Such coarse sampling is generally sufficient to locate all of the sky patches where the earliest timestep peculiar velocity error (compared to linear theory expectations given the potential at that time) drops down to low ($<$100 \kms) values.
Once a rough direction is selected, a procedure that minimizes on the errors can be used to improve the precision in direction.  
For galaxies with proper motion measurements, one needs to scan only within a restricted range of directions.


In some cases, one wishes to backtrack a giant galaxy, say because one is looking for a specific proper motion solution or to force an interaction with another galaxy.   
Often one knows that two galaxies have been interacting with a specific configuration by telltale H\,I bridges or tails.  
It is possible to alternate the backtracking between two galaxies until they are maneuvered into the desired configuration.
This must be followed up with a NAM relaxation to find self-consistent mutual dynamics for the entire ensemble with the new orbits for the pair.
Backtracking is written as a subroutine to the NAM procedure to easily go between the two. 
For example, using backtracking in this manner, we forced an interaction between NGC\,300 and NGC\,55  that improved their redshifts.  

\section{Numerical Results}
\label{sec:Results}
\subsection{Large Scales}
 \label{sec:LS}

It is important for the reconstruction of orbits to have a proper description of the gravitational potential
due to large scale structures.  Nearby entities such as the IC\,342/Maffei Group and the Centaurus A Group
induce significant tidal fields important to the history of the Local Group.  Extending in scale, the
Local Group lies in a thin plane of galaxies, the Local Sheet, itself lying in the
equatorial plane of the Local Supercluster (and the equatorial plane in supergalactic coordinates).
Orthogonal to this plane to the supergalactic north and abutting the plane is the Local Void.  Expansion
of the Local Void gives the Local Sheet a deviant velocity of 260~\kms\ toward the supergalactic south pole
\citep{Tully2008}.  A gradient in Local Void expansion velocities is seen.  The Local Sheet is catching
up with galaxies below (south of) the equatorial plane while galaxies above (north of) the equatorial plane
are catching up to us \citep{Tully+2013}. {\it Evacuation of the Local Void and compression of the Local Sheet must play a
fundamental role in the formation of the Local Group.}
\begin{figure}
	\includegraphics[width=0.45\textwidth]{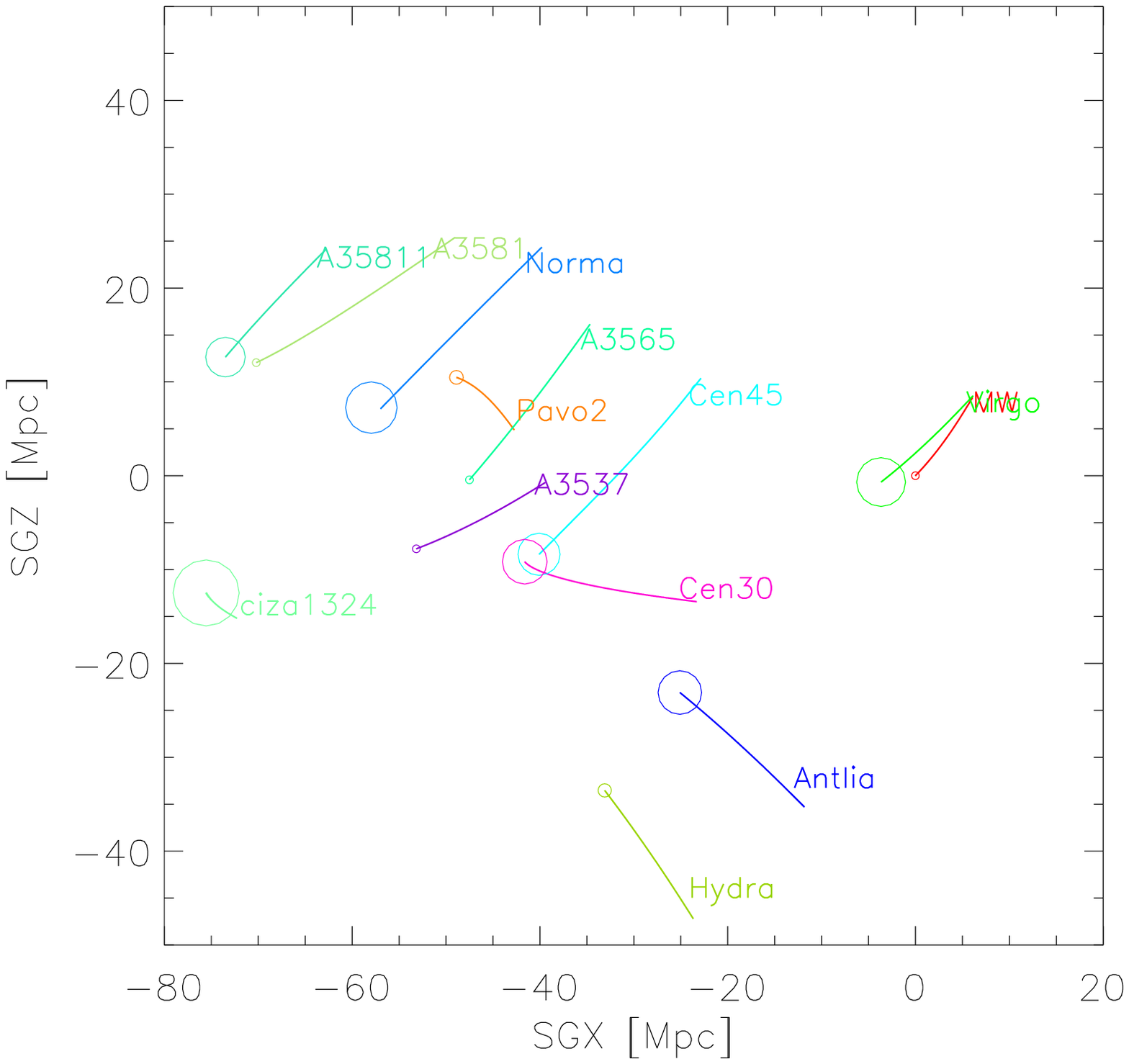}	\\			
	\hspace{0.1cm}
	\includegraphics[width=0.45\textwidth]{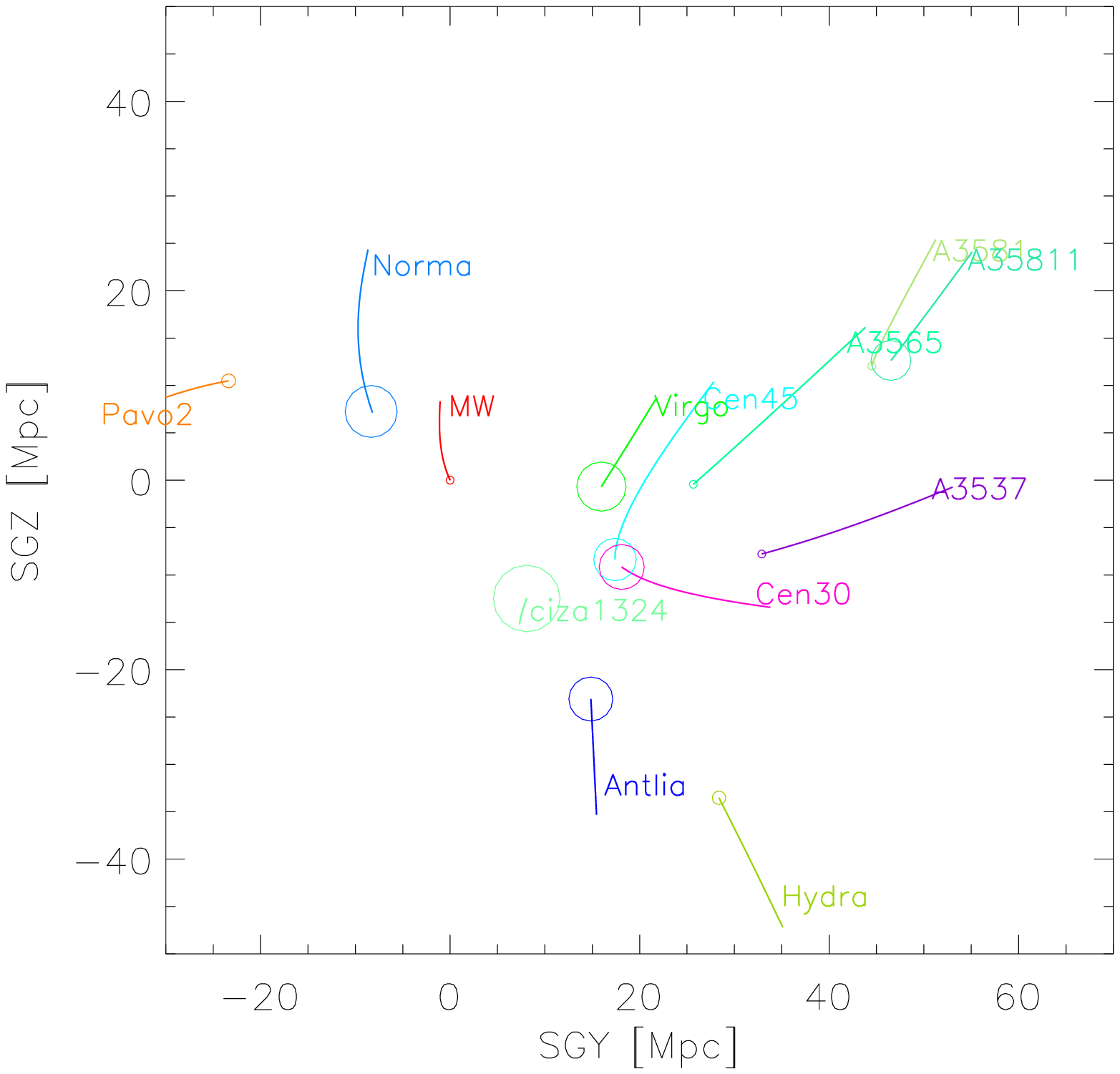}
	\caption{\it Trajectories in comoving coordinates from z=4 to the present for a portion of the clusters of galaxies in our large scale  NAM reconstruction.  Masses of each cluster were varied to solve for best fit to redshifts.  Present distances were fixed. Circles are placed at present positions, and size is proportional to the logarithm of mass. }
	\label{fig:orbs100Mpc}
\end{figure}

\end{twocolumn}
\begin{onecolumn}
\begin{landscape}

\begin{figure}
	\includegraphics[width=1.05\textwidth]{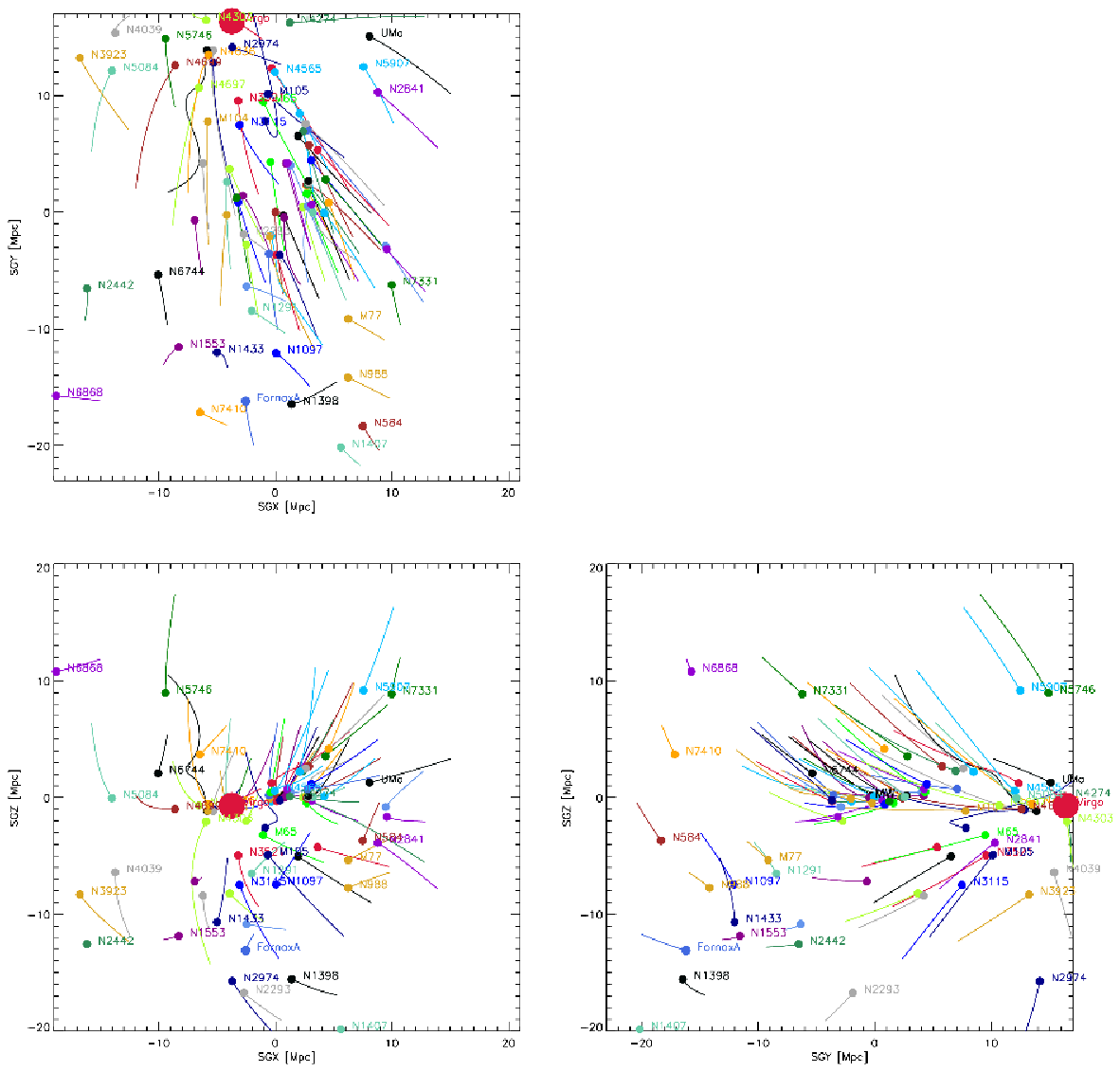}				
	\caption{\it Trajectories in comoving coordinates of galaxies in the local region of the Local Supercluster after the mass of the Virgo Cluster was optimized for infall of 9 galaxies along its line of site. }
	\label{fig:LSC}
\end{figure}
\end{landscape}
\begin{figure}
	\includegraphics[width=1\textwidth]{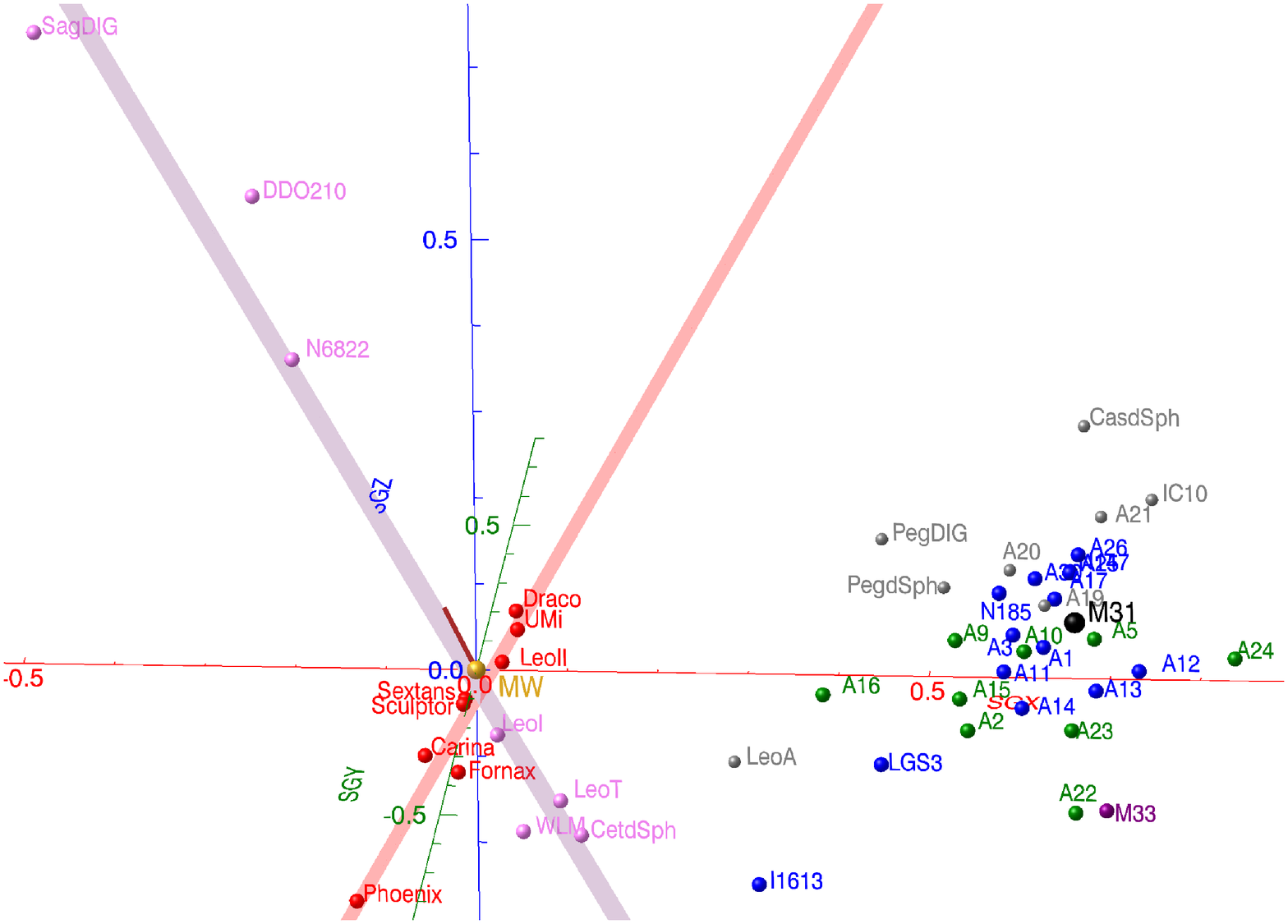}	\\		
	\includegraphics[width=1\textwidth]{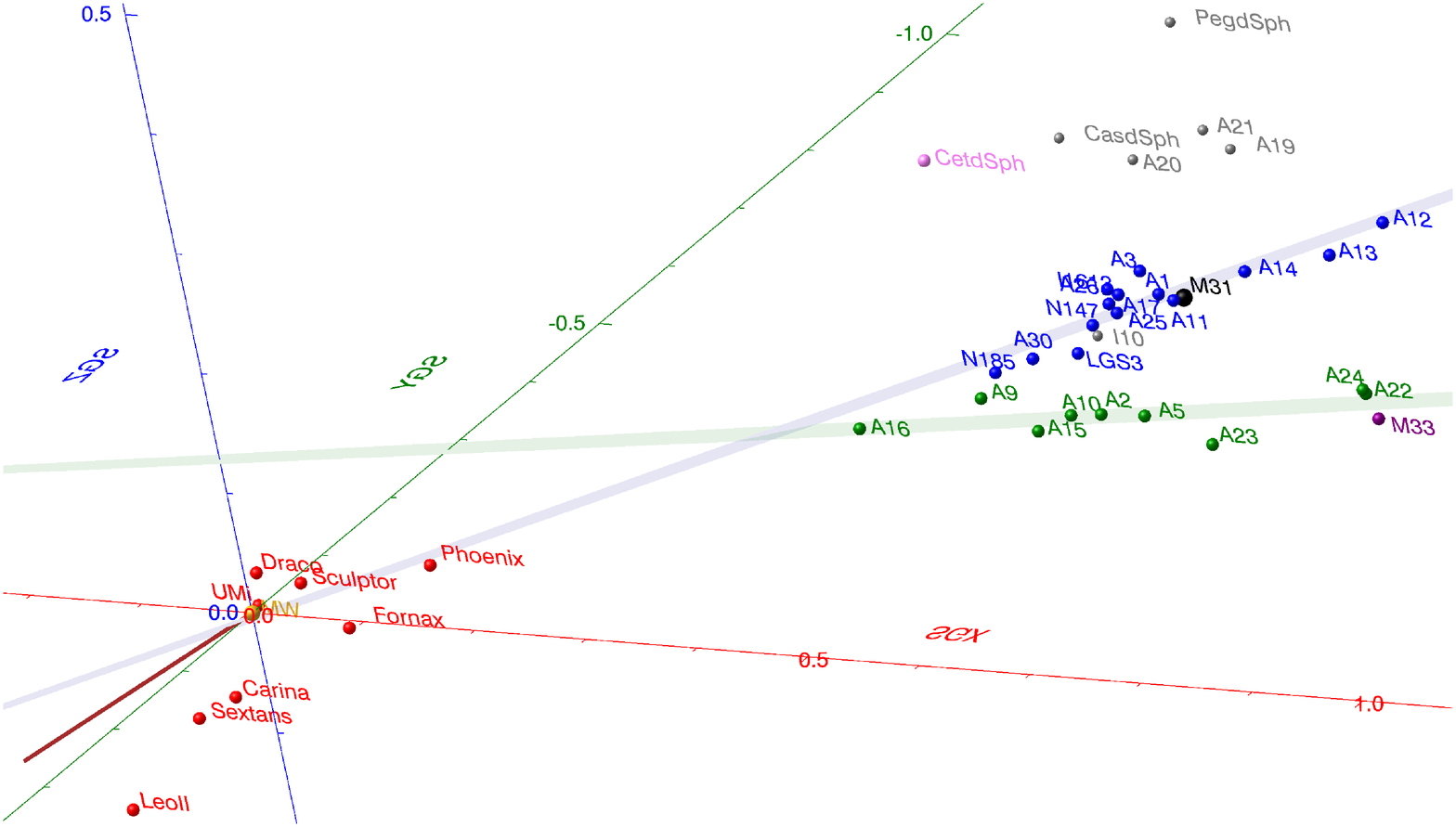}
	\caption{\it Present Positions of 4 planes of galaxies in the Local Group in two projections.  A vector pointing from the MW toward Virgo of length 0.25 Mpc is colored brown. SGX,Y, and Z-axes are red, green, and blue respectively. Top) Edge on view of the outer MW galaxies (Plane\,3, violet) and the inner MW galaxies (Plane\,4, red) . Bottom) Edge on to both planes near M\,31, \textbf{Plane\,1}  (blue) and \textbf{Plane\,2}  (green).  Other LG galaxies are grey.  Information on the fits to these plane are given in Table~\ref{table:planes}.}
 	\label{fig:presentplanes}
\end{figure}
\end{onecolumn}
\begin{twocolumn}

Continuing to larger scale, the Virgo Cluster exerts a significant tidal field on the Local Group, but to
determine the mass of the Virgo Cluster one must correctly model the tidal field of the Norma/Hydra/Centaurus
complex.  In particular, to model the Local Supercluster peculiar velocity field the component of the MW
peculiar velocity toward the Virgo Cluster must be deconstructed, and this nearby cluster provides
significant compression along the MW/Virgo Cluster line.
The motions of galaxies out to 3,000~\kms\ were studied using NAM (Shaya, Peebles, \& Tully 1995).
However, to provide a better potential field from larger scales, that analysis has now been extended with
more distances and improved quality \citep{Tully+2013}.   

In a preliminary analysis of the peculiar motions of clusters on large scales, we solved for the masses of 27 major clusters at $<$ 100 Mpc within a potential from 170 clusters out to 210 Mpc.  
The best set of masses was the one that provided best fits to the peculiar velocities of clusters within 100 Mpc as given from Numerical Action solutions. 
The resulting trajectories of a subset near the Hydra/Centaurus/Norma Supercluster are shown in Figure~\ref{fig:orbs100Mpc}.
The overall flow in this region is toward the two most massive clusters, Norma and an X-ray discovered cluster CIZA\,1324.
The supercluster has a total mass of $\sim 1.2 \times 10^{16} \Msun$.  

After inserting the above GA mass at the position of the Norma Cluster, we solved for the mass of the Virgo Cluster by modeling the motion of nine galaxies along lines of site within $15\degree$ of the Virgo Cluster center plus the MW. 
We determined a mass for Virgo of $\sim 6 \times 10^{14} \Msun$, about twice the virial mass, and a value for the smooth density $\Omega_s$ of 0.11.
The K-band mass-to-light ratios that worked best here and also in the Local Group analysis are $M/L_K = 40 \Msun/\Lsun$ for spirals and $M/L_K = 50 \Msun/\Lsun$ for ellipticals.
We are presently obtaining additional TRGB distances of more galaxies in the direction to the Virgo Cluster using HST and will publish separately a more 
detailed account of this analysis.

The trajectories of galaxies in the 20 Mpc region of this NAM calculation are presented in Figure~\ref{fig:LSC}. 
Pronounced streaming of peculiar velocities are obvious in these diagrams.  
Flow toward the Virgo Cluster, at SGY=16 Mpc, is seen throughout this volume.   
The paths trace out several V-shapes in these diagrams as  giant galaxies come together for the first time forming groups, while all move toward the Virgo Cluster.

In Figure~\ref{fig:LSC}, one sees that by the first time step at z=4 a wall of material was already formed at the boundary of the Local Void which then traveled several Mpc toward  -SGZ and +SGY, forming the plane of the Local Sheet seen today.
The nearby groups are therefore composed almost exclusively from material evacuated from the Local Void.
\textit{It is out of this thin structure that the Local Group formed. }  

\subsection{Local Group}
 \label{sec:LG} 
Recent high quality distances to Local Group galaxies provide an accurate 3-D mapping of the LG.  In Figure~\ref{fig:presentplanes}, we show that 84\% of the LG satellite galaxies can be placed in four thin (axial ratio $>10$) planes.  
Considering that there are four convincing vast thin planes in the LG and evidently two in the nearby Centaurus\,A Group, it may well be that planes of satellite galaxies are common.
Rather than using the term Vast Thin Plane of Galaxies \citep{Ibata_etal2013} which connotes some rare and special structure, we will simply refer to it as \textbf{LG Plane\,1} .
The other plane near M\,31 we call \textbf{LG Plane\,2} , and two near the MW are \textbf{LG Planes\,3 and 4}.
In fact when examined closely in 3-D, \textbf{Plane\,1}  and \textbf{Plane\,2}  appear to really be a single system in the shape of a 
vast V or U wrapping around M\,31.
The top projections in Figure~\ref{fig:presentplanes} shows the thin side of the MW pair of planes by viewing down the line common to both planes. 
The bottom of the figure does the same for the M\,31 pair of planes.
The planes are highlighted by drawing in the best fit plane through them colored as follows.
Plane\,1 is blue, \textbf{Plane\,2}  is green,  \textbf{Plane\,3}  is violet, and \textbf{Plane\,4}  is red.
For \textbf{Plane\,1} ,  seen in the PAndAS survey, we add IC1613 and LGS3 and we remove And\,27 and And\,9.
And\,27 has such a high negative velocity that it must originate outside the LG, as discussed below.
And\,9, we find, has a trajectory similar to other \textbf{Plane\,2}  galaxies and unlike \textbf{Plane\,1} galaxies. 
Note that both M\,31 and the MW are in \textbf{Plane\,1}.  
\textbf{Plane\,2} is offset from M\,31 by about 100 kpc.
Its near side ends at the center of light of the LG.
\textbf{Plane\,3}, the MW outer satellites, and \textbf{Plane\,4} , the MW inner satellites, are nearly perpendicular to the two M\,31planes and the direction to Virgo is nearly in \textbf{Plane\,3} . 
Fits to these planes and the directions of normals to these planes are presented in Table~\ref{table:planes}.
The table provides, for each plane, the constants in the plane equation:
\begin{equation}
\hat{n}_x x + \hat{n}_y y + \hat{n}_z (z - z_0) = 0
\end{equation}
where $\bold{\hat{n}}$ is the unit normal to the plane and $z_0$ is the z-value in the plane at $x=y=0$.

\subsubsection{MW, M\,31, \& M\,33} 
Our fixed present position NAM code was used to find orbits for 180 galaxies/groups/clusters within 50 Mpc and with 14 clusters from 50 -- 80 Mpc contributing only to a first order external tidal field tensor. 
We used paths with 200 timesteps beginning at z=4.  
We ran some experiments in the most complicated cases with 400 time steps and found that the results were not significantly different.
Since most of the mass of the Hydra/Centaurus/Norma supercluster was put into Norma and therefore into the external tidal field tensor, the flow pattern of the model does not generate much motion in that direction.  
Within the Local Group only MW, M\,31, and M\,33 were included.
The masses were set by multiplying the luminosity by a mass-to-light ratio and were held constant.    

As mentioned,  a recent measurement of M\,31's proper motion by  \citet{vdM_etal2012} indicates a fairly radial motion into the MW 
(transverse velocity $< 60\ \kms$ assuming the Sun's LSR is $v_c = 240\ \kms$).
Our initial runs typically ended up with substantial proper motion for M\,31. 
Part of the reason for this was that the early time position of IC\,342 was usually trailing M\,31, near where it would need to be if its motion was directly toward the MW.  
With backtracking, we found a different solution for IC\,342 with a substantially lower early time SGZ position.  
Then, backtracking M\,31, constrained to the range of directions consistent with its proper motion, resulted in a mutually consistent scenario  that fully relaxed well in NAM.
But another part of the problem was that certain orbits for M\,33 destabilized this solution.

\textbf{Backtracking M\,33}, constrained to the range of directions consistent with the VLBI measurements of two of its H$_2$O masers \citep{Brunthaler_etal2005},
resulted in many tentative solutions.
However, only two of these remained after relaxing it together with all other mass tracers in NAM and requiring M\,31 to maintain its observed low proper motion. 
In one solution (Figure \ref{fig:M33_1}), M\,33 was $\sim$2 Mpc lower than the MW in SGY at early times, and in the other  (Figure \ref{fig:M33_2})
it began $\sim$1 Mpc higher in SGY and SGZ before pulling over to its present position just below M\,31 in SGZ.   
The solutions were not consistent with the idea that the ``H\,I bridge" seen between M\,31 and M\,33 \citep{BraunThilker2004} and  resolved into small knots \citep{Wolfe_etal2013} was formed by M\,33 passing near to M\,31.
Models of the H\,I bridge \citep{Bekki2008, McConnachie_etal2009} as well as models explaining the warp in the M\,33 outer H\,I disk \citep{Putnam_etal2009} require that the encounter impact parameter was $<$ 100  kpc, but for our two acceptable orbits the closest approach is happening now.
However, as we will see, other possibilities to explain these phenomena have arisen.
In all that follows, we use only the second solution.
The other solution will be studied in future work.
\begin{figure}
	\includegraphics[width=.45\textwidth]{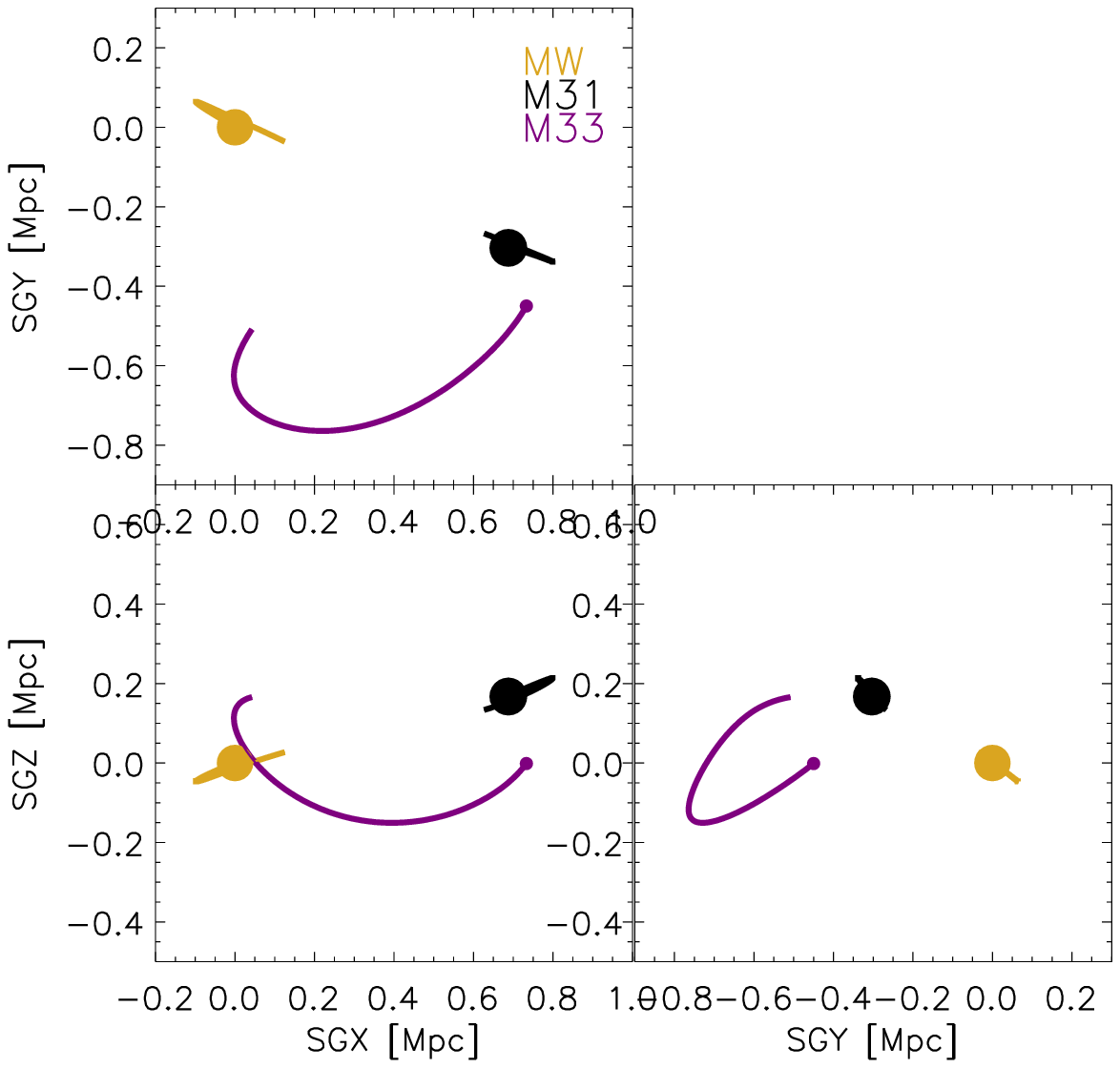}	\\			
	\includegraphics[width=.45\textwidth]{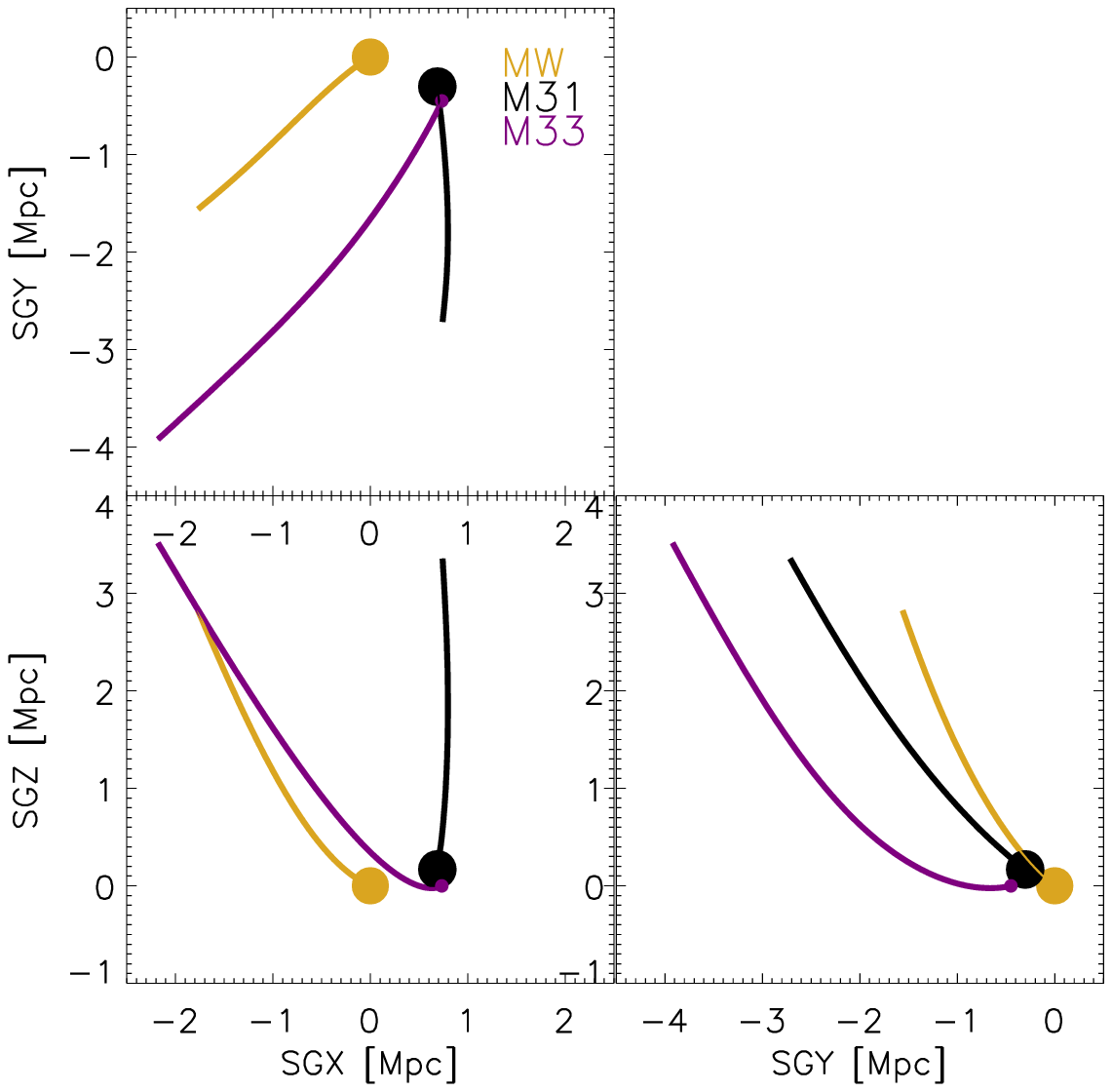}
	\caption{\it A solution for the orbit of M\,33 after backtracking within the proper motion error box and then relaxing with the other galaxies in the sample in NAM.   Supergalactic coordinates.  Top) set of three plots in the LG center of mass frame and Bottom) set of three plots in comoving coordinates and the Local Supercluster (LSC) frame. }
	\label{fig:M33_1}
\end{figure}
\begin{figure}
	\includegraphics[width=.45\textwidth]{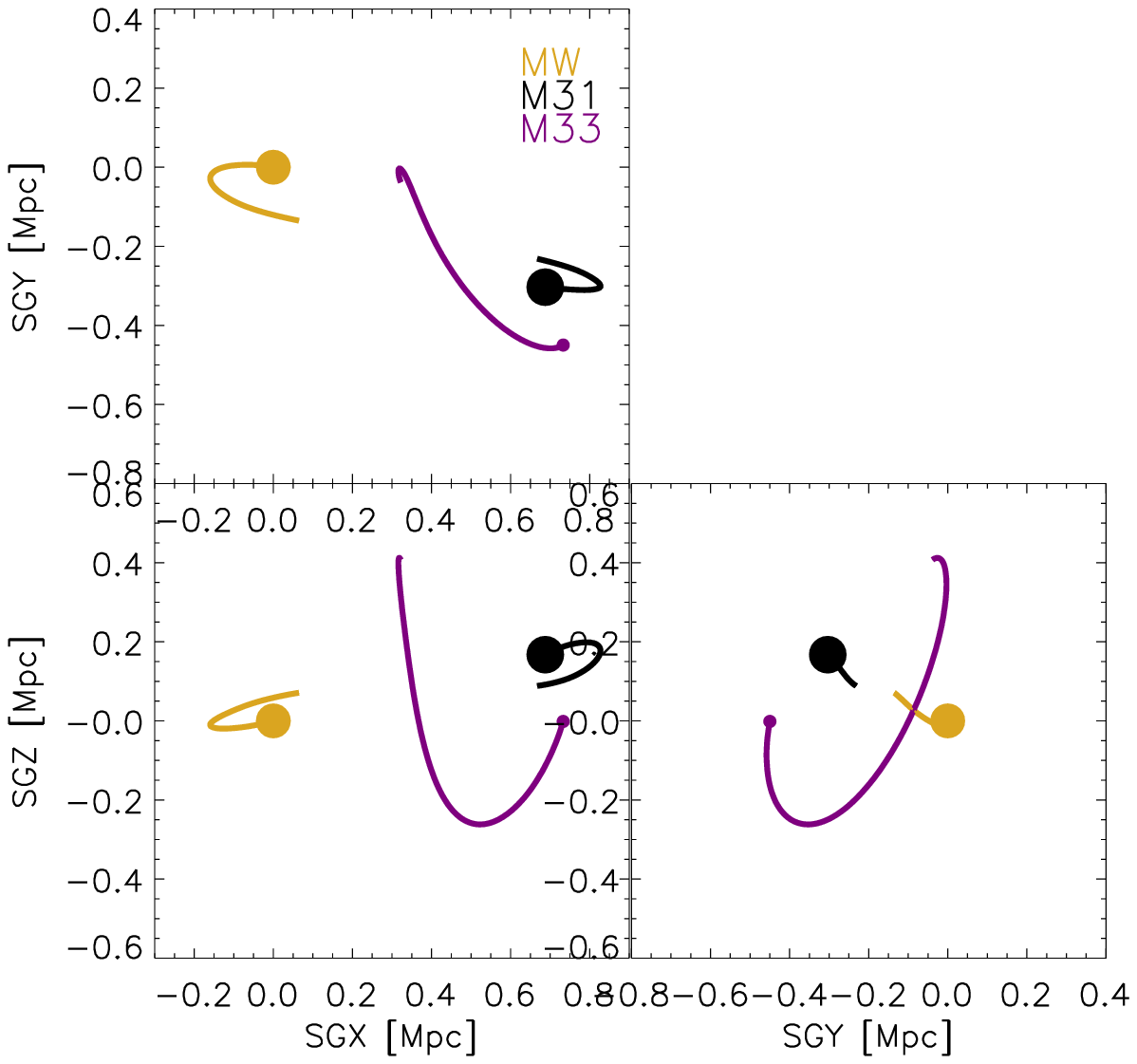}		\\		
	\includegraphics[width=.45\textwidth]{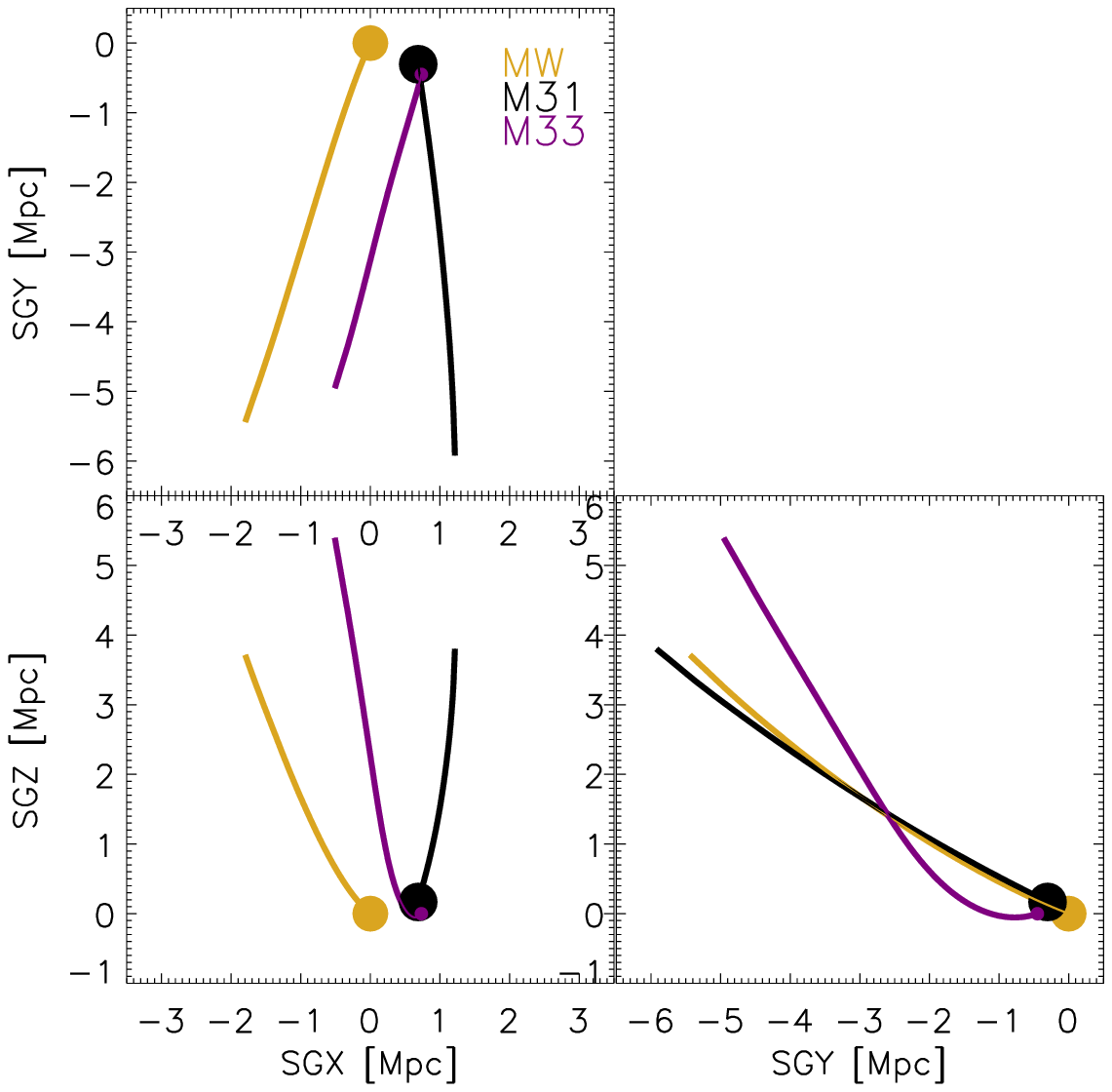}
	\caption{\it Same as previous figure, but this is the other acceptable solution for the orbit of M33.  This solution is used in the rest of the analysis }
	\label{fig:M33_2}
\end{figure}

When backtracking galaxies near a giant galaxy, each would have a dozen or so solutions with acceptable initial velocities, but most of these had the galaxy falling in close to a giant galaxy and then executing many very tight orbits.  
Since, we expect that orbital plunging from frictional dissipation and tidal field disruption would occur in such cases, we considered these not to be acceptable solutions. 

Typically, there would be 1 to 5 less complex orbits to choose from if we allowed for $\sim 25$ \kms\ errors in initial velocities.
To select from among these simple solutions, we chose the one that had present motion consistent with how the plane developed.  
For \textbf{Planes\,1 and 2}, this means having motion parallel to the plane in the M\,31 reference frame,
and for \textbf{Planes\,3 and 4}, this means having motion parallel to its plane in the MW reference frame.
In rare cases where there were two solutions with motion in the plane, then the one with the lowest error at the first time step was selected.
To keep this report  at a reasonable length, we only describe here our final choice paths. 
After all, the point here is not to be certain about the particular path of any given galaxy, but rather to look for physical processes that could explain the existence of thin planar galaxy structures so late in the development of a group.
Besides And\,12, no galaxy needed to have its distance adjusted, and small velocity changes of 1 or 2 \kms\ arose from inaccuracies in the procedure, not from attempts to improve the fits.
And\,12, which had no solution at its TRGB assigned distance, fit in well after being adjusted from 940 kpc to 920 kpc. 

When we use the terms `leading' or `ahead', it refers to objects that have run ahead of
M\,31 (or the MW for satellites of the MW) and `trailing' or `behind' refer to those that are catching up.
After backtracking the smaller galaxies in the LG, the single best solution had nearly all of the LG galaxies in a single plane at early times slightly trailing the plane of the larger galaxies (Figure~\ref{fig:initial}) in the Local Supercluster (LSC) frame.
If one rotates the coordinate system by $25\degree$ around the SGY-axis,  one finds that many \textbf{Plane\,1}  galaxies have been just moving back and forth with respect to M\,31 in this single plane.
It should be noted that this was not a criterion in deciding which solutions to choose. 
Rather, restrictions on the motion at the present time so as to keep the plane in existence for an extended time, were the selection criteria. 

In most cases, the initial positions happened to be outside of our graphic viewing port, and therefore could not have influenced our decisions. 
\begin{figure}
	\includegraphics[width=.5\textwidth]{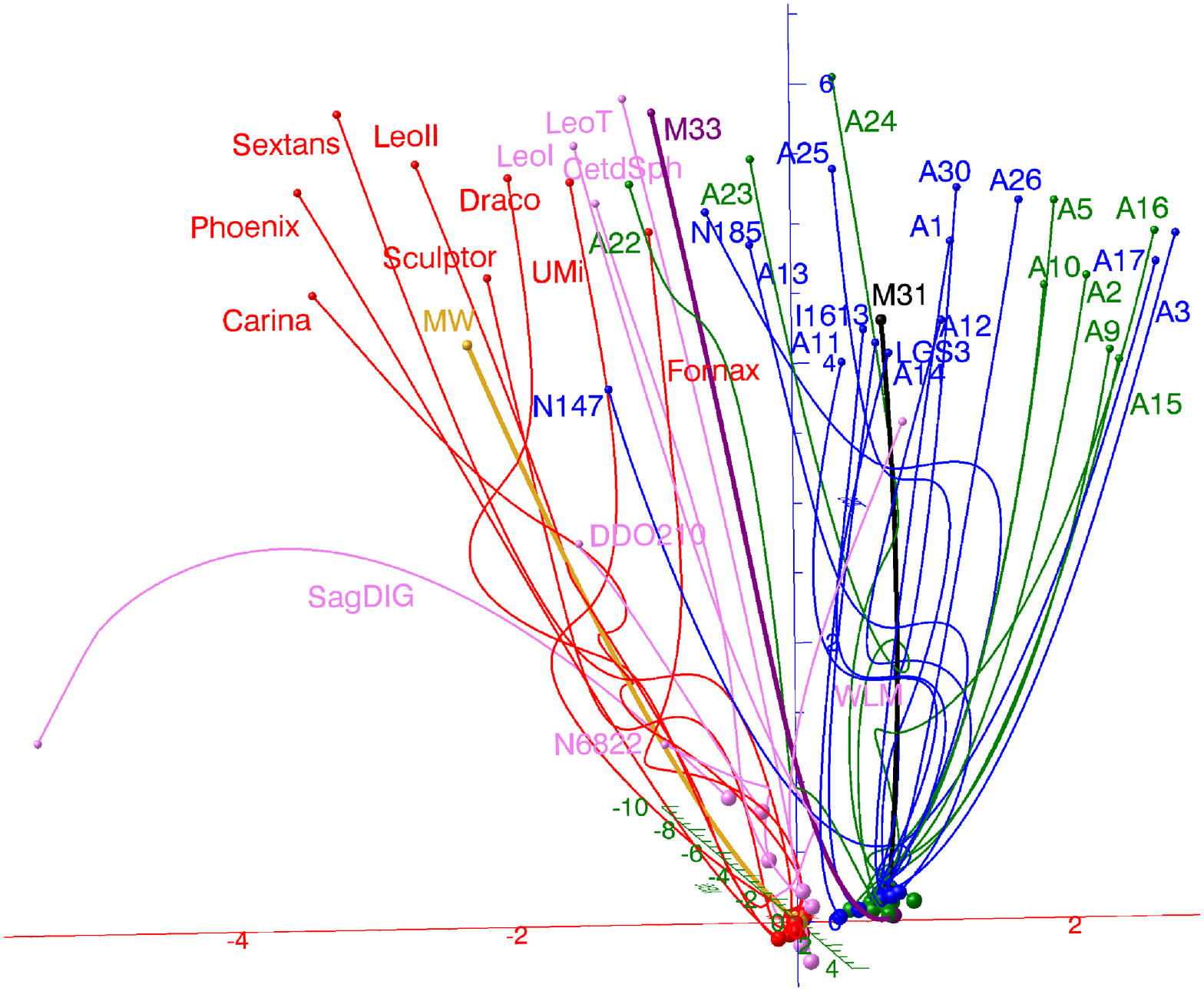}	\\			
	\includegraphics[width=.5\textwidth]{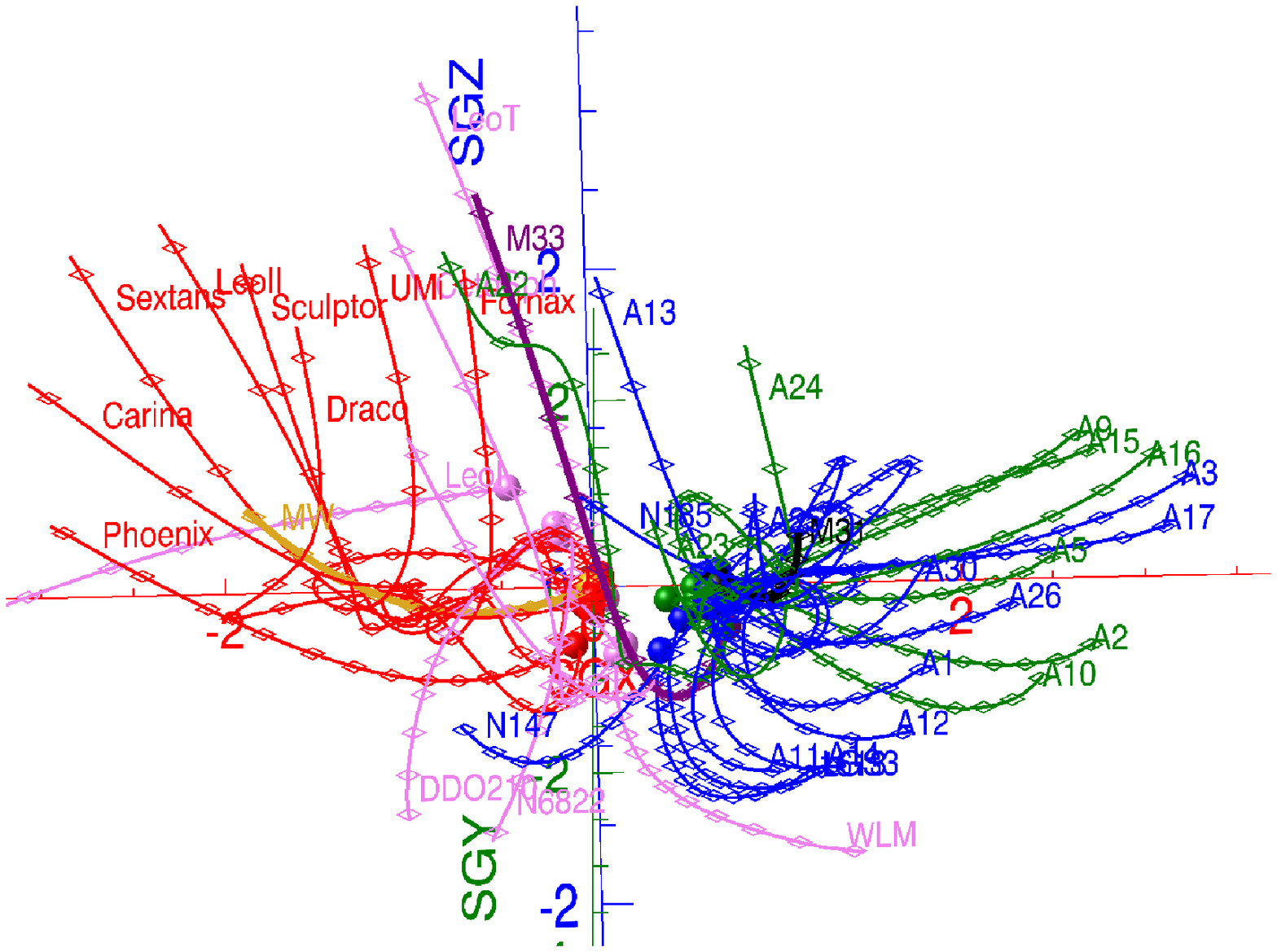}
	\caption{\it  Trajectories in the Local Supercluster  frame and comoving coordinates - Large spheres are at the present locations and small spheres are at the z=4 location for galaxies  in the 4 planes. 
Plane\,1- blue, \textbf{Plane\,2}  - green, \textbf{Plane\,3}  - violet, \textbf{Plane\,4}  - red, MW - goldenrod, M31 - black, M33 -  purple. 
Top) Nearly SGX-SGZ projection, but tilted to show the "initial" plane that most of the galaxies were in at z=4.   Note that the MW and M31 are below in SGZ, but the MW$-$M31 line intersects the initial plane of the dwarfs at positive SGX. 
Bottom) View is roughly down in the direction of motion of M\,31and MW.   Intervals of 1 Gyr are marked with squares.
The initial configuration of the dwarf galaxies was either a pair of  filaments or a single ring shaped filament contained within the wall of the Local Void.}
	\label{fig:initial}
\end{figure}

\textbf{Leo\,I}'s  proper  motion has recently been detected based on two epochs of HST ACS/WFC images separated by $~5$ years
\citep{Sohn_etal2013}. 
Within the errors (the  proper motion of ($\mu_W$, $\mu_N$) = (0.1140 $\pm$ 0.0295, -0.1256 $\pm$ 0.0293) mas/yr), we find several solutions that require it to pass near M\,31 before passing near the MW and one solution in which it interacts with M\,33 but no solution in which it just interacts with the MW.
The path with an interaction with M\,33 (Figure~\ref{fig:i10_leoi}) was chosen here, but it may well have gone past M\,31 instead. 
In this solution, at z=4, Leo\,I was already close to M\,33 and traveling more or less in the same direction.
About 5.3 Gyr ago, it received a large kick in the $-$SGX direction by interacting with M\,33, reaching a minimum distance of 21 kpc from M\,33, and then was caught by the gravity of the MW.
About 1 Gyr ago, it passed $\sim$45 kpc behind the MW.
When Leo\,I reached the MW path, it was trailing the MW in the LSC frame and made a right angle turn to follow the MW.  
It then went from nearly directly trailing MW to  leading it in the LSC frame, 
similar to what several other \textbf{Plane\,3}  galaxies are doing.

For \textbf{IC\,10}, the range of present motion is restricted by the error bars in the proper motion measurement from an H$_2$0 maser \citep{Brunthaler_etal2007}.    
There were 4 acceptable solutions within the error bars of the proper motion measurements.
In all but one of these,  IC\,10 does 2 or 3 passes by M\,31.  
In its simplest orbit, it executes a wide sweep at negative SGZ around M\,31 and is headed back up.  
We chose the simplest one to present here for our best guess of the ensemble of orbits.

\end{twocolumn}
\begin{onecolumn}
\begin{deluxetable}{l|r|rrr|r|rr}
\tablecolumns{7}
\tabletypesize{\small}
\tablecaption{Best Fits to Planes  \label{table:planes}}
\tablewidth{0pt}
\tablehead{
&&\multicolumn{3}{c|}{Normal (Cartesian)\tablenotemark{b}} && \multicolumn{2}{c}{Normal (Spherical)\tablenotemark{d}}\\
\colhead{Plane} \vline&  \colhead{RMS\tablenotemark{a}}\vline&\colhead{$SGX_{N}$} & \colhead{$SGY_{N}$} & \colhead{$SGZ_{N}$} \vline& \colhead{$z_0$\tablenotemark{c}} \vline& \colhead{$SGL_{N}$} & \colhead{$SGB_{N}$}\\
\colhead{$\cdots$}\vline&\colhead{kpc}\vline& & \colhead{Mpc} &  &\colhead{Mpc} \vline &  \colhead{degrees} & \colhead{degrees}}
\startdata

Plane\,1 & 11.9& -0.339  &  -0.234 &    0.912&   -0.0041    &   -145.4&       65.7 \\
Plane\,2  &11.5& -0.108 &    -0.411 &   0.905 &  -0.1285  &    -104.7&       64.8\\
Plane\,3 &  34.2&  0.849 & 0.225 &  0.478  &   -0.0065   &   14.8  &    28.6\\
Plane\,4& 11.3   &  0.872   & -0.083& -0.483 &  0.0267&         -5.4&-28.9\\
\enddata
\tablenotetext{a}{Thickness of plane (rms of distances perpendicular to plane).}
\tablenotetext{b}{Normal to plane in supergalactic Cartesian coordinates.}
\tablenotetext{c}{Offset of plane in z at SGX=SGY=0.  MW is origin of system.}
\tablenotetext{d}{Normal to planes in supergalactic spherical coordinates.}

\end{deluxetable}

\begin{deluxetable}{|l|rrr|rrr|}
\tablecolumns{7}
\tabletypesize{\small}
\tablecaption{Input/Output for Proper Motions  \label{table:pm}}
\tablewidth{0pt}
\tablehead{
\colhead{}&\multicolumn{3}{c|}{Observed Proper Motion} & \multicolumn{3}{c|}{Model Proper Motions}\\
\colhead{Galaxy} \vline &  \colhead{$\mu_{SGL}$} & \colhead{$\mu_{SGB}$} & \colhead{$\abs{\mu}$} \vline & \colhead{$\mu_{SGL}$} & \colhead{$\mu_{SGB}$} & \colhead{$\abs{\mu}$} \vline\\
\colhead{$\cdots$}\vline & \multicolumn{3}{c|}{$\muasperyr$} & \multicolumn{3}{c|}{$\muasperyr$} 
}
\startdata
Leo\,I &   -81.87 &  -33.57  &   88.49 &   -62.10   & -31.42 &    69.59 \\
M\,31 &     -8.26 &      -3.48&      8.96 &  -7.23   &  -4.15  &    8.34   \\
IC\,10    &    20.46&     38.79&     43.86&  34.71    &  60.22   &   69.51 \\
M\,33 &  32.92  &    28.87 &     43.79&    11.45  &   48.72   &  50.04  \\
\enddata
\end{deluxetable}

\begin{deluxetable}{lrrrrrr}
\tabletypesize{\scriptsize}
\tablecaption{Output Data for Local Group Galaxies  \label{table:output}}
\tablewidth{0pt}
\tablehead{
\colhead{Galaxy} & \colhead{Mass\tablenotemark{a}} & \colhead{$cz$\tablenotemark{b}}
& \colhead{SOS\tablenotemark{c}} & \colhead{$\mu_{SGL}$\tablenotemark{d}} & \colhead{$\mu_{SGB}$\tablenotemark{e}} & \colhead{$|\mu|$\tablenotemark{f}}\\
\colhead{$\cdots$} & \colhead{$\Msun$} & \colhead{$\kms$} &\colhead{$\cdots$} &  \colhead{$\muasperyr$} & \colhead{$\muasperyr$} & \colhead{$\muasperyr$} 
}
\startdata
UMi &1.22E+05 & -54 &5.87E-03 &-200.62 & -54.45 &  207.88\\
Draco &1.09E+06 & -71 &2.55E-04 & -80.51 &-219.72 &  234.00\\
Sculptor &4.48E+06 &  71 &7.18E-03 & 757.32 & 723.03 & 1047.04\\
Sextans &3.91E+05 &  55 &5.91E-04 & -85.41 & 142.39 &  166.04\\
Carina &1.50E+06 & -11 &2.80E-02 & -21.04 &-141.22 &  142.78\\
Fornax &4.73E+07 & -53 &6.64E-04 &-301.36 &-172.04 &  347.01\\
Leo\,II &1.74E+06 &  15 &2.32E-03 & -10.06 & -60.79 &   61.61\\
Leo\,I &2.29E+07 & 160 &5.43E-03 & -62.10 & -31.42 &   69.59\\
Leo\,T &2.15E+07 & -72 &1.18E-03 &  34.53 & -62.93 &   71.78\\
Phoenix &7.35E+06 &-115 &2.23E-03 &  -2.91 & -45.98 &   46.07\\
And\,16 &7.05E+07 &-191 &8.67E-04 & -56.91 &  -7.87 &   57.45\\
NGC\,6822 &4.76E+09 &  51 &2.22E-03 &  -3.67 & -60.28 &   60.39\\
And\,9 &4.44E+07 &  -9 &1.19E-03 &  11.46 &  -2.49 &   11.72\\
And\,15 &9.20E+07 &-142 &2.43E-03 &   8.91 & -20.65 &   22.49\\
NGC\,185 &1.06E+10 & -18 &6.86E-04 &  -3.25 &   8.44 &    9.04\\
LGS3 &3.80E+05 &-133 &3.10E-03 &  54.00 &  18.84 &   57.19\\
And\,2 &2.17E+08 & -24 &1.62E-03 & -15.57 &  -8.43 &   17.70\\
And\,10 &6.65E+07 &  33 &5.33E-03 & -39.61 &   5.47 &   39.99\\
And\,30 &5.00E+06 &  72 &1.96E-03 & -64.70 &  23.86 &   68.96\\
And\,3 &2.24E+08 &-135 &4.44E-03 &-100.49 &   8.78 &  100.87\\
And\,11 &1.28E+07 &-241 &7.01E-03 &  85.32 &  32.32 &   91.24\\
And\,17 &2.60E+07 & -65 &9.96E-03 & -78.15 & -45.75 &   90.56\\
Cetus\,dSph &7.85E+06 & -20 &2.17E-02 &  -3.17 &  10.44 &   10.91\\
NGC\,147 &7.65E+09 &  21 &2.89E-03 & -21.17 & -11.41 &   24.04\\
And\,5 &1.70E+08 &-201 &2.54E-03 & -13.38 &  -0.66 &   13.39\\
And\,20 &7.45E+06 &-248 &2.33E-03 &-112.43 &  -2.92 &  112.47\\
And\,25 &1.52E+08 & 101 &6.54E-03 &-104.81 &  14.33 &  105.78\\
Leo\,A &3.15E+07 & -16 &5.62E-03 &  15.23 & -25.96 &   30.09\\
And\,1 &9.45E+08 &-180 &1.83E-03 & 143.89 &  28.32 &  146.65\\
And\,23 &3.02E+08 & -72 &3.75E-03 &  89.15 &  20.58 &   91.50\\
And\,26 &1.61E+08 & -45 &1.18E-03 & -17.76 &   5.20 &   18.51\\
IC\,1613 &2.28E+09 &-150 &6.93E-04 &   4.69 &   7.30 &    8.68\\
M\,31 &2.46E+12 &-103 &6.58E-04 &  -7.23 &  -4.15 &    8.34\\
Cas\,dSph &2.12E+09 & -71 &2.66E-03 & -16.46 &   0.19 &   16.46\\
And\,14 &5.90E+07 &-308 &5.90E-03 &  36.95 &  17.12 &   40.72\\
IC\,10 &3.48E+10 &-127 &1.02E-03 &  34.71 &  60.22 &   69.51\\
Peg\,dSph &8.10E+08 &-148 &9.44E-03 & -30.73 &  57.82 &   65.48\\
And\,19 &6.90E+07 &  89 &6.28E-03 &  68.28 &  24.11 &   72.41\\
And\,21 &2.13E+08 &-146 &5.08E-03 &  -7.17 &  86.45 &   86.75\\
And\,22 &1.30E+07 & -19 &3.60E-02 &  -3.24 &  13.96 &   14.33\\
M\,33 &1.62E+11 & -19 &1.17E-10 &  11.45 &  48.72 &   50.04\\
And\,13 &2.17E+07 &  -9 &3.97E-04 &  -7.79 &  -1.34 &    7.91\\
And\,12 &1.72E+07 &-371 &4.90E-03 &  -2.07 &  -1.28 &    2.43\\
And\,24 &6.30E+07 &  50 &2.38E-03 &  49.18 & -20.25 &   53.19\\
Tucana &1.96E+06 &  88 &2.00E-03 &  -7.32 &   2.80 &    7.83\\
Peg\,DIG &3.66E+07 &   3 &7.46E-03 & -23.52 &  -0.26 &   23.52\\
DDO\,210 &3.69E+07 &  -3 &1.93E-02 & -12.22 &   6.51 &   13.84\\
WLM &1.07E+09 & -60 &7.32E-04 & -18.34 &   6.34 &   19.41\\
Sag\,DIG &3.71E+07 &  17 &5.66E-03 &   0.13 &   2.23 &    2.23\\
And\,18 &1.51E+08 &-112 &9.64E-04 & -30.26 &  -4.03 &   30.53\\
And\,27 &8.45E+07 &-329 &3.49E-02 & -35.33 & -45.84 &   57.88\\
NGC\,3109 &1.43E+09 & 170 &8.64E-03 &   4.58 &   7.19 &    8.52\\
Antlia &3.50E+06 & 124 &3.76E-03 &   3.05 &  11.31 &   11.71\\
UGC\,4879 &1.42E+08 &  21 &2.20E-02 &   7.49 &  18.54 &   19.99\\
Sextans\,B &1.08E+09 & 152 &9.92E-04 &   7.74 &   4.08 &    8.75\\
Sextans\,A &1.04E+09 & 144 &2.58E-03 &   7.61 &   6.10 &    9.76\\
\enddata
\tablenotetext{a}{Mass used in calculations of orbits is simply luminosity times Mass-to-Light Ratio for galaxy type}
\tablenotetext{b}{Redshift in frame of the MW after backtracking. Small changes from input result from discreteness of time intervals in paths.}
\tablenotetext{c}{Sum of Square of gradients measures disagreement in the peculiar velocity at first timestep between backtracked orbit and linear theory expectation from the potential at that time.  Described in Appendix A3.}
\tablenotetext{d}{Output proper motion in the supergalactic longitudinal direction}
\tablenotetext{e}{Output proper motion in the supergalacitic latitudinal direction}
\tablenotetext{f}{Output amplitude of motion in the plane of the sky}

\end{deluxetable}
\end{onecolumn}
\begin{twocolumn}

\begin{landscape}
\clearpage
\begin{figure}
	\includegraphics[width=1.4\textwidth]{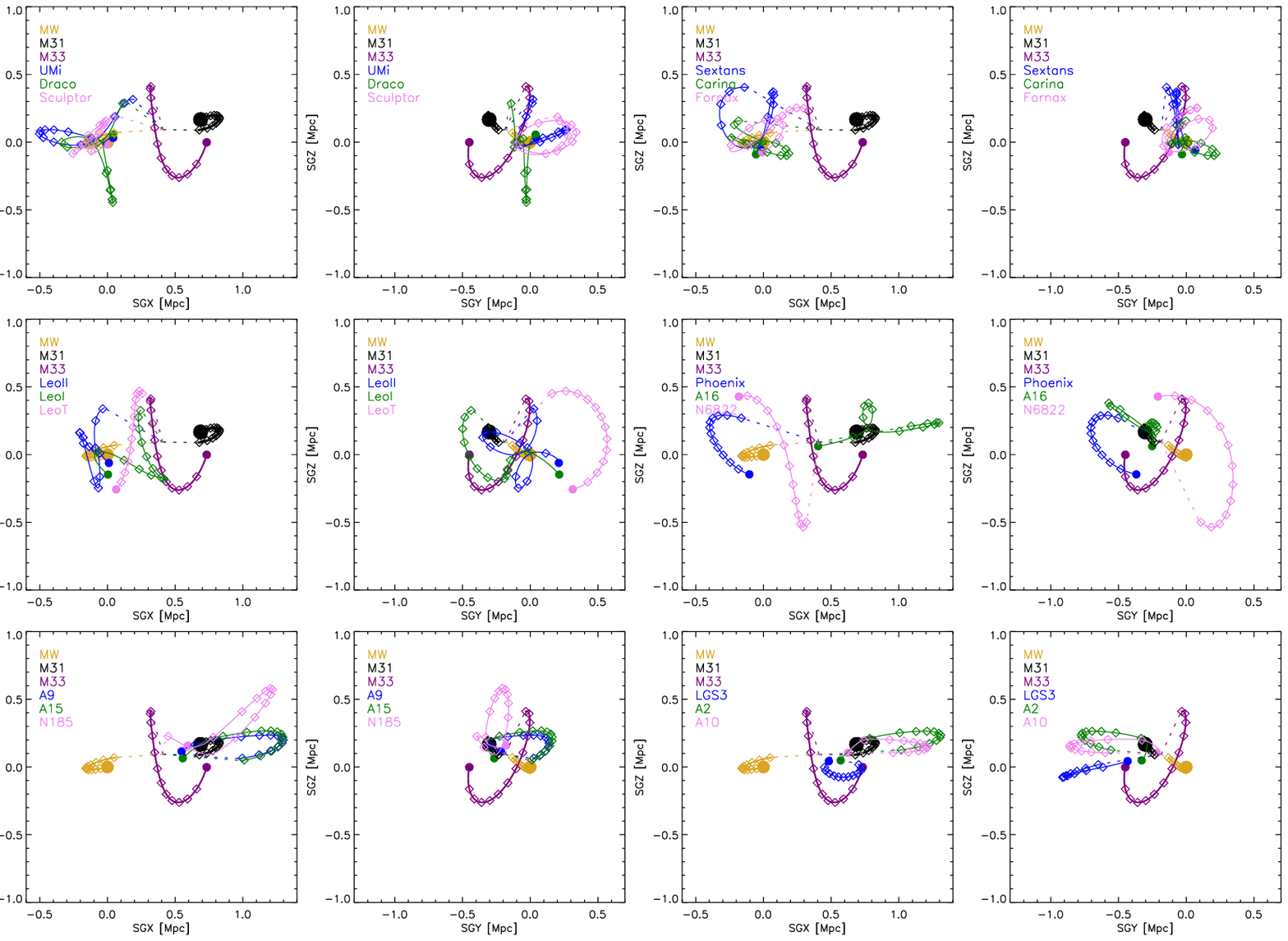}
	\caption{\it Orbits of Local Group Galaxies (d $<$ 1.5 Mpc) in physical coordinates and the LG frame.  There is a pair of plots (XZ and YZ projections) for a few galaxies at a time.  1 Gyr intervals are marked as boxes. Data range is the same for all except for the last 3 sets.}
\label{fig:eachc}
\end{figure}
\clearpage
\begin{figure}
	\includegraphics[width=1.4\textwidth]{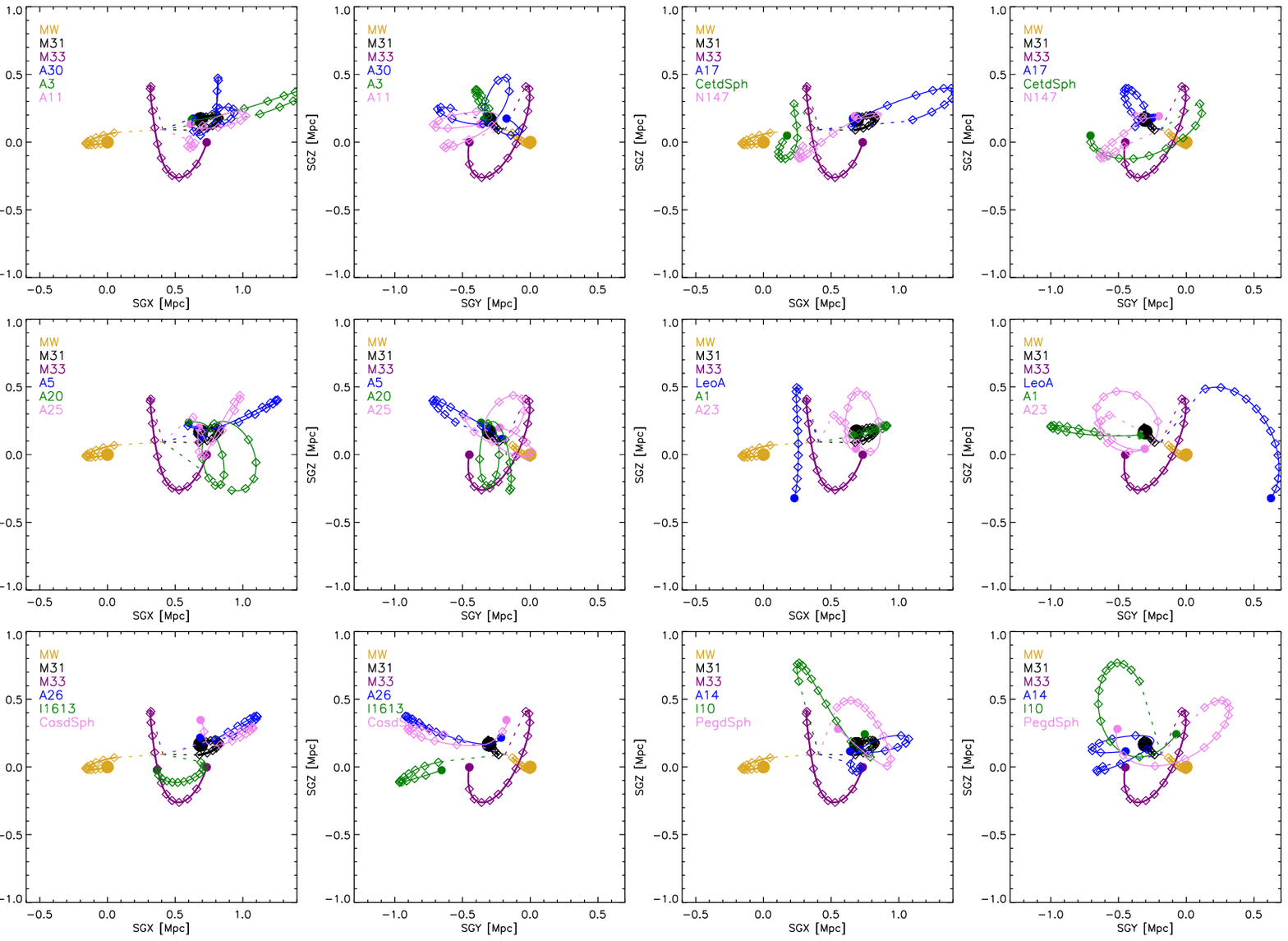}
\contcaption{\it Orbits of Local Group Galaxies (d $<$ 1.5 Mpc) in physical coordinates and the LG frame.  There is a pair of plots (XZ and YZ projections) for a few galaxies at a time.  1 Gyr intervals are marked as boxes. Data range is the same for all except for the last 3 sets.}
\end{figure}
\clearpage
\begin{figure}
	\includegraphics[width=1.4\textwidth]{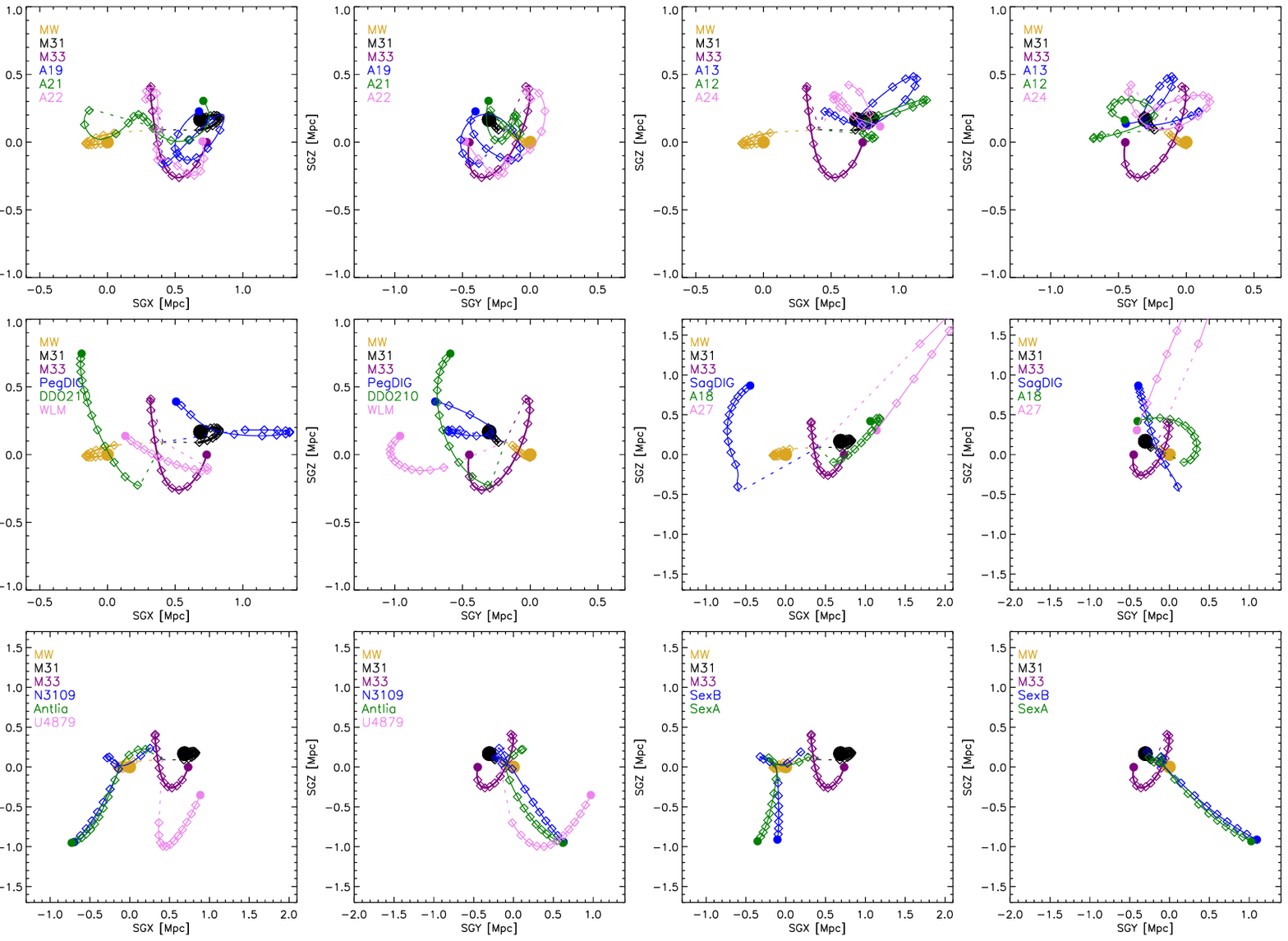}	
	\contcaption{\it Orbits of Local Group Galaxies (d $<$ 1.5 Mpc) in physical coordinates and the LG frame.  There is a pair of plots (XZ and YZ projections) for a few galaxies at a time.  1 Gyr intervals are marked as boxes. Data range is the same for all except for the last 3 sets.}
\end{figure}
\end{landscape}

\begin{figure}
	\includegraphics[width=.5\textwidth]{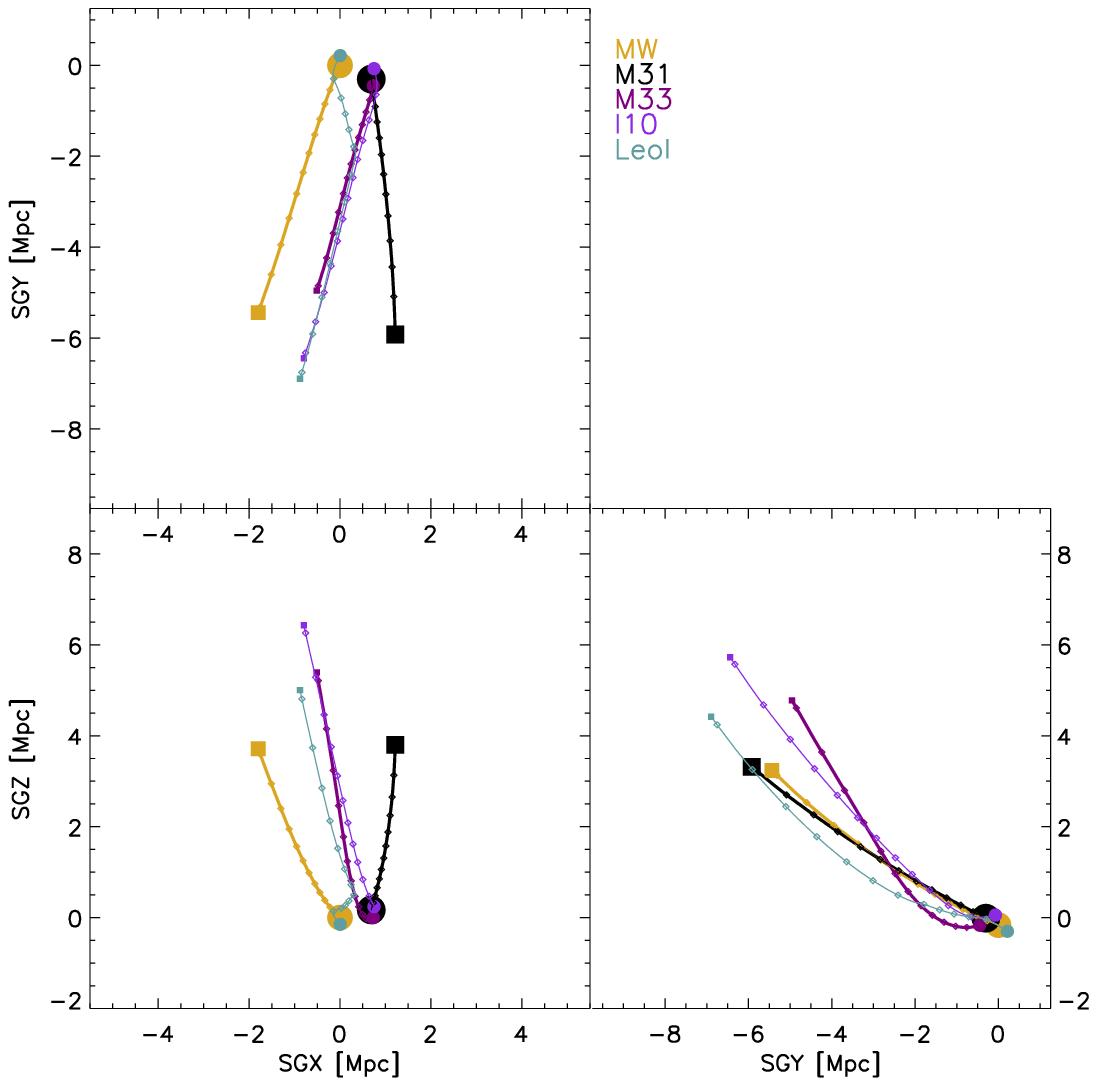}\\	
	\includegraphics[width=.5\textwidth]{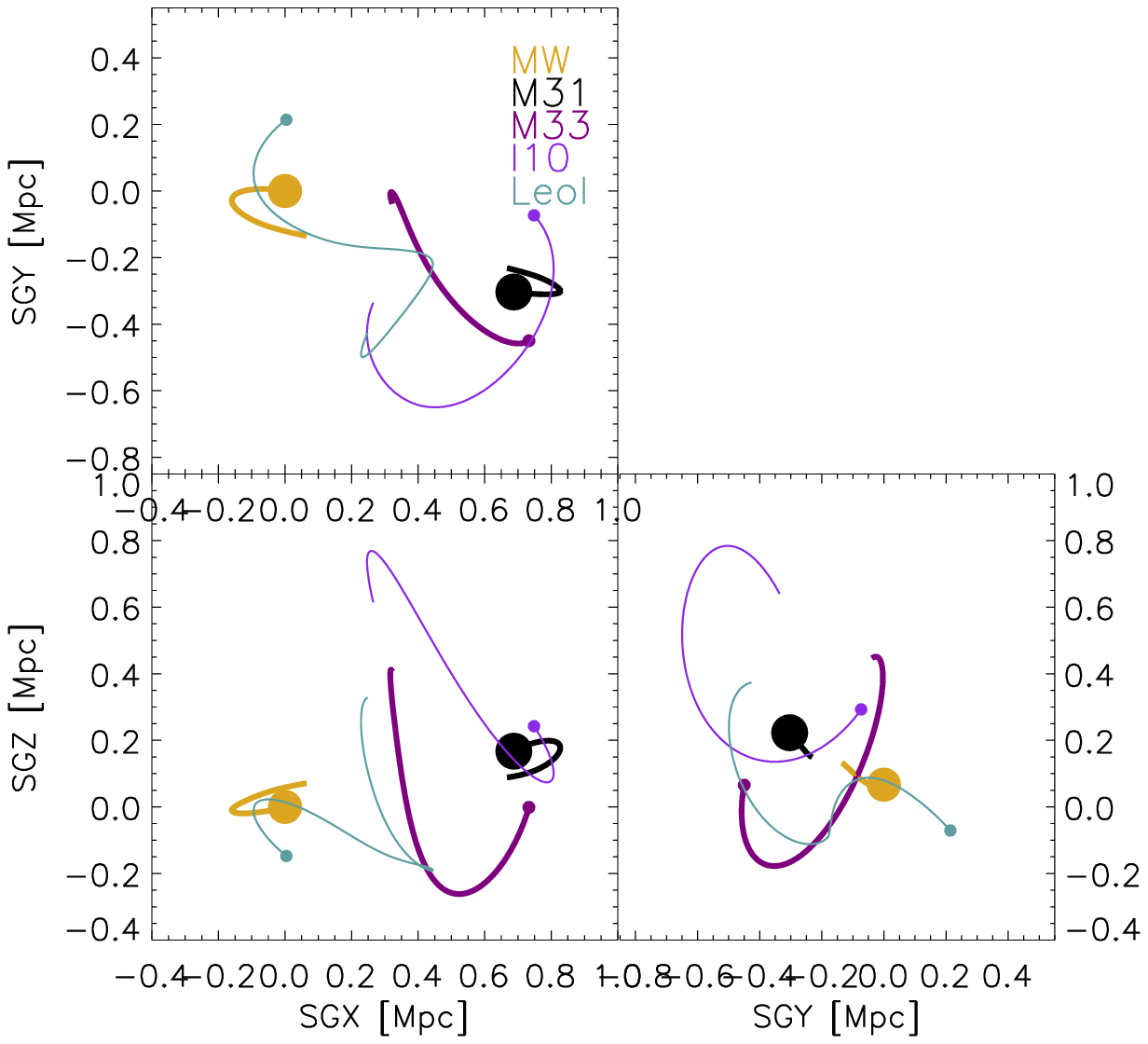}	
	\caption{\it Orbits for Leo\,I and IC\,10, galaxies with known proper motions.  Leo I has an interaction with M\,33 which then sends it towards an interaction with the MW.  Acceptable solutions with Leo\,I interacting with M\,31 also exist.  Top) Comoving coordinates and LSC frame, Bottom) Proper coordinates and LG frame of rest.  }
		\label{fig:i10_leoi}
\end{figure}

In Table~\ref{table:output}, we present some of the output results for each LG galaxy.
These include distance, recessional velocity, assumed mass,  and SGL and SGB component of proper motions. 
In Figure~\ref{fig:eachc}, we present, in physical coordinates and Local Group frame of rest, the trajectories of all LG galaxies, a few at a time, with SGX-SGZ diagrams on the left side and SGY-SGZ diagrams on the right.  

\subsubsection{M\,31 Satellites}.

\textbf{LG Plane\,1} is a thin plane of dwarf galaxies, nearly edge on to our line of site and tilted only 25\degree\ from the supergalactic equator.
M\,31 resides in it and  one can see in Figures~\ref{fig:plane1blue}, \ref{fig:plane1red}, and \ref{fig:plane1} that, in our model,   the MW quite recently moved nearly precisely into this plane. 
The galaxies with greater redshift than M\,31 are predominantly at higher SGL than M\,31, and the galaxies with lower redshifts than M\,31 are predominantly  at lower SGL and lower SGY. 
This redshift/blueshift split has been taken as evidence that this is a disk in rotation. 
In our model the story is a bit more complicated.
\begin{figure}
	\includegraphics[width=.48\textwidth]{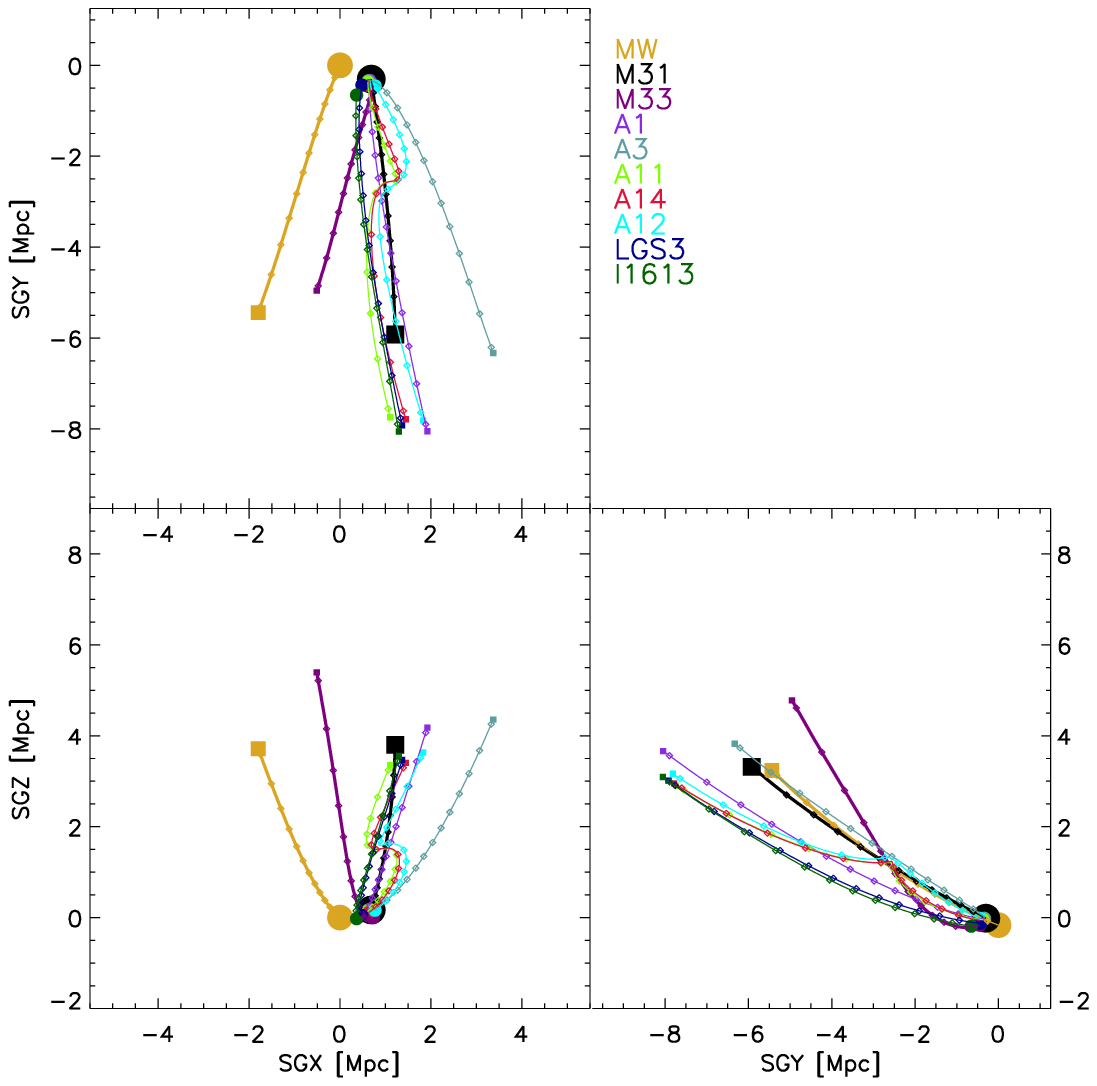}\\	
	\includegraphics[width=.48\textwidth]{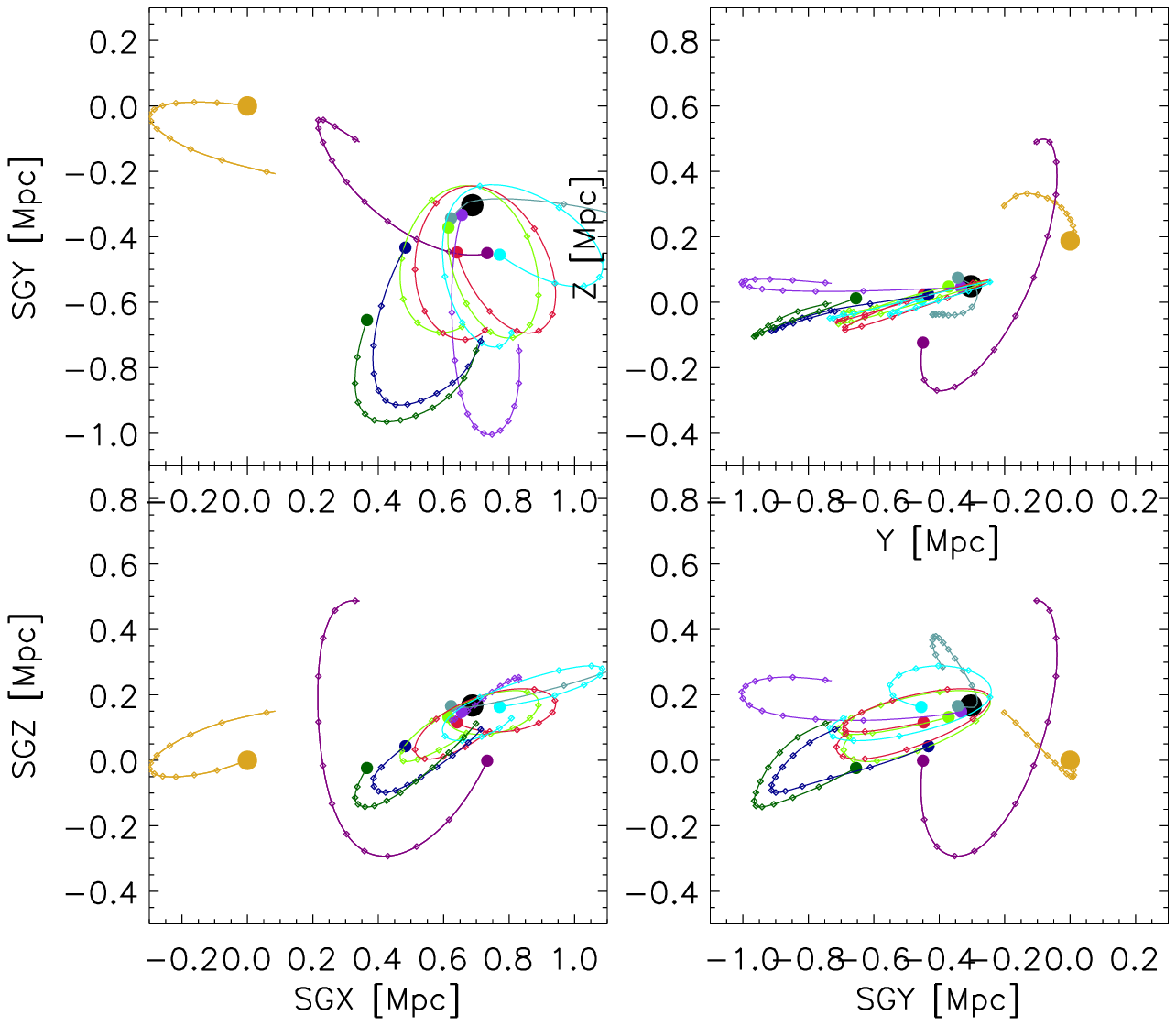}	
	\caption{\it Top) Orbits for \textbf{Plane\,1}, blueshifted only: comoving supergalactic coordinates, LSC frame. Bottom) Proper coordinates and in the M\,31 frame.  A fourth plot in the upper right is close to SGZ-SGY projection but there is a $25\degree$ rotation around the SGY-axis.}
		\label{fig:plane1blue}
\end{figure}
\begin{figure}
	\includegraphics[width=.48\textwidth]{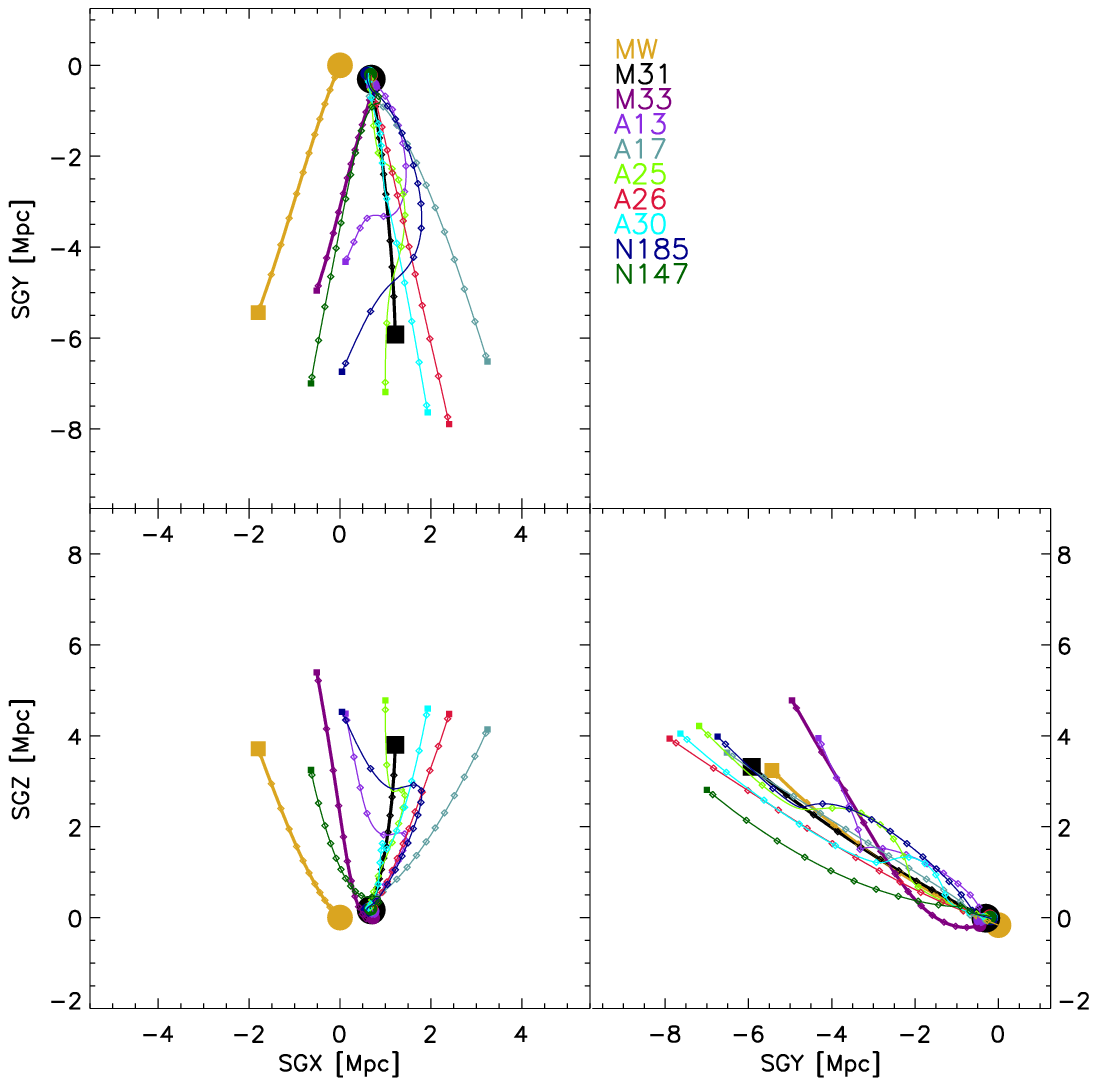}\\	
	\includegraphics[width=.48\textwidth]{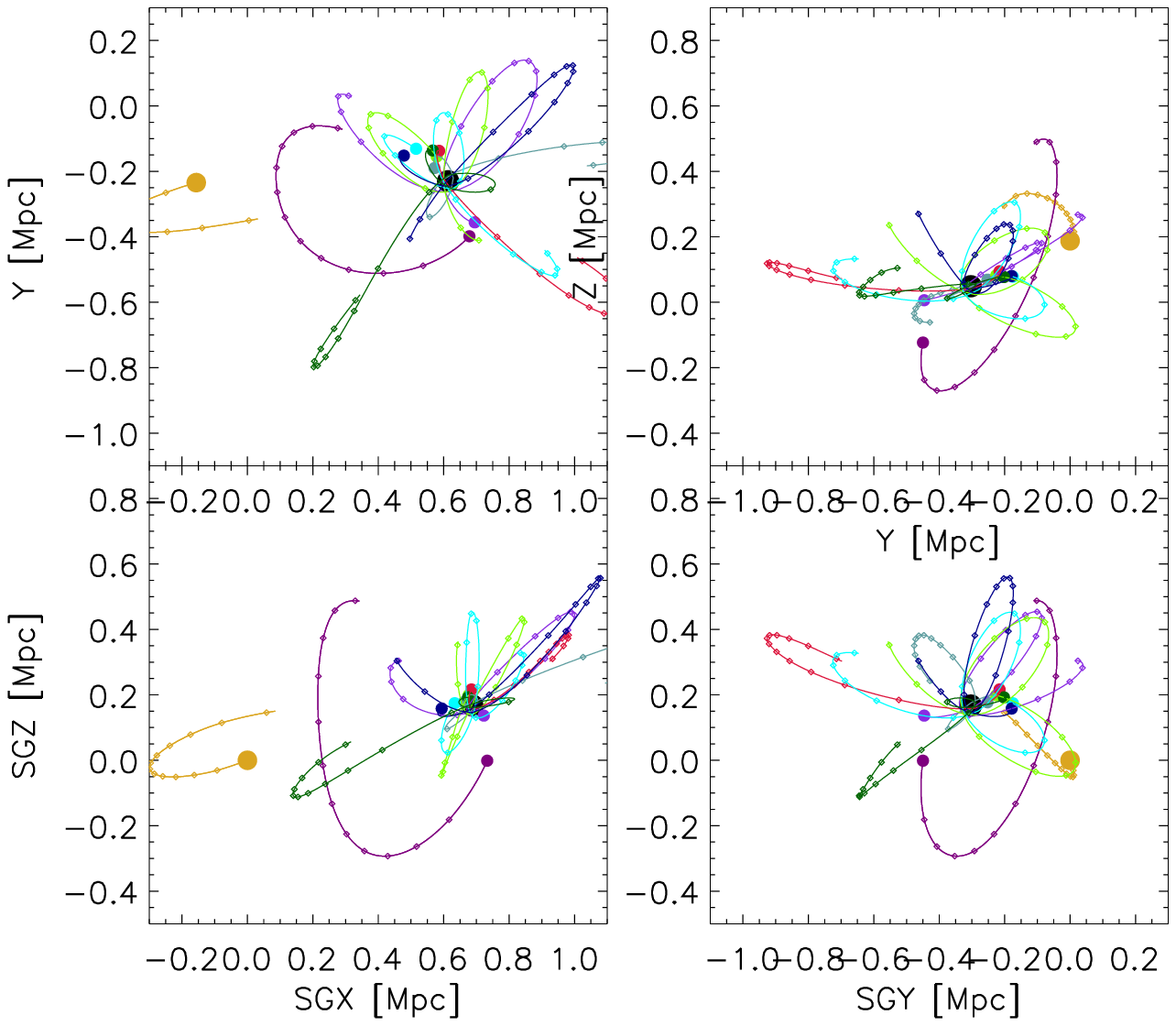}	
	\caption{\it Top) Orbits for \textbf{Plane\,1}, redshifted only: comoving supergalactic coordinates, LSC frame. Bottom) Proper coordinates and M\,31 frame.   A fourth plot in the upper right is close to SGZ-SGY projection but there is a $25\degree$ rotation around the SGY-axis.}
		\label{fig:plane1red}
\end{figure}   
In Figure~\ref{fig:i10_leoi} the 'best' orbit of IC\,10 is shown in physical coordinates and the LG frame. 
\citet{Tully2013} noted that IC\,10 resides in \textbf{Plane\,1} , however, since its orbit is highly inclined to \textbf{Plane\,1}  we conclude that it probably is not a real member of the plane and will pass through quickly.  
It would require a fairly large error in either its proper motion or that of M\,31 for it to be reconsidered for membership.
Table~\ref{table:pm} summarizes the relevant observed proper motions for the key galaxies and compares them to what we derive in our modeling.  All proper motions have been transformed to supergalactic coordinates.

The blueshifted subset of the \textbf{Plane\,1} galaxies, shown in Figure~\ref{fig:plane1blue},  started in the plane with a very narrow range in SGZ and most went out between 800 and 1000 kpc from M\,31 in the $-$SGY direction.
Some have already been around M\,31 and  gone back to the original side while others  
remained on the original side, but they are now moving together headed toward encounters  with M\,31.
On the top of Figure~\ref{fig:plane1blue}, we show the paths in the M\,31 frame of reference and comoving coordinates. 
On the bottom, proper coordinates are used. 
In the upper right of the bottom set, we show the SGY-SGZ plane after rotating by $25\degree$ around SGY to orient \textbf{Plane\,1} edge-on.
Most of the motion has been restricted to the plane.  
One can also see in the SGX-SGY projection, 
the effects of the strengthening tidal field from the MW on the orbits of And\,12 and And\,14,
the most extreme blue-shifted satellites,
whose orbits deviated far from ellipses after they flew by M\,31.
On the second half of their first orbit, the orbits begin to align with the MW$-$M\,31 line, and on their second descent they become very well aimed at both M\,31 and MW.  
This occurs, we surmise, because of the rapidly growing tidal field of the MW near M\,31.
The tidal field provides more time and distance for their velocities and paths to come into alignment.
However all of the motion has been taking place within \textbf{Plane\,1},  hence this phenomenon was not needed, in this case, to  bring these galaxies into a single plane, although it helped to further blueshift their velocities.
Currently all are located at lower SGY than M\,31 and have fairly high SGY velocities,
which account for most of their blueshift with respect to M\,31, even though, in SGX, most are moving away from us.
And\,1, LGS\,3 and IC\,1613 are still  on their first approaches to M\,31.

\begin{figure}
	\includegraphics[scale=1.2]{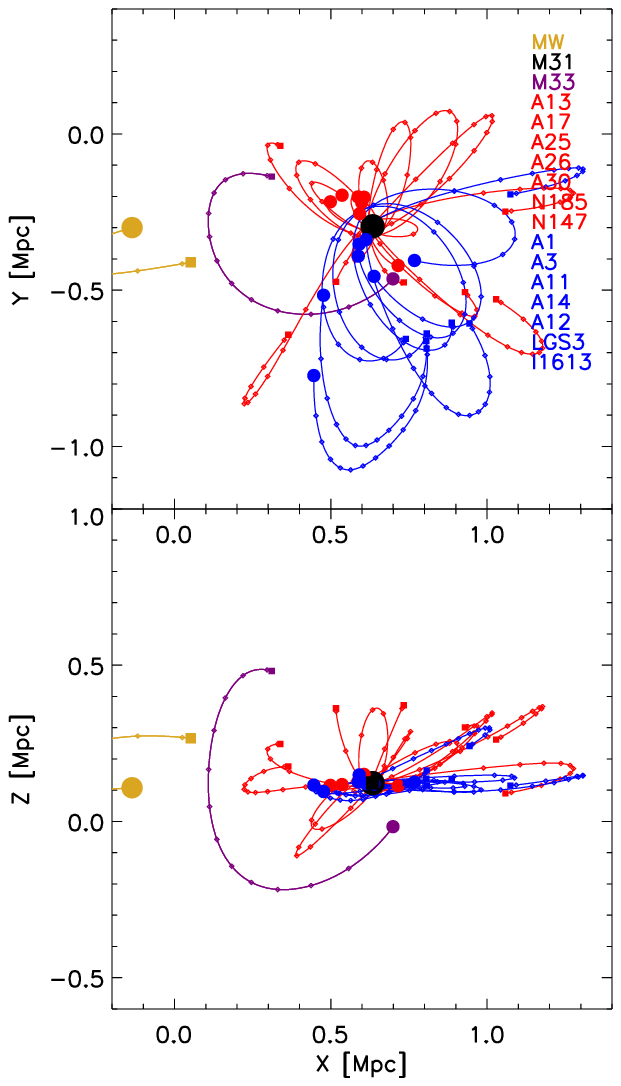}
	\caption{\it Orbits for Plane1 in physical coordinates and M\,31 frame of rest: Blueshifted relative to M\,31 are blue, redshifted are red.  Plot is rotated to place the MW$-$M\,31 line parallel to the X-axis, and the normal to \textbf{Plane\,1} in projection is parallel to the Z-axis.}
	\label{fig:plane1}
\end{figure}
The redshifted set is shown in Figure~\ref{fig:plane1red} and can be seen together
with the blueshifted set in Figure~ \ref{fig:plane1} .  
In the $25\degree$ rotated plot, one sees that only two, And\,13 and And\,17,  in the redshifted set have spent the entire Hubble time in the present plane.
They have looped around in various ways to positive SGY and are now falling away from us in SGY, but they  remained in the plane.  
Most of the redshifted set came from higher SGZ than the blueshifted ones and began outside of \textbf{Plane\,1}.  
NGC\,185, And\,25, and And\,30 run a fancy pattern that lets them drop in SGZ.  
They pass under M\,31 and then shoot up in SGZ.  
After they reach their apocentric points, the rest of their paths resemble those of And\,12 and And\,14 of the blueshifted set. 
They precess into a loop executed just below the MW-M\,31 line but, having been pulled away from M\,31, they have the time and path length to become well aimed along the MW-M\,31 line.
But these galaxies end up moving away from us in SGY and that gives them a positive velocity with respect to M\,31.


\textbf{LG Plane\,2} is obvious in an SGY-SGZ projection as a line of ten galaxies, including M\,33,  running below M\,31 in SGZ.
Four of its members, And\,2, And\, 5, And\,10, and And\,16, were in a line at greater SGX than the \textbf{Plane\,1}  galaxies at z=4 (Figure~\ref{fig:plane2}).
And\,22 has been bound to M\,33 from the beginning and the pair seem to have taken a unique route to the plane. 
And\,24 and And\,23 dropped down from higher SGZ than other \textbf{Plane\,2}  galaxies.  

\begin{figure}
	\includegraphics[width=.5\textwidth]{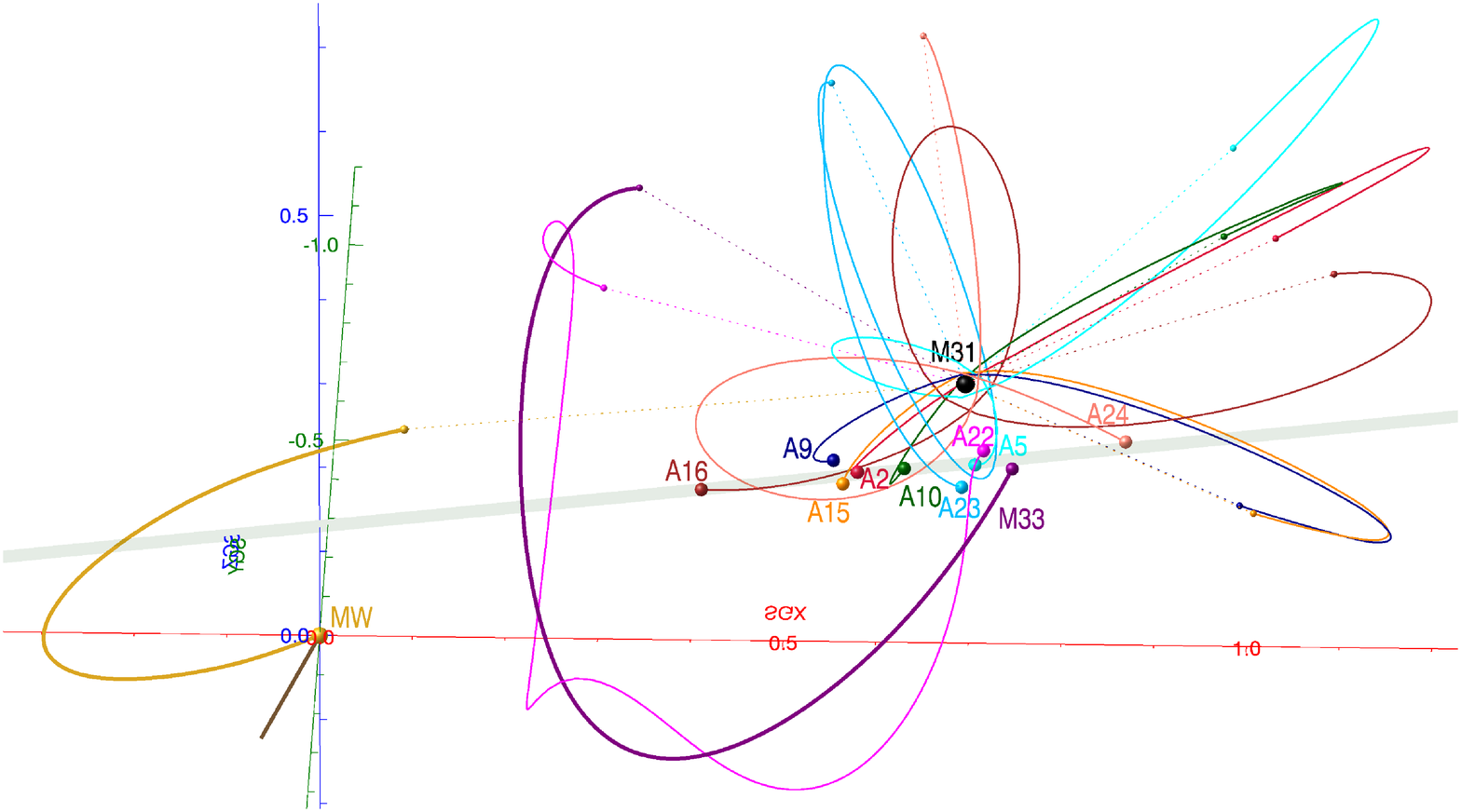}\\	
	\includegraphics[width=.5\textwidth]{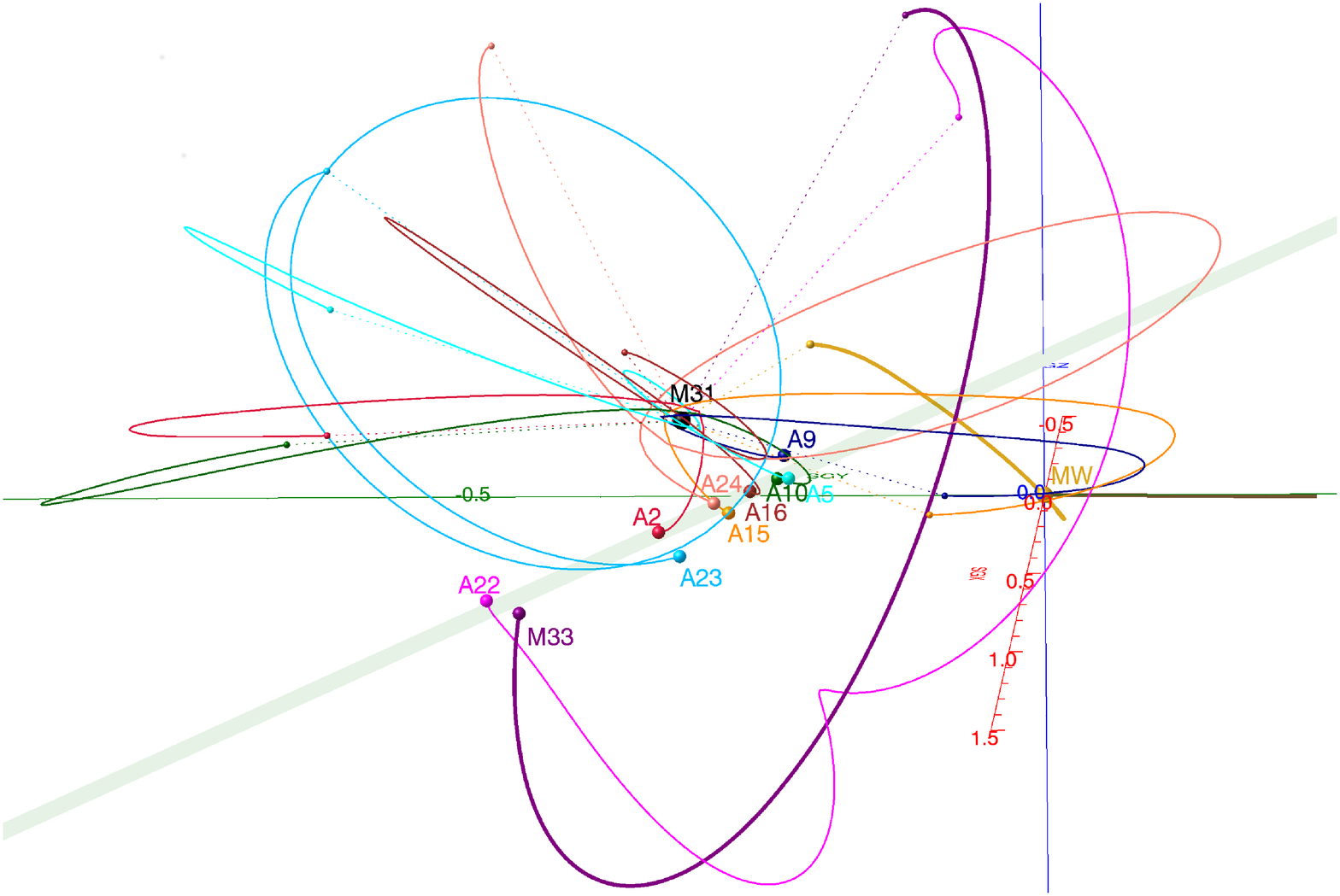}	
	\caption{\it Orbits for \textbf{Plane2}, physical coordinates in the M\,31 frame.  SGX,Y,Z-axes are red, green, blue respectively.  Light green stripe is plane fit to the galaxies.  A brown line is vector from MW to Virgo Cluster and is 250 kpc long.  Top) Axes are oriented close to XZ projection.  Bottom) Axes are close to YZ projection.}
		\label{fig:plane2}
\end{figure}

The \textbf{Plane\,2}  galaxies executed either a single or double passby of M\,31. 
Those that have executed a double pass reached peri-galactic points on the $-$SGZ side of M\,31, which sent them to high SGZ. 
They then passed M\,31 headed back to the negative SGZ side.  
Those executing a single pass, go over the top of M\,31 in SGZ, which pulls them simply towards the negative SGZ side.
Their distribution has been focused by M\,31's gravity to under 100 kpc distances from M\,31.
 They are all close to their turnaround in their SGZ component.  
This results in them gathering at the SGZ location of the plane and to spread out in the other components, much like the top of a water spout that has reached its maximum height. 
In other words, \textbf{Plane\,2}  may be a density cusp that occurs at turnaround points when objects fall into a potential from similar heights.

\textbf{Plane\,2} and \textbf{Plane\,1}  intersect at the leading sides of both, forming a 'V'.
Consequently, And\,16 and And\,9, near the vertex, could be considered to be  in \textbf{Plane\,1}  as well.
The gap of the 'V' appears to  be due to the fact that galaxies spend little time in that region since they pass through there immediately after getting a downward boost as they pass M\,31.

\textbf{The M\,31-M\,33 H\,I Bridge}  may be comprised of debris scattered from interactions of M\,31 with 
 And\,2 and perhaps And\, 15.   
We plot the paths of these galaxies on the plane of the sky overlaid on a PAndaS image of Andromeda with H\,I contours, taken from \citet{Lewis_etal2013} in Figure~\ref{fig:bridge}.  
And\,15 and And\,2 have passed through the disk of M\,31 in the past 0.7 Gyr.  
Tidal tails plus ram pressure driven gas from either of them and M\,31 may have created the H\,I bridge and the Giant Stellar Stream \citep{Ibata_etal2001} of M\,31.
The warped outer gas disk of M\,31 has been given as evidence of an M\,31$-$M\,33 encounter.  
However, And\,22 is bound to M\,33 and their interactions could explain the highly distorted outer H\,I contours of M\,33.

\begin{figure}
	\includegraphics[width=.5\textwidth]{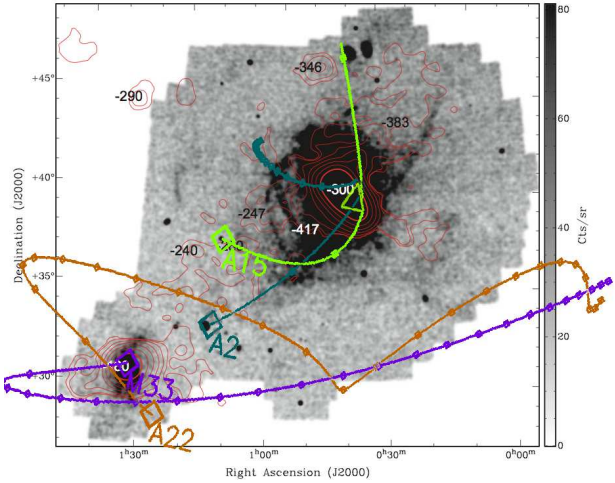}	
	\caption{\it The orbits of M\,33, And\,22, And\,2, and And\,15 projected onto the plane of the sky are overlayed on an image from \citet{Lewis_etal2013} which is itself an overlay of H\,I contours on top of greyscale image of PAndAS survey of M\,31.   Numbers on H\,I contour lines are heliocentric velocities.  Intervals of 1 Gyr are marked as small squares along the galaxy paths.   This implicates And\,22 and And\,15 in raising the Giant Stellar Stream seen along their paths.  And it implicates And\,22 in the distorted H\,I contours and stellar tail seen in M\,33  }
		\label{fig:bridge}
\end{figure}

\textbf{Leo A} (see Fig~\ref{fig:eachc}) has gone out far (700 kpc) in SGY and has recently passed the MW$-$M\,31 plane and is far enough away that it has been primarily attracted to the center of mass of the Local Group rather than either of the giants.
It thus has substantial tangential motion to us and perpendicular to \textbf{Plane\,3} , so it will eventually cross it.
Its present location happens to mark  the point of convergence of the M\,31 and the MW in the LSC frame. 

\textbf{And\,27} and the more distant galaxy NGC\,6789 have  too highly blueshifted velocities to be explained without invoking some interaction outside of the Local Group.  Both, it turns out, are in the direction of NGC\,6946, and we have found solutions that involve the former two in tight interactions with it (Figure~\ref{fig:And27}).  And\,27 was considered a \textbf{Plane\,1}  member by \citet{Ibata_etal2013}, but we conclude that, like IC\,10, it is just passing through and have removed it from \textbf{Plane\,1}  membership.  However,  the redshift measurement for And\,27 is not of high signal-to-noise and may be flawed.
\begin{figure}
	\includegraphics[width=.47\textwidth]{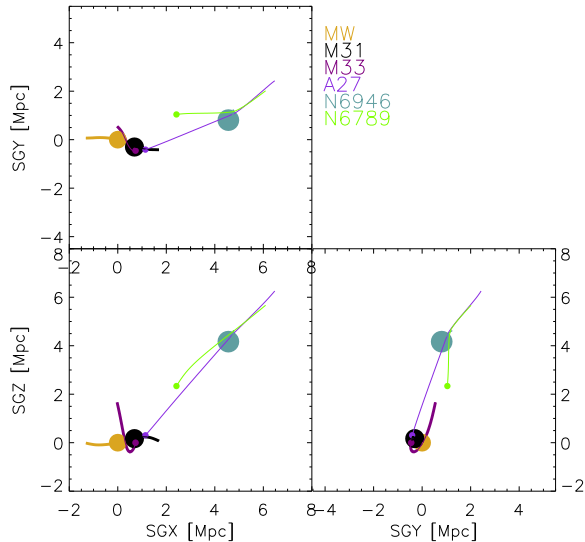}	
	\caption{\it Orbits And27 and N6789 getting gravity assist from N6946.  LG frame of reference and comoving coordinates were used.}
	\label{fig:And27}
\end{figure}

\subsubsection{MW Satellite Galaxies}

The galaxies around the MW divide up into two thin planes.  The outer dwarfs (Leo\,I, Cetus Dwarf Spheroidal, WLM, Leo\,T, NGC\,6822, DDO\,210, and SagDIG) are in a plane with rms width of 34 kpc and 1.2 Mpc in its longest extent.
We will refer to this structure as \textbf{LG Plane\,3}.
The set of inner dwarf galaxies,  \textbf{LG Plane\,4} , out to Leo\,II, at 220 kpc, plus Phoenix, at 410 kpc has an rms in the distances  from the best fit plane of only 11.3 kpc.  
Planes with slightly different sets of satellite galaxies, have been discussed previously \citep{LyndenBell1976, Metz_etal2007, Pawlowski_etal2012}, except that we do not include the LMC or SMC, for reasons discussed below.  
These two MW planes were presented in Figure~\ref{fig:presentplanes}.  

Our best model orbits for \textbf{Plane\,3}  are presented in Figure~\ref{fig:plane3}, and for \textbf{Plane\,4},  in Figure~\ref{fig:plane4}.  Motions in comoving coordinates and the LSC frame for both \textbf{Plane\,3}  and 4 are shown in Figure~\ref{fig:plane3d}.
The view is edge-on to \textbf{Plane\,3}  and nearly normal to \textbf{Plane\,4}.
The location of the starting points, at z=4, for most of the galaxies in the two planes was the tilted initial plane shown in Figure~\ref{fig:initial}.   

\begin{figure}
	    \includegraphics[width=.47\textwidth]{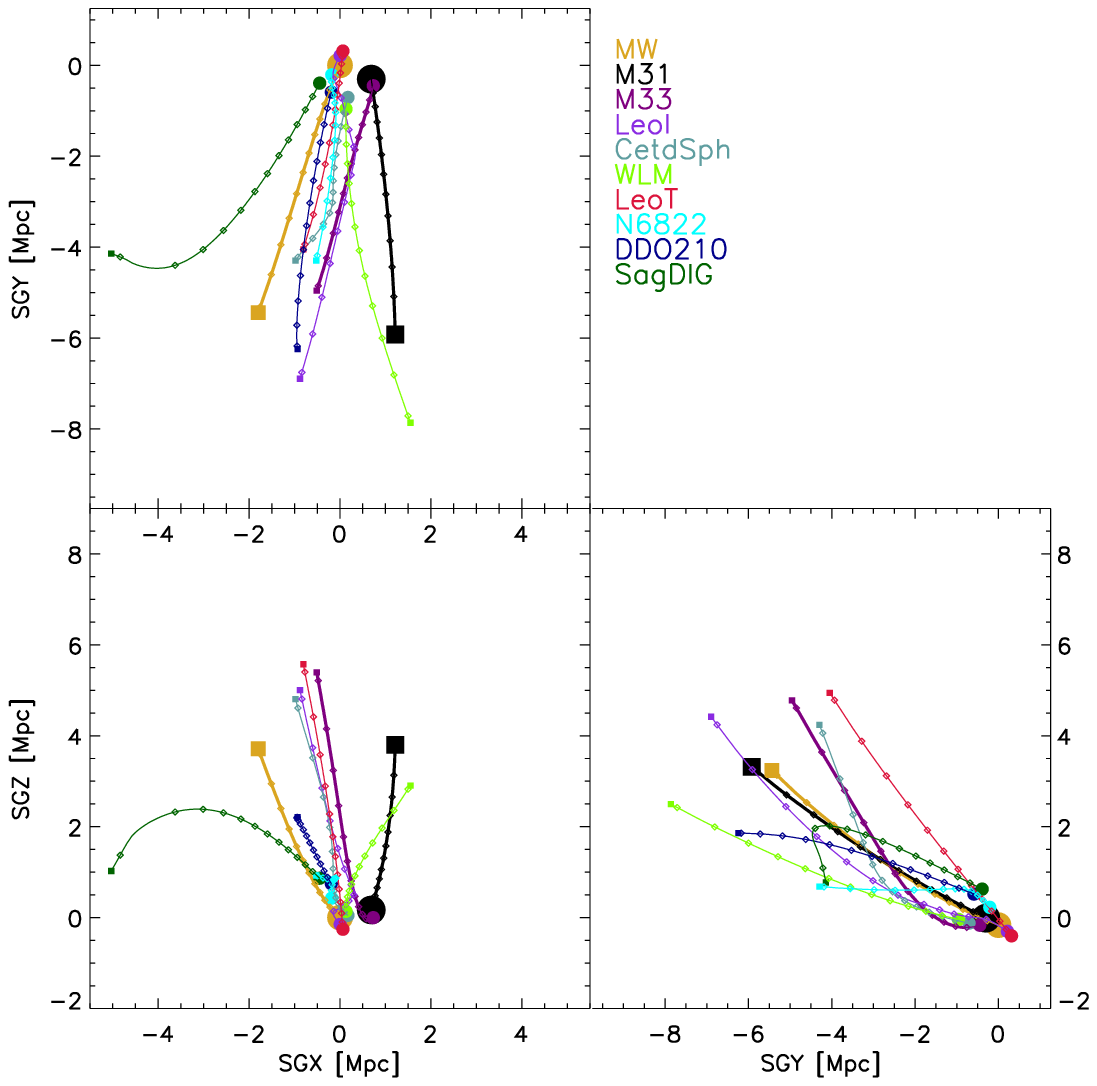}\\	
	      \includegraphics[width=.47\textwidth]{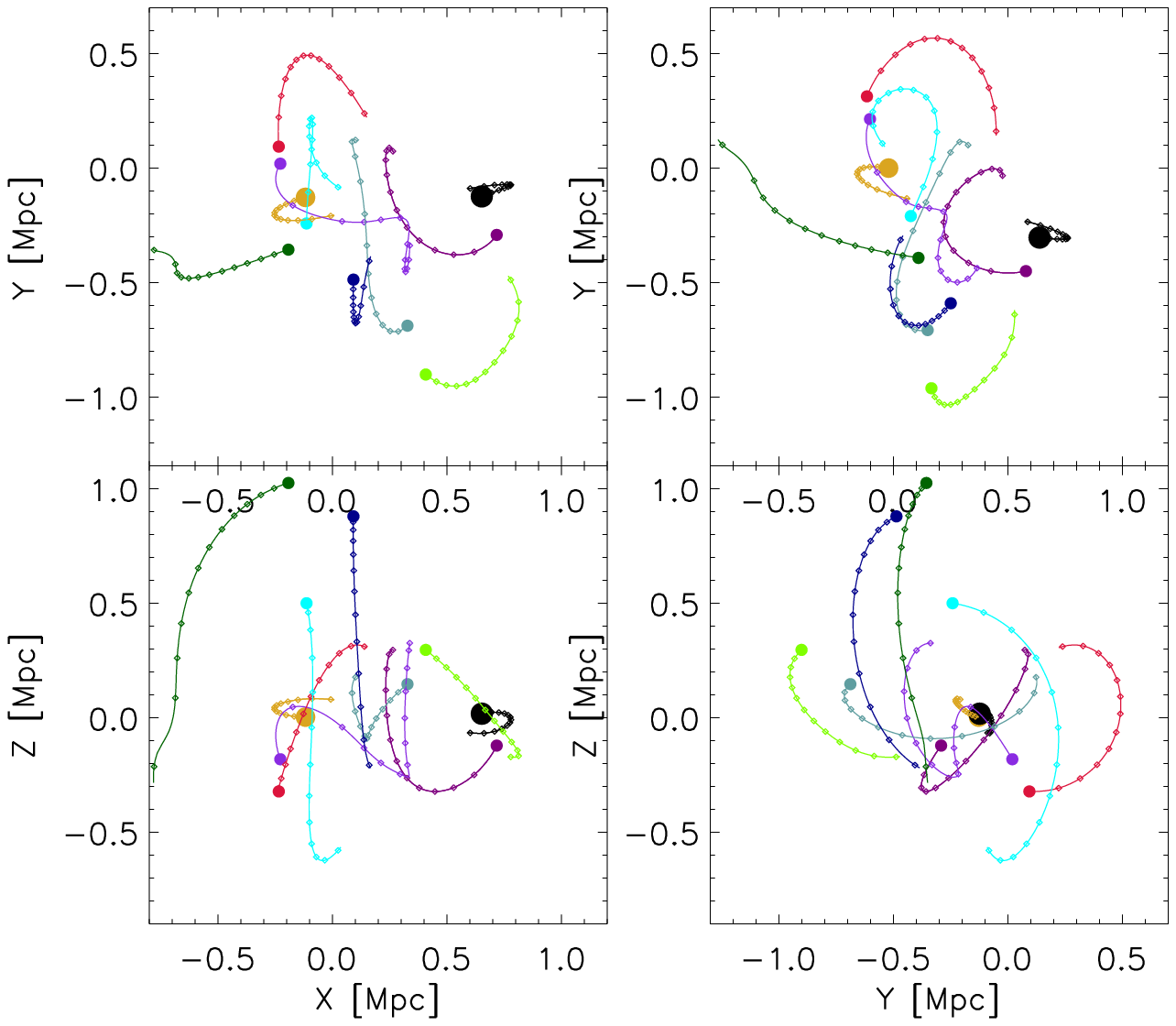}	
	\caption{\it Orbits for \textbf{Plane\,3} . Top) Co-moving coordinates, LSC frame. Bottom) Proper coordinates, LG frame. Upper right panel is tilted from one to the left to an orientation edge on to \textbf{Plane\,3} .}
	\label{fig:plane3}
\end{figure}

\begin{figure}      
                \includegraphics[width=.5\textwidth]{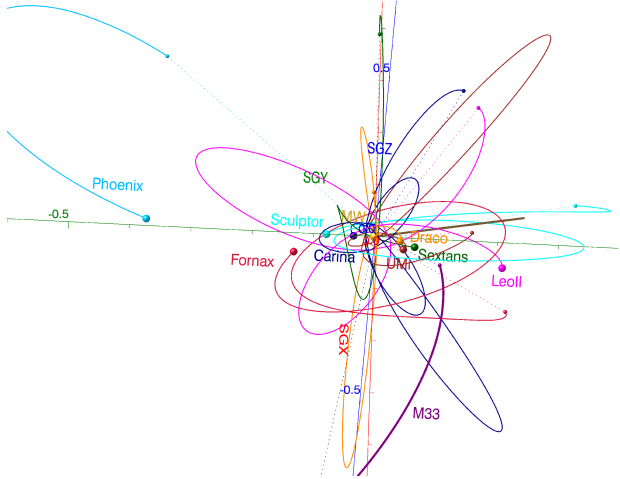}	
	\caption{\it Trajectories of \textbf{Plane\,4}  galaxies in physical coordinates and the MW frame of rest.  The view is projected onto a plane that includes the Virgo Cluster.  A dark brown line stretches out from the MW in the direction to the Virgo Cluster and is 250 kpc long.  Large spheres are placed at present positions and small spheres are at z=4 positions.  
Most of these galaxies started at 100 to 250 kpc to positive SGZ of the MW and have been interacting closely with it.
The direction of greatest elongation of \textbf{Plane\,4} 's  distribution is close to the direction to Virgo. }
	\label{fig:plane4}
\end{figure}
\begin{figure}      
	      \includegraphics[width=.5\textwidth]{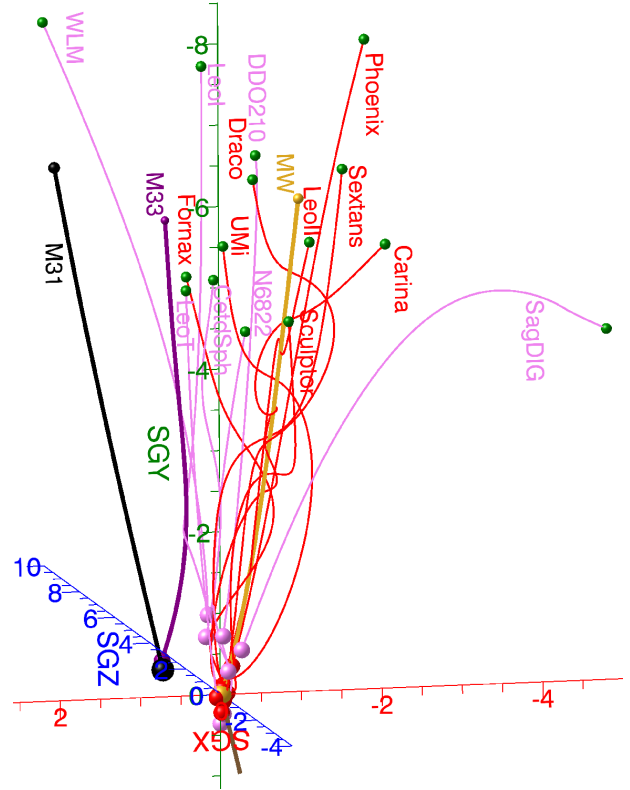}	
	\caption{\it The trajectories of \textbf{Plane\,3}  galaxies (violet) and \textbf{Plane\,4}  (red) galaxies in comoving coordinates and in the Local Supercluster frame.  Small spheres are at the z=4 positions (green for \textbf{Plane\,3}  and \textbf{Plane\,4}  galaxies) and larger spheres are at the present positions. A brown vector stretches 1 Mpc from the MW toward the Virgo Cluster and is in the plane of the plot.  \textbf{Plane\,3}  galaxies were either between M\,31 and the MW or at large distances.  They followed the LG center of mass and only recently did they distinctly fall toward the MW.    Most remain behind the MW, but are catching up.
For \textbf{Plane\,4} , most have been interacting more closely with the MW.  Both sets of galaxies are stretched out along the 
 the MW large scale motion which is toward the Virgo Cluster. }
	\label{fig:plane3d}
\end{figure}

An intriguing question is, why are these planes nearly perpendicular to the M\,31$-$MW line, when M\,31 is presently the dominant tidal stressor?
The MW, in the LSC frame of reference has been mostly moving toward the Virgo Cluster at SGX,Y,Z=($-$3.7,16.3,$-$0.7) and partly moving to lower SGZ.
Interestingly, \textbf{Plane\,3}  galaxies tight-roped in between M\,31 and the MW.
They were either between M\,31 or arriving from large distances and therefore attracted to the LG center of mass rather than either one of the giants.  
Only recently did they begin to clearly fall toward the MW.   
Most are still behind, never having had a close flyby of the MW. 
The angle between the plane and the MW motion is simply due to a change in gravitational dominance from the center of mass at earlier times to the MW more recently.  

For \textbf{Plane\,4} , most started at about the same SGZ level as the MW and so have been interacting closely with it.  
Again, one sees that the direction of greatest elongation of \textbf{Plane\,4} 's distribution is aligned with the MW large scale motion.  
The particular orientation of the best fitting plane, and the high axial ratio for \textbf{Plane\,4}  depends a great deal on Phoenix.  
But, Phoenix is an intermediate to \textbf{Plane\,3}  in that it started fairly far from the MW and still has not had a near interaction with the MW, 
although it has always been on the far side of the MW from M\,31.

In addition, both of the MW planes are oriented close to the neutral points in the tidal field of M\,31, 
where there is a balance between inward forces along the perpendicular and outward forces along the line to the source of the tide.
Perhaps, avoiding either inward or outward forces was helpful for the survival of these galaxies.
Natural selection may be very important to understanding these structures in that satellites that did not evade the M\,31 tidal field might more readily have been destroyed.
In Figure~\ref{fig:plane3}, one notes that the z=4 distribution simply began quite narrow perpendicular to the plane and already wide in SGY.  
The small range in initial SGX for \textbf{Plane\,3}  and 4 galaxies could be attributed to
the fact that the MW sits between M\,31 and the Cen\,A Group.  
Assuming the ratio of masses for M\,31 and the MW is similar to the luminous ratio, then M\,31 would have pulled in galaxies quite close to the MW on its side.
Centaurus\,A, now at 3.6 Mpc, was only 1.6 Mpc away as recently as z = 1 (t$\sim$6 Gyr). 
At this time, M\,31 was just reaching its maximum separation from the MW of $\sim$1 Mpc. 
The Roche limit radii at the MW of M\,31 and Cen\,A were, therefore, comparable at z $>$ 1.
 
\textbf{LMC and SMC} have been examined but do not have any definitive solution that we can find.  
All allowed solutions, given the MW mass being used in this model of $1.8 \times 10^{12} \Msun$,
result in these two satellites remaining tightly bound satellites.  
For an assumed V$_c$ of 240 \kms, the tangential velocity is 313 \kms\  and the radial motion is 54.7 \kms, for a ratio of 0.17.  Thus, the ratio of the energies is 3\%.  
This is unique among the satellite galaxies and is probably an indication that the LMC has undergone circularization from dynamical friction.
The SMC has less total energy than the LMC and simply does not have the energy that would be expected if it fell from a substantial distance from the MW.  
It too must have undergone substantial orbital dissipation.

As a check that there has been enough time for dynamical friction to have played an important role in the Magellanic clouds we use Equations 3 and 5 in \citet{Zhao2004}, which together give for the dynamical friction time for angular momentum loss of a massive object inside a host halo as :
\begin{equation} \label{eq:tdyn0}
t_{dyn}^{-1} = 4 \pi G \rho \frac{Gm}{v_c^3} \frac{\ln \Lambda}{\frac{4}{3} + u^3}
\end{equation}
where m is the satellite mass, $\rho$ is the density of the host at the position of the satellite, $v_c$ is circular velocity  in the rotation curve of host at this point, $\ln \Lambda$ is the dimensionless Coulomb logarithm, and  $u$ is the velocity of the satellite divided by $v_c$.

Assuming the host is an isothermal spherical galaxy, we can substitute $\rho = M_r/(4\pi r^3)$ and $v_c= 2 \pi r /P$, where $M_r$ is the mass interior to r of the host and $P$ is orbital period around the center of the host at r,  to find the convenient formula for dynamical time:
\begin{equation} \label{eq:tdyn}
t_{dyn} = \frac{P}{2 \pi} \frac{M_r}{m} \frac{\frac{4}{3} + u^3}{\ln \Lambda}
\end{equation} 

As a concrete example, take a satellite orbiting the MW with velocity 1.5 times the local circular velocity and  $\ln \Lambda = 2.5$. 
If we set $t_{dyn}$ = 10 Gyr, then we find $m=3.0 \times 10^9 \Msun$ at 50 kpc.  
Since $P$ and $M_r$ grow linearly with r, this mass limit scales as $r^2$. 
If a MW satellites has mass below this estimate, 
it probably has not yet lost much orbital energy from dynamical friction.  
But for galaxies with masses near or above this mass limit, one needs to consider how much time has been spent at small Galactic radii.  
At 50 kpc and with tangential velocities much greater than radial velocities, a reasonable case can be made that both the LMC, with stellar mass of $2.7 \times 10^9 \Msun$, and the SMC, with total mass of $2.4 \times 10^9 \Msun$ \citep{BekkiChiba2008}, have undergone significant circularization of their orbits.
On the other hand, all of the other satellites in this study appear to be safely below this red line in mass, provided that a majority of  initial mass has not been lost.

\subsubsection{NGC\,3109 Association}

The NGC\,3109 Association, including Antlia, Sextans\,A, and Sextans\,B, has always been a bit of a puzzle.  It is $\sim$1.3 Mpc from the MW,  near to the LG turn around radius, yet the redshifts of these four galaxies in the MW frame are between 120 and 160 \kms\ or effective Hubble constant of $>90$ km/s/Mpc rather than an expectation of nearly zero. 
Until now it was thought that the explanation required that the  MW has a substantial velocity away from their direction.  
But, if the proper motion measurement of M\,31 is correct, then the motion of the MW is nearly perpendicular to their direction.
The group simply has to come through or around the LG at very high velocities.
Our solutions from backtracking  (Figure~\ref{fig:n3109}) have these galaxies beginning very close to the MW's line of motion and interacting with the MW to gain their high velocities through gravitational assists.  
It appears that the underlying cause of the gravity assist is the tidal fields from M\,31 and Cen\,A and the entire Local Sheet.   
At early times, M\,31 and Cen\,A provided a large tidal field that accelerated the association inward toward the MW, but by the time the galaxies passed the MW, the gravity fields from M\,31 and Cen\,A had subsided considerably.  
The time dependent asymmetry provided the extra energy they needed to get away from the MW with some excess velocity. 
For Antlia, Sextans A and Sextans B, a single pass from trailing to leading the MW $\sim$ 7 Gyr ago at  $\sim$ 25 kpc from the Galactic center generated the needed impulse. 
For NGC\,3109,  there were two passes by the MW separated by 1.5 Gyr and getting as close as 24 kpc. 
\begin{figure}       
	\includegraphics[width=.48\textwidth]{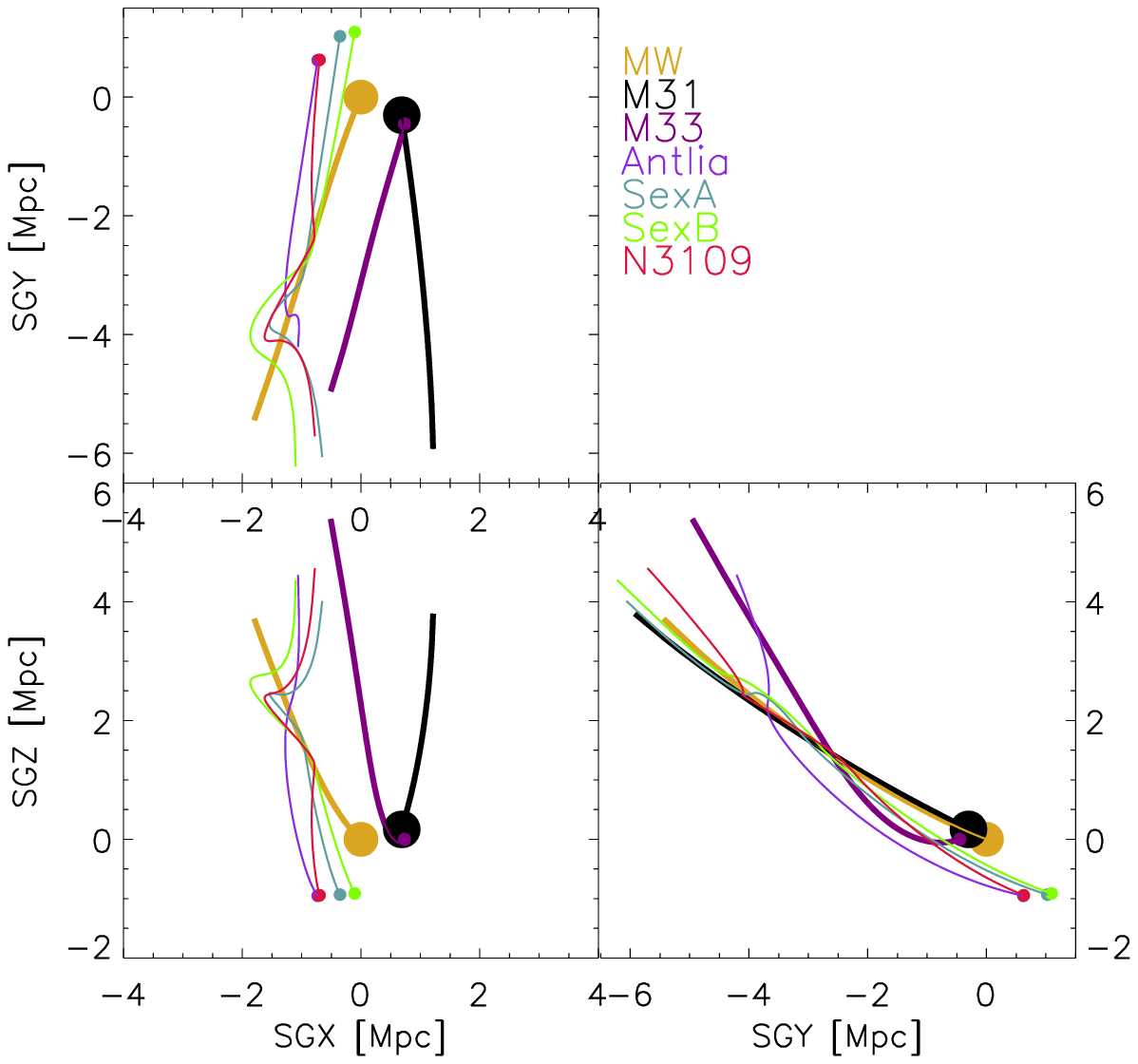}\\	
	\includegraphics[width=.48\textwidth]{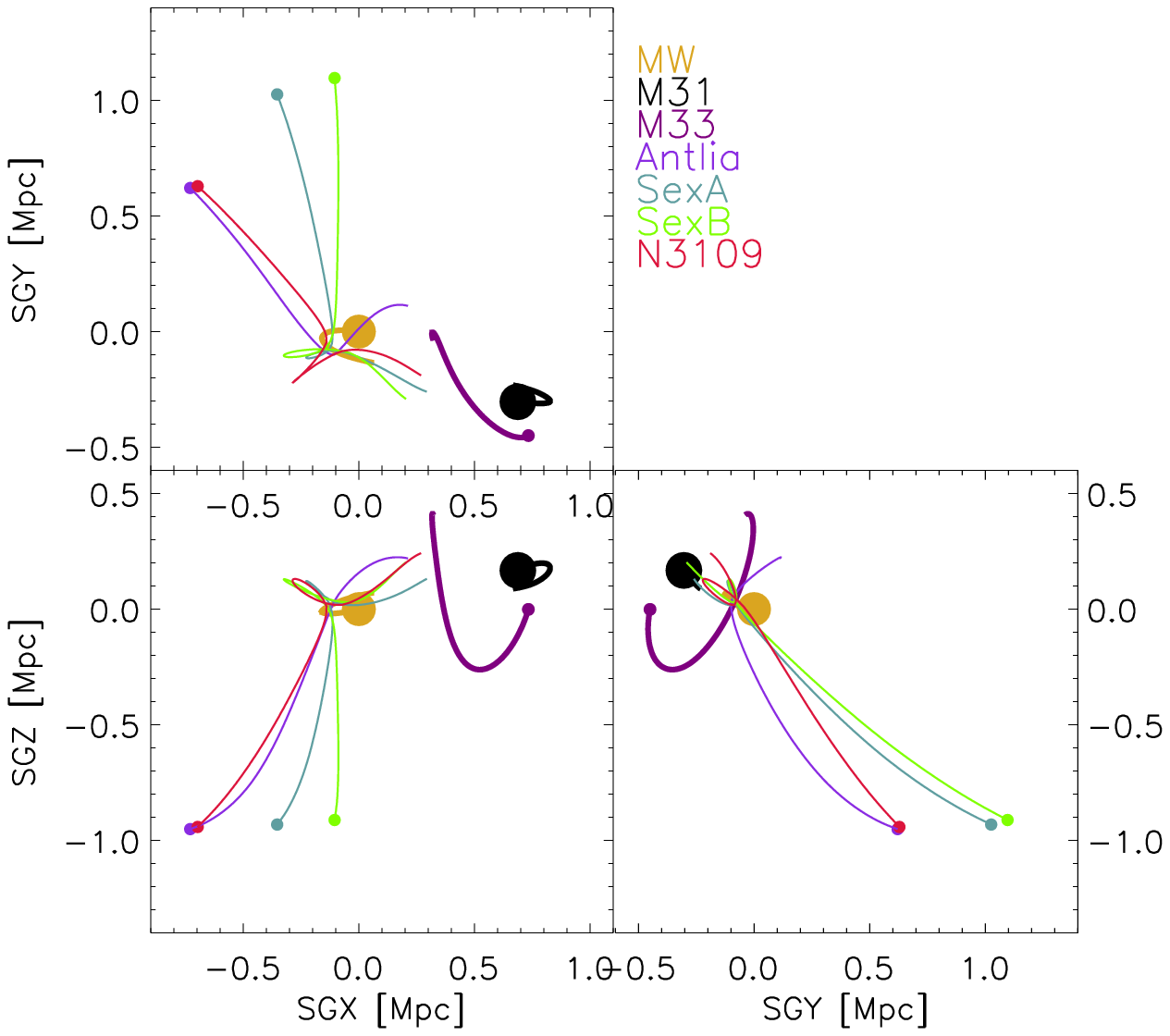}	
	\caption{\it Orbits for the NGC\,3109 Subgroup of galaxies: 
	Top) Comoving Supergalactic Coordinates, Large Scale Frame, Bottom) Proper Coordinates, LG frame.  
Note that these 4 galaxies lie in an SGZ plane as well.  
The motion is well aligned with the Virgo Cluster whose tidal field probably played a role. }
\label{fig:n3109}
\end{figure}
   
This discussion settles an issue over whether these dwarf galaxies form a separate group.  They should be considered Local Group galaxies since one needs to consider them when doing the dynamics of the Local Group.  
NGC 3109, Sextans\,A, and Sextans\,B are probably dominated by dark matter halos \citep{JobinCarignan1990, Huchtmeier_etal1981} with dark matter to baryon ratios out to the last measured points of between 4 and 9. 
Therefore, these are probably not tidal remnants from a merger or disruptive encounter with the MW, 
and they made these close flybys without losing their dark matter.

\section{Discussion}\label{sec:Discussion}

It should be kept in mind that this work is only a first cut at solving orbits for these planes of galaxies in that none of the relevant parameters that would normally be adjusted to create a better looking model were adjusted. 
Specifically, we chose not to make use of the $\sim$5\% distance uncertainties, the nearly completely unconstrained mass-to-light ratio for individual galaxies, or a possible several \kms\  disparity between measured baryonic motion at the center of a galaxy and the center of mass motion of a galaxy's dark matter.  
Only one galaxy needed its distance altered to find a solution consistent with the potential from the early time distribution of galaxies out to large distance.  
A single mass-to-light ratio for all spiral galaxies and a slightly higher value for all elliptical galaxies in the entire sample were used.  
By not taking liberties with the free parameters, we hopefully have explored a larger set of possible islands in phase-space that could create these planes.  
It could be that for each plane only one of these islands was actually populated and the other trajectories are spurious, but it is not too unreasonable that galaxies originating from two nearby initial islands happen to form a single plane.
In particular, the redshifted and blueshifted \textbf{Plane\,1}  galaxies do suggest two slightly different starting locations and different histories. 
This may have been theoretically predicted.  In a statistical analysis of numerical simulations of cold flows, \citet{Goerdt_Burkert2013} find that when there are 4 or more cold flows into a halo, the odds become favorable for two of them to share a common plane to within a few degrees.

The picture here is somewhat different from the N-body+Hydro code models that form several cold thin sheets in a group \citet{Goerdt_Burkert2013}.
We are finding that by z=4 nearly all of the LG dwarf galaxies were in a plane of filaments slighly tilted to the M\,31$-$MW line and this was all part of the wall to the Local Void.
In fact, the overall appearance of the LG dwarf galaxies distribution at z=4 is that of a single closed filament or ring.  
Then, in the process of falling into and orbiting the nearby giant galaxies, the ring broke up and assempled into four separate but related planes. 
A useful, but not necessary, ingredient to forming a very thin sheet is having a giant galaxy in the plane.
The solution here for \textbf{Plane\,2}  shows that some planes could be temporary turnaround cusps from an initial filamentary set of galaxies.
In the end, the thinness of these planes reminds us of the thinness of the larger scale Local Sheet and other walls around voids which also typically have high axial ratios.
Given that the standard LCDM model is essentially scaleless, it should not be too surprising to find structures on group scales as thin as those on a higher scale.
It is encouraging, to see NAM type calculations succeeding in producing early time filaments and sheets with such high axial ratios.  

Most of the satellites in the Local Group  appear to have been just 100-200 kpc higher in SGZ at  z=4 than the MW and M\,31 but still within the young Local Sheet.  
It seems reasonable to assume that the center of mass of the MW and M\,31 defined the center of the sheet and absorbed structures from both above and below them in nearly equal amounts.
This would imply that nearly all of the satellites that were at or below the two giant galaxies have been consumed, and the existing satellites are the rare ones that managed to avoid that fate, so far. 
These last surviving satellite galaxies had special histories that allowed them to avoid the destruction that met hundreds of extinguished dwarf galaxies. 
One survival mechanism found in these solutions is to start at the far edges of the LG's portion of the Local Sheet, at the divide between the LG and other groups, where gravitational balancing slowed progress inward.
It also appears that the tidal fields of the Virgo Cluster and the MW played roles in keeping  many \textbf{Plane\,1}  and \textbf{Plane\,2}  galaxies at bay from M\,31 until recently.
In our calculations, the early time LG distribution consisted of just 2 or 3 filaments, 
but there may have been others, at smaller radii from the giant galaxies, that were cannibalized.

A video of the motions of the LG galaxies is available at our data site (http://www.astro.umd.edu/$\sim$eshaya/LG.  
The degree of complexity is high and it takes time to understand all of the dynamical processes at play in the development of the planes.  
One notes, when watching the video, that the final structures developed rapidly in the last  Gyr or so. 
However, at each epoch there were other ``Great Planes" of satellites in other parts of the LG.  
We are led to the impression that it is not that a few planes of galaxies persisted throughout time, but rather that sets of galaxies went in and out of phase with each other at various places and times.  
Some galaxies in \textbf{Plane\,1}  did just move within a single plane, but  a few executed strongly precessing orbital loops in which their semi-major axes dropped strongly and their motions came into alignment only in their last orbit.  

Perhaps the `Missing Satellite Problem' and the `Excess Planar Structure Problem', the absence of planes reported in N-body simulations, are highly related. 
If the galaxies are constrained to very thin planes that preferentially go near the centers of giant galaxies, then satellites will be queued up, in lines of order 10 kpc thickness, to be very efficiently swallowed by their host galaxies.  
The LCDM N-body simulations could well be correct about the number of subhalos created, but highly underestimating the rate at which they are destroyed by their hosts.
 
\section{Conclusions}
\label{sec:Conclusions}

We have found orbits for 57 galaxies in the Local Group, including the galaxies of the NGC\,3109 subgroup,
consistent with the time dependent gravitational forces from the MW, M\,31, M\,33 plus galaxies,
groups and clusters on scales out to 50 Mpc. Concordant proper motions are found for those
with precision observed proper motions: M\,31, M\,33, IC\,10, and Leo\,I. From these trajectories, possible
non-dissipational explanations are beginning to emerge for four large planar structures of Local
Group dwarf galaxies.
Near M\,31 there are two planes of galaxies, both with rms thicknesses of about 13 kpc.  The first plane, described by \citet{Ibata_etal2013}, includes M\,31  and 14 dwarf galaxies.  It extends 600 kpc between the two most separated galaxies.  It would also include the MW, if extended that far.  
\textbf{Plane 2} includes 9 members and is offset by 100 kpc from M\,31.  Its maximum extent is 460 kpc.
For the MW, we again see two thin planes, an inner one goes out to about 0.4 Mpc (\textbf{Plane 4}) and \textbf{Plane\,3} goes out to $\sim$1 Mpc. 
Both include the MW, within the errors. 
\textbf{Plane\,3} has 7 members and is a bit wider than the other planes, 34 kpc, but it is also longer with maximum extent of 1.42 Mpc.
\textbf{Plane\,4} has 8 members, is about 12 kpc thick and there are 600 kpc between the two most separated galaxies.
The normals to the MW planes differ in SGL value by about $20\degree$, but the SGB values differ by $60\degree$, so the smallest axes
 are not at all parallel, nevertheless the longest axes of the two distributions  are nearly parallel to each other and to the LSC motion
 of the MW.

 In our delineation of the four planes, 41 of the 50 galaxies within 1.1 Mpc, including
the three major ones, are in these planes and can reasonably be assumed to have been physically
related to them since early epochs.  Of the nine not in planes, IC\,10 and, at larger distance, And\,27 are located in \textbf{Plane\,1},
but the proper motion of the former and the redshift of the latter indicate, in the context of our modeling,
that they are not physically related to \textbf{Plane\,1}.

We found two trajectories for M\,33 with concordant proper motions and chose one of these for the analysis presented here.  
After settling on the motions for M\,31, M\,33 and the MW, we used orbit integrations from the present back to z = 4 for the much lighter remaining LG galaxies.   
Solutions with many tight orbits around a giant galaxy were passed over in favor of simpler ones.  
The fact that the satellites lie in planes today provided sufficient guidance to select between multiple solutions in most cases.  By looking for commonality in the proper motions of galaxies within a plane, such as motion within the plane in the appropriate frame of rest, a best choice out of the few simple solutions was nearly always clear.   

Nearly all of the dwarf galaxies began, relative to the MW and M\,31, in the upper (+SGZ) part of the sheet of material that forms the wall of the Local Void.
The dwarf galaxy region was roughly 1 Mpc x 400 kpc x 80 kpc in physical coordinates at z=4, stretching  
up to 300 kpc  trailing  M\,31 and the MW center of masses in its LSC motion.  
This may indicate that a lower SGZ section of the wall was completely absorbed by M\,31 and the MW and all remaining dwarfs came from the upper edge of an earlier wall.

For \textbf{Plane\,1} galaxies, the blueshifted ones and some of the redshifted ones  have spent a Hubble time in the plane. 
As they fell toward M\,31, their distribution was gravitationally compressed and their motions were focused into the direction of motion of M\,31 which, in the LG frame, is falling directly into the MW.  
As M\,31 is accelerated toward the MW, dwarf galaxies fall toward where it used to be and constantly realign toward where it is;  thus they line up with the same direction of motion as M\,31. 
Other galaxies shifted by about 100 - 200 kpc in the SGZ direction by making 1.5 or 2.5 orbits that were compressed into the M\,31-MW line by the tidal field of the MW and, to a lesser extent, the entire Local Supercluster.
  
The \textbf{Plane\,2} galaxies began at about 300 kpc from M\,31 in  the SGX-SGZ projection.
We have solutions in which some executed one pass by M\,31 and some made two. 
But, all have, in the last Gyr, made a last pass which sent them to negative SGZ through points quite near to M\,31.
They are all close to their turnaround in their SGZ component and this results in a gathering at the SGZ location of the plane and a spreading in the other components.
The tidal field of the MW accentuates the spreading in the SGX direction.

The MW's companion galaxy distribution is partly explained by the fact that the z=4 distribution was quite narrow in SGX.
This may have been because the MW sits between M\,31 and the Cen\,A Group, and Cen\,A was nearly as close as M\,31 at z $>$ 1.  
Also, the distributions could have resulted from the tidal field: particularly the elongation in SGY could be due to the influence of the Virgo Cluster, which must have been quite strong at high z.

The procedure used here is guaranteed to produce the present day structures correctly and so one may well ask what was proven.  
We cannot, by our methods, show that such structures are common or expected. 
Rather, we have shown that there exists solutions within the standard model that have reasonable initial conditions, i.e. a sheet of material coming out of a void that broke up into filaments and then formed the interesting structures of the Local Group.   

\section*{Acknowledgements} 
We would like to acknowledge many years of discussions, enlightenment and friendship that we received from Prof. P.J.E. Peebles without which this work would not have been possible.
This research has been undertaken with support from the NASA Astrophysics Data Analysis Program award NNX12AE70G and from a series of awards from the Space Telescope Science Institute, most recently associated with programs AR-11285, GO-11584, and GO-12546.

\bibliographystyle{mn2e}

\begin{thebibliography}{}
\bibitem[Aaronson et al. (1982)]{Aaronson_etal1982} Aaronson, M, Huchra, J, Mould, J.R, Schechter, P.L. \& Tully, R.B. 1982, \apj, 258, 64
\bibitem[Bekki (2008)]{Bekki2008}{Bekki, K. 2008, MNRAS, 390, L24}
\bibitem[Bekki and Chiba (2008)]{BekkiChiba2008}Bekki, K. \& Chiba, M. 2008, \apjl, 679, L89–L92
\bibitem[Bonanos et al. (2006)]{Bonanos_etal2006}Bonanos et al. 2006, ApJ, 652, 313
\bibitem[Branchini \& Carlberg (1994)]{BranchiniCarlberg1994} Branchini, E. \& Carlberg, R.G. 1994, \apj, 434, 37
\bibitem[Branchini et al. (2002)]{Branchini_etal2002}Branchini, E., Eldar, A., Nusser, A. 2002,   MNRAS, 335, 53
\bibitem[Braun \& Thilker (2004)]{BraunThilker2004} Braun, R., \& Thilker, D. A. 2004, \aap, 417, 421
\bibitem[Brunthaler et al. (2005)]{Brunthaler_etal2005} Brunthaler, A., Reid, M. J., Falcke, H., Greenhill, L. J., Henkel, C. 2005,  Science, 307, 1440
\bibitem[Brunthaler et al. (2007)]{Brunthaler_etal2007} Brunthaler, A., Reid, M.~J., Falcke, H., Henkel, C. \& Menten, K.~M. 2007, \aap, 462, 101
\bibitem[Cheng et al. (2012)]{Cheng_etal2012} Cheng, L., Guhathakurta, P., Kirby, E., Yang, L: SPLASH collaboration 2012, AAS, 219, 244.18 
\bibitem[Conn et al. (2012)]{Conn_etal2012} Conn, A.R. et al. 2012, \apj, 758, 11
\bibitem[Conn et al. (2013)]{Conn_etal2013} Conn, A.R. et al. 2013, \apj, 766, 120
\bibitem[Danovich et al. (2012)]{Danovich_etal2012} Danovich, M.,  Dekel, A., Hahn, O., Teyssier, R., 2012, \mnras, 422, 1732
\bibitem[Dalcanton et al. (2009)]{Dalcanton_etal2009}Dalcanton et al. 2009, ApJS, 183, 67
\bibitem[Dolphin et al. (2002)]{Dolphin_etal2002}Dolphin et al. 2002, MNRAS, 332, 91
\bibitem[Dolphin et al. (2003)]{Dolphin_etal2003}Dolphin et al. 2003, AJ, 125, 1261
\bibitem[Dressler et al. (1987)]{Dressler_etal1987} Dressler, A., Faber, S.M., Burstein, D., Davies, R.L., Lynden-Bell, D., Terlevich, R.J., \& Wegner, G. 1987, \apj, 313, L37
\bibitem[Freedman et al. (2001)]{Freedman_etal2001}Freedman et al. 2001, ApJ, 553, 47
\bibitem[Goerdt \& Burkert (2013)]{Goerdt_Burkert2013}Goerdt, T. \& Burkert, A. 2013, eprint arXiv:1307.2102
\bibitem[Huchtmeier et al. (1981)]{Huchtmeier_etal1981}Huchtmeier, W. K., Seiradakis, J. H. \&  Materne, J. 1981, \aap, 102, 134
\bibitem[Ibata et al. (2013)]{Ibata_etal2013} Ibata, R. et al. 2013, Nature 493, 62
\bibitem[Ibata et al. (2001)]{Ibata_etal2001} Ibata, R., Irwin, M., Lewis, G., Ferguson, A., Tanvir, N., Nature 412, 49
\bibitem[Jacobs et al. (2009)]{Jacobs_etal2009}Jacobs et al. 2009, AJ, 138, 332
\bibitem[Jarrett et al. (2003)]{LGA} Jarrett, T.H., Chester, T., Cutri, R., Schneider, S.E., Huchra, J.P. 2003, \aj, 125, 525
\bibitem[Jobin \& Carignan (1990)]{JobinCarignan1990} Jobin, M. \& Carignan, C. 1990, \aj, 100, 648
\bibitem[Karachentsev et al. (2013)]{Karachentsev_etal2013} Karachentsev, I.D., Makarov, D.I., \& Kaisina, E.I. 2013, \aj, 145, 101
\bibitem[Klypin et al. (1999)]{Klypin_etal1999} Klypin. A., Kravtsov, A.V., Valenzuela, O., \& Prada, F. 1999, \apj, 522, 82
\bibitem[Kunkel \& Demers (1976)]{Kunkel_Demers_1976} Kunkel W. E. \& Demers S., 1976, R. Greenwhich Obs. Bull., 182, 241
\bibitem[Lewis et al (2013)]{Lewis_etal2013} Lewis, G., Braun, R., McConnachie, A., Irwin, M., Ibata, R., Chapman, S., Ferguson, A., Martin. N., Fardal. M., Dubinski, J., Widrow, L., Mackey. A.D., Babul, A., Tanvir, N., and Rich, M. 2013, \apj, 763, 1
\bibitem[Libeskind et al. (2011)]{Libeskind_etal2011}Libeskind, N., Knebe, A., Hoffman, Y., Gottl¨ober, S., Yepes, G \& Steinmetz, M. 2011, \mnras, 411, 1525
\bibitem[Lynden-Bell (1976)]{LyndenBell1976} Lynden-Bell, D. 1976, \mnras, 174, 695 
\bibitem[Makarov et al. (2006)]{Makarov_etal2006} Makarov, D., Makarova, L., Rizzi, L., Tully, R. B., Dolphin, Andrew E., Sakai, S. \& Shaya, E. 2006, \aj, 132, 2729
\bibitem[McConnachie et al. (2009)]{McConnachie_etal2009}McConnachie, A., Irwin, M. J.,\& Ibata, R. A. et al. 2009, Nature, 461, 66
\bibitem[Metz et al. (2007)]{Metz_etal2007} Metz, M., Kroupa, P., \& Jergen, H. 2007, \mnras, 374, 1125
\bibitem[Moore et al. (1999)]{Moore_etal1999} Moore, B. et al. 1999, \apj, 524, L19	
\bibitem[Nusser \& Branchini (2000)]{NusserBranchini2000}Nusser, A. \& Branchini, E. 2000, \mnras, 313, 587
\bibitem[Pawlowski et al. (2012)]{Pawlowski_etal2012} Pawlowski, M.S, Pflamm-Altenburg, J. \& Kroupa, P. 2012, \mnras, 423, 1109
\bibitem[Peebles (1989)]{Peebles1989} Peebles, P. J. E. 1989, \apj, 344, L53
\bibitem[Peebles (1990)]{Peebles1990} Peebles, P. J. E. 1990, \apj, 362, 1
\bibitem[Peebles (1994)]{Peebles1994} Peebles, P. J. E. 1994, \apj, 429, 43
\bibitem[Peebles (1995)]{Peebles1995} Peebles, P. J. E. 1995, \apj, 449, 52
\bibitem[Peebles (1996)]{Peebles1996} Peebles, P. J. E. 1996, \apj, 473, 42
\bibitem[Peebles \& Tully (2013)]{PeeblesTully2013} Peebles, P.J.E., \& Tully, R.B. 2013, arXiv:1302.6982
\bibitem[Peebles et al. (2001)]{Peebles_etal2001} Peebles, P.J.E., Phelps, S.D., Shaya, E.J., \& Tully, R.B. 2001, \apj, 554, 104
\bibitem[Phelps (2002)]{Phelps2002} Phelps, S.D. 2002, \apj, 575 ,1
\bibitem[Phelps et al. (2006)]{Phelps_etal2006} Phelps, S.D., Desjacques, V., Nusser, A. \& Shaya, E. 2006, \mnras 370, 1361
\bibitem[Pietrzynski et al. (2009)]{Pietrzynski_etal2009}Pietrzynski et al. 2009, AJ, 138, 459
\bibitem[Richardson et al. (2011)]{Richardson_etal2011}Richardson et al. 2011, ApJ, 732, 76 
\bibitem[Putman et al. (2009)]{Putnam_etal2009}Putman, M. E., Peek, J. E. G., \& Muratov, A. et al. 2009, ApJ, 703, 1486
\bibitem[Skrutskie et al. (2006)]{XSC}Skrutskie M.F. et al. 2006, \aj 131, 1163
\bibitem[Sharpe et al. (2001)]{Sharpe_etal2001}	Sharpe, J., Rowan-Robinson, M., Canavezes, A., Saunders, W., Branchini, E., Efstathiou, G., Frenk, C., Keeble, O., McMahon, R. G., Maddox, S,; Oliver, S. J., Sutherland, W., Tadros, H., White, S. D. M. 2001, MNRAS, 322, 121
\bibitem[Shaya (1984)]{Shaya1984} Shaya, E.J. 1984, \apj, 280, 470
\bibitem[Shaya, Peebles, \& Tully (1995)]{Shaya_etal1995} Shaya, E. J., Peebles, P. J. E., \& Tully, R. B. 1995, \apj, 454, 15
\bibitem[Sohn et al. (2013)]{Sohn_etal2013}Sohn, S. T., Besla, G., van der Marel, R. P., Boylan-Kolchin, M.; Majewski, S. R. \& Bullock, J. S. 2013, \apj, 768,139S
\bibitem[Tully (2013)]{Tully2013} Tully, R. B. 2013, Nature 493, 31
\bibitem[Tully et al. (2008)]{Tully2008}Tully et al. 2008, ApJ, 676, 184 
\bibitem[Tully et al. (2013)]{Tully+2013}Tully et al. 2013, AJ, in press (arXiv:1307.7213)
\bibitem[Tully \& Fisher (1987)]{TullyFisher1987} Tully, R.B. \& Fisher 1987, J.R. Nearby Galaxies Atlas (Cambridge University Press)
\bibitem[Sohn et al. (2012)]{Sohn etal 2012} Sohn, S., Anderson, J, \& van der Marel, R. 2012. \apj, 753, 7
\bibitem[van der Marel et al. (2012)]{vdM_etal2012} van der Marel, R.., Besla, G., Cox, T. , Sohn, S., \& Anderson, 2012 \apj, 753, 9
\bibitem[Wolfe et al. (2013)]{Wolfe_etal2013}Wolfe, S., Pisano, D., Lockman, F., McGaugh, S., \& Shaya, E. 2013 accepted to Nature
\bibitem[Zhao (2004)]{Zhao2004}Zhao, H. 2004, \mnras, 351, 891



\end{thebibliography}

\end{twocolumn}
\begin{onecolumn}
\appendix
\numberwithin{equation}{section}
\section{More on Numerical Action}
\subsection{Equation of motion}
In a cosmologically flat universe the expansion parameter satisfies
\begin{equation}
\frac{\dot{a}^2}{a^2} = \frac{H_0^2\Omega_m}{a^3} + (1-\Omega_m)H_0^2, 
\quad\frac{\ddot{a}}{a} = \frac{H_0^2\Omega_m}{2a^3} + (1-\Omega_m)H_0^2,
\end{equation}

with present value $a_o$ = 1. The equations of motion for the paths $\vec{r_i}$ of $i$ mass tracers in physical length units are
\begin{equation}
\frac{d^2 r_{i,k}}{dt^2} = \sum_{j \neq i} \frac{Gm_j(r_{j,k}-r_{i,k})}{|\vec{r}_i-\vec{r}_j|^3} + (1-\Omega_m)H_0^2 r_{i,k}
\end{equation}
Changing variables to the comoving coordinates $x_{i,k} = r_{i,k}/a(t)$ used here brings eq. (A.2) to
\begin{equation}
\frac{d}{dt}a^2\frac{dx_{i,k}}{dt} =  
\frac{1}{a}\left[\sum_j Gm_j\frac{(x_{j,k} - x_{i,k})}{|\vec{x}_i - \vec{x}_j|^3}  + \frac{1}{2}\Omega_m H^2_o x_{i,k} \right]
\end{equation}
This is derived from the action
\begin{equation}
S   = \int_0^{t_0} dt 
 \left[ \frac{1}{2} \sum_i m_i a^2 \dot{x_i}^2 + \frac{1}{a}\left(\sum_{j \neq i} \frac{Gm_im_j}{|\vec{x}_i - \vec{x}_j|} 
 + \frac{1}{4}\sum_i m_i \Omega_m H_0^2 x_i^2\right) \right]
\end{equation}

when the present positions are  $\delta x_i(t_0) = 0$, initial conditions satisfy

\begin{equation}
a^2\dot{x}_i \rightarrow 0 ~\mathrm{at}~ a(t) \rightarrow 0.
\end{equation}

Physical trajectories occur at the extrema of S or where $\delta S = 0$ for small variations of the paths of mass tracers.  The Numerical Action technique is to begin with all paths taking random walks and make changes in the paths that reduce the derivatives of S with respect to coordinates of positions at discrete times until they are sufficiently close to 0.

\subsection{Discrete representation}
In a discrete representation the coordinates are $x_{i,k,n}$ where i labels the particles, k = 1, 2, 3 the Cartesian coordinates, and $1 \leq n \leq n_x+1$ the time steps. The present positions $x_{i,k,nx+1}$ are fixed and given. The relevant derivatives of the action are
\begin{align}
S_{i,k,n} &= \frac{\partial S }{\partial x_{i,k,n}},\quad \notag\\
S_{i,k,n;j,k^\prime,n^\prime} &= \frac{\partial^2 S}{\partial x_{i,k,n}\partial x_{j,k\prime,n^\prime}},\quad
1 \leq n,n^\prime \leq n_x.
\end{align}
If S is close to quadratic in the $x_{i,k,n}$ then position shifts $\delta x_{i,k,n}$ to a solution at an extremum of S satisfy
\begin{equation}
S_{i,k,n} + \sum_{j,k^\prime,n^\prime} S_{i,k,n;j,k^\prime,n^\prime} \delta {x_j,k^\prime,n^\prime} = 0
\end{equation}
If the $x_{i,k,n}$ are not close to a solution, S will not be close to quadratic in the $x_{i,k,n}$, but experience
shows that these coordinate shifts walk toward a solution.  
However, where the first derivative is positive and the second derivative changes from positive to negative or where the first derivative is negative and the second derivative changes the other way, it is possible to get stuck at the point where the second derivative changes sign.  A simpler walk up or down the gradient is advised in these circumstances.

Approximate the action (Eq. A.4) integral discretely as
\begin{align}
S & = \sum_{i,k,n=1,n_x} \frac{m_i}{2} \frac{(x_{i,k,n+1} - x_{i,k,n})^2}{a_{n+1}-a_n} \dot{a}_{n+\sfrac{1}{2}}a^2_{n+\half} \notag\\
& + 
\sum_{i,j,n=1,n_x}\frac{t_{n+\half} - t_{n-\half}}{a_n}
\left[ \sum_{j < i} \frac{Gm_im_j}{|\vec{x}_{i,n}-\vec{x}_{j,n}|} 
+ \frac{1}{4} \sum_i m_i \Omega_m H_0^2 x^2_{i,n} \right]
\end{align} 
The times $t_{n \pm \half}$ interpolate between the time steps at $n$ and $n \pm 1$ in leapfrog fashion. 
The approximation to the kinetic energy in (Eq A.8) is motivated by linear perturbation theory,
where $dx=da$ is nearly independent of time, so $\frac{x_{n+1}-x_n}{a_{n+1}-a_n}$ is a good approximation to $dx/da$ at $a_{n+\half}$. 
The earliest time at which positions are computed is at $a_1 > 0$. 
The leapfrog back in time from $a_1$ is to $a_{\half} = 0 = t_{\half} $. 
Present positions at $a_{n_x+1} = 1$ are given as $x_{i,k,n_x+1}$.
The derivative of the action with respect to the coordinates $x_{i,k,n}$ for $1 \leq n \leq n_x$, gives

\begin{align}
S_{i,k,n} 
& = -\frac{a^2_{n+\half} \dot{a}_{n+\half}}{a_{n+1} - a_n}(x_{i,k,n+1}-x_{i,k,n}) + \frac{a^2_{n-\half} \dot{a}_{n-\half}}
{a_{n} - a_{n-1}}(x_{i,k,n}-x_{i,k,n-1}) \notag\\
& + \frac{t_{n+\half}-t_{n-\half}}{a_n}
\left[\sum_{j \neq i} G m_j \frac{x_{j,k,n}-x_{i,k,n}}{|\vec{x}_{i,n}-\vec{x}_{j,n}|^3}
 + \frac{1}{2}\Omega_m H^2_o x_{i,k,n} \right] 
\end{align}
When the first order derivatives $S_{i,k,n} = 0$, this is a discrete approximation to the equation of motion (Eq. A.3)

\subsection{The First Time Step and Backtracking}  
In linear perturbation theory for a continuous pressureless 
fluid the unwanted decaying mode
has peculiar velocity that decays as $v = a\dot{x} \propto \frac{1}{a(t)}$, and in the wanted growing mode the
coordinate position of a particle is changing with time as
\begin{equation}
x(t) - x(0) \propto t^\frac{2}{3} \propto a(t)
\end{equation}
This implies that in the growing mode, the left hand side of eq. (A.3) is constant, meaning $a^2 \frac{dx_{i,k}}{dt} \simeq$ constant $\times~ t$, where $t$ is the time measured from a = 0. 
Following this, we choose the positions $x_{i,k,1}$ and $x_{i,k,2}$ at the first two time steps so the rate of change of position satisfies
\begin{equation}
a^2_{\threehalves}\frac{dx_{i,k,\threehalves}}{dt} 
\simeq 
a^2_{\threehalves} \dot{a}_{\threehalves} \frac{x_{i,k,n}-x_{i,k,1}}{a_2-a_1}
=
 \frac{t_{\threehalves}}{a_1} 
\left[
 Gm_j\frac{(x_{j,k,1} - x_{i,k,1})}{|\vec{x}_{i,1} - \vec{x}_{j,1}|^3}  + \frac{1}{2}\Omega_m H^2_o x_{i,k,1} 
\right]
\end{equation}
Recall the half time step earlier than $a_1$ is at $a_{\half} = 0.$
It will be noted that $a_1$ may be much larger than $a_2 - a_1$, meaning the numerical solution
commences at modest redshift with small time steps. But $t_{\threehalves}$ is still the time from $a_{\half} = 0$
to the time midway between $a_1$ and $a_2$.
Eq. A.11 is represented in terms of the derivative of the action as 
\begin{equation}
S_{i,k,1} = -\frac{a^2_{n+\half} \dot{a}_{n+\half}}{a_{n+1} - a_n}(x_{i,k,n+1}-x_{i,k,n})
+ \frac{t_{\threehalves}}{a_1} \left[\sum_{j \neq i} g_{i,k,1,;j} + \frac{1}{2} \Omega_m H^2_o x_{i,k,1} \right]
\end{equation}
To backtrack a galaxy, we use the present velocity and position to obtain $x_{i,k,n_x+1}$ and $x_{i,k,n_x}$ and get all of the other positions by setting the action derivatives to 0 and solve Equation A.9 for $x_{i,k,n-1}$ in terms of $x_{i,k,n}$ and $x_{i,k,n+1}$.  
However, at the first time step one is left with a non-zero value for $S_{i,k,1}$ because there is no $x_{i,k,n-1}$ term to adjust in A.12.
This will generally mean there is a mismatch between velocity and the potential field at that time.  
We therefore vary the present velocity direction holding the redshift and proper motion constant and search for an acceptably low value of the sum of the square of the three components of $S_{i,k,1}$.

The first term on the r.h.s. of Equation A.12 is close to $a^2_{\threehalves}\frac{dx_{i,k,\threehalves}}{dt}$, 
therefore the sum of the square of the derivatives at the first time step is effectively the error in $(a_{\threehalves} |\vec{v}|)^2$ assuming the potential is accurate.  
Thus an error of 25 \kms\ in the velocity at z=4 corresponds to a $2.5\times10^{-3}$ value in this parameter (NAM code uses velocity units of 100 \kms). Errors up to about this amount are considered acceptable given the number of unknowns for the potential field at z=4.

\end{onecolumn}
\end{document}